\documentclass[article,11pt]{amsart}
\usepackage{amsaddr}
\usepackage{amssymb, amsthm, amsfonts, amsgen, stmaryrd}
\usepackage{slashed}
\usepackage[english]{babel}
\usepackage{fullpage}
\usepackage[centertags]{amsmath}
\usepackage{indentfirst}
\usepackage{fancyhdr}
\usepackage[dvips]{graphicx}
\usepackage{psfrag}
\usepackage{enumerate}
\numberwithin{equation}{section}
\usepackage[latin1]{inputenc}

\usepackage{hyperref}
\hypersetup{
pdftitle={},%
pdfauthor={},%
pdfsubject={},%
pdfkeywords={},%
colorlinks=true,%
linkcolor=blue,%
citecolor=red,%
linktocpage=true,%
hyperfootnotes=true,%
pageanchor=true
}
\usepackage{color}

\newcommand{\blu}[1]{ \textcolor{blue}{{#1}}}

\title{Symmetries and reduction \\ Part I --  Poisson and symplectic picture } 
\date{revised 11 august 2021}
\author{Giuseppe Marmo} 
\address{Dipartimento di Fisica ``E. Pancini'', Universit\`a di Napoli Federico II; \\
INFN - Sezione di Napoli, \\ Via Cintia - 80126 Napoli, Italy}
\email{marmo@na.infn.it}
\author{Luca Schiavone}
 \address{Dipartimento di Matematica ed Applicazioni ``R. Caccioppoli'', Universit\`a di Napoli Federico II \\ 
Via Cintia - 80126 Napoli, Italy; \\ Departamento de Matem\'aticas, Univ. Carlos III de Madrid \\ Av.da de la Universidad 30 -- 28911 Legan\'es, Madrid, Spain}
\email{luca.schiavone@unina.it}
\author{Alessandro Zampini}
 \address{Dipartimento di Matematica ed Applicazioni ``R. Caccioppoli'', Universit\`a di Napoli Federico II; \\ 
 INFN - Sezione di Napoli \\ Via Cintia - 80126 Napoli, Italy } 
 \email{alessandro.zampini@unina.it}

\newtheorem{theo}{Theorem}[section]

\newtheorem{prop}[theo]{Proposition}

\newtheorem{example}{Example}[section]

\newcommand{\nn}{\nonumber}

\newcommand{\dd}{{\rm d}}

\newcommand{\figureheight}{8cm}
\newcommand{\putfig}[2]{\begin{figure}[htp]
        \special{isoscale c:/itex/texfig/#1.wmf, \the\hsize \figureheight}
        \vspace{\figureheight}
        \caption{#2}\label{fig:#1}
        \end{figure}}
\newcommand{\pictureheight}{4cm}
\newcommand{\putpicture}[2]{\begin{figure}[htp]
        \special{isoscale c:/itex/texfig/#1.wmf, \the\hsize \pictureheight}
        \vspace{\pictureheight}
        \caption{#2}\label{fig:#1}
        \end{figure}}

\newcommand{\beqa}{\begin{eqnarray}}
\newcommand{\eeqa}{\end{eqnarray}}
\newcommand{\beq}{\begin{equation}}
\newcommand{\eeq}{\end{equation}}

\newcommand{\del}{\partial}

\newcommand{\R}{{\mathbb{R}}}

\newcommand{\N}{{\mathbb{N}}}
\newcommand{\C}{\mathbb{C}}

\newcommand{\D}{\mathcal{D}}

\newcommand{\g}{\mathfrak{g}}

\begin{document}

\thispagestyle{empty}

\begin{abstract}
Coherently with the principle of analogy suggested by Dirac, we describe a general setting for reducing a classical dynamics, and the role of the Noether theorem -- connecting symmetries with constants of the motion -- within a reduction. This is the first of two papers, and it focuses on the reduction within the Poisson and the symplectic formalism. 
\end{abstract}


\maketitle
\tableofcontents

\section{Introduction}
\label{sec:intro}

Symmetries and invariants for differential equations have been  already considered  by  F. Klein, S. Lie and E. Noether, both in mathematics (for classification purposes) and physics (in terms of conservation laws). The interest for explicit solutions to evolutionary physical equations (mostly described by non-linear equations) induced a widespread interest for completely integrable systems and the generation of infinite  conservation laws in field theories. To this aim, a consistent theory of bi-Hamiltonian systems was developed, reading  the introduction  of higher order invariant geometrical structures, namely Poisson Brackets, Nambu Brackets and  Moyal Brackets among the others\footnote{see \cite{jbrackets} for a more complete account}. 
When it was  realised that a more coherent description of the external physical world is provided indeed by a quantum theory, Dirac suggested that the classical description of a dynamics should emerge as a suitable limit of the quantum one. As the limiting procedure is yet not rigorously defined, Dirac proposed a \emph{principle of analogy}. In his own words, 

\emph{...the value of classical analogy in the development of quantum mechanics depends on the fact that classical mechanics provides a valid description of dynamical systems under certain conditions, when the particles and bodies composing the systems are sufficiently massive for the disturbance accompanying an observation to be negligible. Classical mechanics must therefore be a limiting case of quantum mechanics. We should thus expect to find that important concepts in classical mechanics correspond to important concepts in quantum mechanics ...}

Within the quantum formalism, commutation relations among observables play a prominent role, both to define the  equations of motion (i.e. the Heisenberg picture) and to introduce uncertainty  relations.
Adopting  Dirac's point of view, it seems quite natural to deal with classical description of dynamical systems from the point of view of Poisson manifolds, Poisson brackets being the analogue of the quantum commutation relations. As the Hilbert space describing the states of a quantum dynamics can be split into  the sum of subspaces carrying an irreducible representation of the canonical commutation relations, so one can see that 
Poisson manifolds are given by the union of symplectic leaves (each one  generated as an integral manifold of Hamiltonian vector fields), i.e. even dimensional manifolds on which the restriction of the Poisson tensor is non degenerate. 
Those symplectic manifolds where the symplectic 2-form has a potential 1-form are called exact symplectic, and they are locally diffeomorphic to a cotangent bundle manifold $T^*Q$ on a configuration space manifold $Q$ associated to a suitable Lagrangian distribution. 
This is in analogy with selecting a maximal set of  commuting observables for a quantum system, each with a continuous spectrum, and representing states as square integrable functions on such a spectrum. Unitary transformations map maximal commutative subalgebras of selfadjoint  operators into another one, isospectrally:  this would classically correspond to a linear symplectic transformation. Among such symplectic transformation, there are those which changes the cotangent bundle structure: an example is given by  the Fourier transform within the quantum formalism, mapping the canonical pair of position and momentum operators $(\hat q,\hat p)\mapsto(\hat p,-\hat q)$, analogous to  the linear symplectic map $(q,p)\mapsto(p,-q)$ on $\R^{2N}$ which results in an alternative cotangent bundle structure. 
Once a cotangent bundle structure has been selected, its dual vector bundle is the tangent bundle $TQ$,  where the Lagrangian and the Newtonian formalisms are  developed, in terms of second order ordinary differential equations.  Alternative cotangent bundle structures result in alternative Lagrangian structures and then alternative tangent bundle structures. 
 In this way we walk in the opposite direction the path  usually followed in text-books, thus  coherently with the analogy principle.

To better unfold the flow of ideas involved, we shall focus on geometrical aspects and refer to the existing literature for analytical results and related theorems.
We shall be using mostly the intrinsic language of differential geometry, having in mind that this approach would pave the way to a transition to infinite degrees of freedom more easily. 
The first part of this paper will deal with Poisson manifolds and the associated symplectic leaves: general properties of the carrier manifolds and of the Poisson algebra, Hamiltonian dynamical systems, symmetries, invariants and reduction.
The second part deals with special symplectic manifolds which are cotangent bundles. The Lagrangian formalism will be considered on the dual tangent bundle. When the Lagrangian function is regular, the Euler-Lagrange equations is  second order and we recover Newton equations.
On both vector bundles we can deal with Hamilton-Jacobi theory. We recall Dirac views of this theory when considering trajectories as solution of the equations of motion,

\emph{...to group the solutions into families (each family corresponding to one principal function satisfying the Hamilton-Jacobi equation). The family does not have any importance from the point of view of Newtonian mechanics; but it is a family which corresponds to one state of motion in the quantum theory, so presumably the family has some deep significance in nature, not yet properly understood.}

In this setting we shall again consider the description of dynamical systems, symmetries and conservation laws, along with the reduction procedure.
Few appendices close the first part, mostly to fix notation.

\section{Dynamical Systems}
\label{sec:uno}
A law of change of the states of any physical system is usually provided by means of a differential equation or of a difference equation. It was first formulated  within Newtonian mechanics in terms of a second order differential equations. A possible way to coherently predict the evolved (in time) state out of an initial one is to abstractly describe the set of states as points of a (smooth) manifold, and the evolution by a system of first order ordinary differential equation on it. When such equations are explicit, they are  described in terms of a 
 (smooth) vector field. In what follows we shall make full use of the formalism of differential geometry, and refer the reader to some of the most used textbooks \cite{AM78, AMP82, gfd, mks_natural,michor-book}.  For the most common notions we have inserted few appendices. 

\subsection{Reduction}
\label{ss:red}
Consider a  dynamical system described by the vector field $\Gamma$ over the smooth ($N$-dimensional, orientable) manifold $M$. The associated o.d.e. system is\footnote{By o.d.e. we mean ordinary differential equation(s). The appendix \ref{app1} recalls the basic notions of exterior differential calculus on a manifold, as well as those of vector fields and maps between manifolds.} 
\beq
\label{equazione1}
\dot x_a\,=\,L_{\Gamma}x_a
\eeq
with respect to a local coordinate chart $\{x_a\}_{a=1,\dots,N}$ on $M$. 
One of the most exploited approaches to solve an o.d.e. system associated to the vector field $\Gamma$ is to analyse whether it is possible (see \cite{mss78,mss79, mssv85}) to \emph{reduce} it, that is whether its flow lies on a suitable \emph{invariant} submanifold $i:V\hookrightarrow M$ or whether it  projects onto a suitable \emph{quotient} manifold  $\pi\,:\,M\,\to\,M/\sim$.  A reduction provides indeed a set of differential equations involving a lower number of variables, whose solutions give partial informations on the dynamics  and enter the remaining equations as (time dependent, in general) parameters. Both an embedding and a quotient space may be described in algebraic terms, i.e. by suitable ideals and derivations of the commutative algebra $\mathcal F(M)$; it is often nonetheless the case, mostly for computational reasons, that one uses adapted local coordinate charts on $M$.

An abstract setting for the reduction of a given dynamical system is as follows. Let $F\,:\,M\,\to\,M'$ be a map between the manifold $M$ and a $k$-dimensional smooth manifold $M'$ with $N>k$. If a vector field $\Gamma_F$ on $M'$ exists, such that it is $F$-related to $\Gamma$, i.e. 
\beq
\label{equazione2}
F_*(\Gamma)=\Gamma_F,
\eeq 
then we say that $F$ \emph{reduces} $\Gamma$. If such a map $F$ exists for a given dynamics $\Gamma$, then we can define 
 \beq
 \label{equa2pp}
 \mathcal A_F\,=\,\{F^*(f')\,:\, f'\in\mathcal F(M')\},\eeq
which is a subalgebra (with respect to the pointwise associative product) in $\mathcal F(M)$  invariant (or stable) under the dynamics, namely $\Gamma(\varphi)\,\in\,\mathcal A_F$ for any $\varphi\in\mathcal A_F$. 
The time evolution of the elements in $\mathcal A_F$, given by the solutions of the differential equation $\dot\varphi=\Gamma(\varphi)$ clearly provides only a partial knowledge on the time evolution generated by $\Gamma$ on the whole $\mathcal F(M)$ algebra. 

Given the map $F$ we can also define  
\beq
\label{3g1}
\mathcal D_F\,=\,\ker\,F_*\,=\,\{Y\,\in\,\mathfrak X(M)\,:\,Y(f)=0\quad\forall \,f\,\in\,\mathcal A_F\}, 
\eeq
which is an infinite dimensional Lie subalgebra in $\mathfrak X(M)$.  It means that $\mathcal D_F$ is an involutive, i.e. (from the Frobenius theorem\footnote{The appendix \ref{app2} recalls the basic notions about distributions, immersions, submersions and the Frobenius theorem.}) an integrable distribution, with  $[\Gamma, Y]\,\in\,\mathcal D_F$ for any $Y\,\in\,\mathcal D_F$. The integral manifolds of the distribution $\mathcal D_F$ can be identified with the level sets of the map $F$, namely 
$$N_{m'}=\{m\in M:F(m)=m'\in M'\}.$$
The quotient space given by identifying points on $M$ belonging to the same $N_{m'}$ (that is the quotient of $M$ by the foliation $\Phi^F$ with leaves $N_{m'}$)  is described by the Gelfand spectrum of $\mathcal A_F$, which is locally homeomorphic to $M'$. The flow generated by $\Gamma$ turns out to map leaves of the foliation $\Phi^F$ into leaves.

To give a more definite example, assume that a  local submersion $F\,:\,M\,\to\,(\varphi_1,\dots,\varphi_{k})\subseteq M'=\R^k$ reduces the dynamics $\Gamma$ on $M$, that is one has 
\beq
\label{intro1}
L_{\Gamma}\varphi_j\,=\,f_j(\varphi)
\eeq where  $f_a(\varphi)$ denotes an arbitrary function of the $\varphi_j$.
The distribution $\mathcal D_F$ can be written as  
\beq
\label{equazione2p}
\mathcal D_{F}=\{Y\in\mathfrak X(M):i_Y\dd \varphi_j=0\}
\eeq
Under suitable regularity conditions on $F$, the quotient $M/\Phi^{F}$ coming upon identifying points on $M$ belonging to the same leaf $N_{m'}$ has a manifold structure and is (locally) diffeomorphic to $M'$. The algebra of functions on this quotient is given by 
\beq
\label{equa2p}
\mathcal F(M/\Phi^F)\,=\,\{f\,\in\,\mathcal F(M)\,:\,Y(f)=0\quad\forall \,Y\,\in\,\mathcal D_F\}\,\simeq\,\mathcal A_F.
\eeq
The vector field $\Gamma$ is projectable onto the quotient $M/\Phi^F$: the evolution infinitesimally generated by  $\Gamma_F$ describes the dynamics transversal to the leaves, the dynamics along each leaf is still to be determined, and may depend on a family of time dependent parameters associated to the transversal evolution. To clarify this,  consider 
an open subset $U\subseteq M$, and select a local coordinate system $\{\varphi_j, y_s\}_{s=1,\dots,N-k}$ \emph{adapted} to the submersion\footnote{Notice that an adapted coordinate system exists by the canonical submersion theorem.}: the $\varphi_j$ coordinates identify a leaf, the $y_s$ coordinates give a local chart on each leaf.   
The o.d.e. system \eqref{equazione1} can now be written as 
\begin{align}
&\dot\varphi_j\,=\,f_j(\varphi), \label{equazione3} \\ 
&\dot y_s\,=\,L_{\Gamma}y_s.
\label{equazione4}
\end{align}
The equations \eqref{equazione3} describe the dynamics \emph{transversal} to the  leaves, their solutions give the integral curves of $\Gamma_F$ on $\R^k$. When such solutions $\varphi_j(t)$ are inserted into \eqref{equazione4}, they appear as time dependent parameters for the dynamics along each leaf $N_{m'}$ with $m'=(\varphi_1(t), \dots,\varphi_k(t))$. Of particular interest is the case when the $y_s$ coordinates can be defined in such a way that the conditions 
$$
f_j\frac{\del}{\del\varphi_j}(L_{\Gamma}y_s)\,=\,0
$$
hold\footnote{In this paper we assume the convention that repeated indices are summed over.}. Under such a condition, the vector field $\Gamma$ can be decomposed as the sum of two commuting vector fields\footnote{Such a splitting is not associated to $\Gamma$ intrinsically, it depends on the chosen system of local coordinates on $M$ adapted to $F$.}. The dynamics on $M$ can then be recovered upon \emph{composing} (in Newton's words) two independent motions. 

The general problem of determining, for a given vector field $\Gamma$ on $M$, a suitable manifold $M'$
and a map $F:M\to M'$ that reduces it is highly non trivial. Our previous analysis suggests two (to some extent equivalent) strategies to solve it. 

The first one consists in looking for a subalgebra (with respect to the associative pointwise product) $\mathcal A\subset\mathcal F(M)$ which is invariant under $\Gamma$, that is $\Gamma(\varphi)\in\mathcal A$ for any $\varphi\in\mathcal A$. 
The set of derivations $X$ of $\mathcal F(M)$ satisfying the condition $X(\varphi)=0$ for any $\varphi \in\mathcal A$ is a Lie algebra with respect to the natural commutator structure\footnote{As we shall recall also in section \ref{sec:poisson}, the Willmore's theorem proves that if $X$ is a derivation operator on $\mathcal F(M)$, then $X$ can be identified with a vector field on $M$, so that the commutator structure within the set of derivations for $\mathcal F(M)$ is equivalently given by the commutator of vector fields.}, thus providing an involutive distribution $\mathcal D_{\mathcal A}$. The Gelfand spectrum of the commutative algebra $\mathcal A$ defines $M'$ and corresponds to the elements of the quotient given by identifying the points in $M$ on the same integral submanifold of $\mathcal D_{\mathcal A}$.

When the distribution $\mathcal D_{\mathcal A}$ is regular, it gives a regular foliation $\Phi_{\mathcal A}$ whose leaves can be identified with submanifolds  of constant (say $N-k$) dimension. The map $F$ is then recovered as the submersion associated to the foliation $\Phi_{\mathcal A}$. The spectrum of the algebra $\mathcal A$ corresponds to the points of the quotient manifold $M/\Phi_{\mathcal A}$, so that one has $\mathcal A\simeq\mathcal F(M/\Phi_{\mathcal A})$, paralleling the identification between the algebras given in \eqref{equa2p} and \eqref{equa2pp}.

The second strategy origins by noticing that, given an involutive distribution  $\mathcal D$ on  $M$ (i.e. $[Y,Y']\in \mathcal D$ for any pair $Y,Y'\in\mathcal D$) one can define its \emph{normaliser} $$\mathrm N_{\mathcal D}=\{X\in\mathfrak X(M)\,:\,[X,Y]\in\mathcal D\quad\forall \,Y\,\in\, \mathcal D\},$$  so to have the short exact sequence of Lie modules
\beq
\label{intro1.3}
0\quad\rightarrow\quad\mathcal D\quad\rightarrow\quad {\rm N}_{\mathcal D}\quad\rightarrow\quad{\rm R}_{\mathcal D}\quad\rightarrow\quad 0:
\eeq
vector fields in ${\rm N}_{\mathcal D}$ can be reduced with respect to the distribution $\mathcal D$, the elements in ${\rm R}_{\mathcal D}$ 
give the equivalence classes of vector fields on $M$ whose projections onto the quotient manifold $M/\Phi_{\mathcal D}$ coincide. A dynamics $\Gamma$ can be reduced via the involutive distribution $\mathcal D$ if $\Gamma\in\rm N_{\mathcal D}$. In such a case, the set 
$$
\mathcal A_{\mathcal D}=\{f\in\mathcal F(M)\,:\,Y(f)=0\,\, \forall\,\,Y\, \in\,\mathcal D\}
$$
is an invariant subalgebra for $\Gamma$, whose Gelfand spectrum defines $M'$ and clearly coincides with the space given by indentifying  elements in $M$ on the same integral submanifold of $\mathcal D$. When $\mathcal D$ is regular, $M'$ has a manifold structure and 
the map $F$ is recovered as the submersion associated to the foliation generated by $\mathcal D$.

\subsection{Symmetries and constants of the motions}
\label{ss:syco}
Since, as discussed in the previous section, the general problem of determining maps  reducing a given dynamics is non trivial, a class of  solutions may be more easily found if the involutive distribution one looks for is generated by a finite dimensional Lie algebra whose action integrates to an action of a Lie group on  $M$ under which the dynamics is invariant, or if constants of the motions exist. Involutive distributions providing a meaningful reduction indeed  appear as the kernel of closed exterior forms invariant along $\Gamma$. We introduce now these notions. 

A function $F\,:\,M\,\to\,M'$ such that the relation \eqref{equazione2} can be solved with $\Gamma_F=0$ is called a \emph{generalised} constant of the motion for the dynamics given by $\Gamma$. 
 In such a case, the dynamics $\Gamma$ is replaced by a family of parameter depending dynamics, each one defined on a leaf of the foliation. 
The set of elements in $\mathcal F(M)$ given as $f=F^*(u)$ where $u\in\mathcal F(M')$ gives  an algebra of invariant functions for the dynamics $\Gamma$, i.e. functions which are constant along the integral curves of $\Gamma$,
\beq
\label{defc00}
L_{\Gamma}f\,=\,i_{\Gamma}(\dd f)\,=\,0. 
\eeq

A diffeomorphism $\Phi\,:\,M\,\to\,M$ defines a  (global) \emph{symmetry} for the given dynamics if $\Phi$ maps integral curves of $\Gamma$ (i.e. parametrised solutions of the differential equation) into integral curves of $\Gamma$ (i.e. if $\Phi$ maps solutions of the first order o.d.e. formulated in terms of  $\Gamma$ into solutions). This definition can be expressed upon using the associated push-forward map  as 
\beq
\label{defsymm}
\Phi_*(\Gamma)=\Gamma,
\eeq
 and this  has an infinitesimal version (which is often used in specific problems): a vector field $X$ provides an (infinitesimal) symmetry for the dynamics described by $\Gamma$ if the condition 
 \beq
 \label{defsymmi}
 [X,\Gamma]=0
 \eeq upon their commutator holds. It is immediate to see that the set of  symmetries for a dynamics is a group under composition, infinitesimal symmetries constitute a Lie algebra. Conversely, only when a Lie algebra of infinitesimal symmetries is given by  complete vector fields, can it be integrated to a group of symmetries for the dynamics. 
 
Since infinitesimal symmetries are described by  vector fields $X\in\mathfrak X(M)$, while   constants of the motion are  described in terms of  functions $f$ defined up to a constant\footnote{ This is again a suitable Lie subalgebra of the algebra of functions on $M$:  infinitesimal symmetries give indeed  a Lie-module over the ring of constants of the motion.}, that is in terms of the differential $\dd f$, (i.e. an exact 1-form  $\dd f=\alpha\in\Lambda^1(M)$) a link between such sets  for a given dynamics can be formulated  by a  map 
\beq
\label{intro00}
\tau\,:\,\Lambda^1(M)\,\to\,\mathfrak X(M),
\eeq
 or viceversa by a map  
 \beq
 \label{intro01}
 \tilde{\tau}\,:\,\mathfrak X(M)\,\to\,\Lambda^1(M).
 \eeq
 When such maps are equivariant with respect to the dynamical evolution, this link can be referred to as a theorem \`a la Noether.  If one requires  such a map $\tau$  to be $\R$-linear and \emph{local}, then its action  can be described as the action of an element $T\,\in\,\mathfrak T^2_0(M)$, that is a  $(2,0)$-tensor $T$ (i.e. twice contravariant) on $M$. Upon setting $\tau\,:\,\alpha\,\mapsto\,X_{\alpha}\,=\,T(\alpha)$, with $\alpha\in\Lambda^1(M)$ and $X_{\alpha}\in\mathfrak X(M)$, it is immediate to compute
\beq
\label{eq3}
[\Gamma, X_{\alpha}]\,=\,L_{\Gamma}X_{\alpha}\,=\,L_{\Gamma}(T(\alpha))\,=\,(L_{\Gamma}T)(\alpha)+T(L_{\Gamma}(\alpha)).
\eeq
Let $\alpha$ be \emph{invariant} along $\Gamma$, i.e. $L_{\Gamma}(\alpha)=0$.  The above relation shows that $X_{\alpha}$ is an infinitesimal symmetry for $\Gamma$ when the tensor $T$ is invariant under $\Gamma$ along the $\Gamma$-invariant 1-forms. Analogously one describes the action of $\tilde{\tau}$ in terms of a tensor $S\in\mathfrak T^0_2(M)$ (i.e. twice covariant) as $\tilde{\tau}\,:\,X\,\mapsto\,\alpha_X\,=\,S(X)$ and computes 
\beq
\label{eq4}
L_{\Gamma}\alpha_X\,=\,L_{\Gamma}(S(X))\,=\,(L_{\Gamma}(S))(X)+S([\Gamma,X])
\eeq
so that, with $X$ an infinitesimal symmetry for $\Gamma$, one has that $\alpha_X$ is $\Gamma$-invariant when the tensor $S$ is invariant under $\Gamma$ along the vector fields which commute with $\Gamma$.

These lines show that, in order to establish a Noether-type theorem between infinitesimal symmetries and constants of the motion for a given dynamics $\Gamma$, one primarily looks  for tensors which are invariant under $\Gamma$, i.e. compatible with the given dynamics. Not any $(2,0)$-tensor turns indeed to be useful for a Noether type theorem. Assume for instance  that an invertible symmetric $(2,0)$-tensor $G$ (i.e. a contravariant metric on $M$) is invariant under the dynamics $\Gamma$, that is $L_{\Gamma}G=0$. The corresponding $\tau:\Lambda^1(M)\to\mathfrak X(M)$ maps a function into a gradient vector field $$\dd f\mapsto X_{\dd f}\,=\,G(\dd f,  ~~)$$ and it is immediate to see that $L_{X_{\dd f}}f=G(\dd f, \dd f)=G^{-1}(X_{\dd f},X_{\dd f})$. Moreover,  this shows that 
if we require the dynamics itself to be written 
as $\Gamma=X_{\alpha}$ for a suitable 1-form $\alpha$,  and we want such a 1-form to be exact $\alpha=\dd f$ and  invariant along $\Gamma$, i.e. $$i_{X_{\alpha}}\alpha=i_{T(\alpha)}\alpha=0$$ (which means we require $f$, the generating functions for the dynamics, to be a constant of the motion),  then $T$ is necessarily  skewsymmetric.  This is one of the reasons why one formulates dynamics in terms of bivectors or of  2-forms on a suitable manifold. It should indeed be stressed that the bracket associated to symmetric $(2,0)$-tensors plays an important role in defining the indetermination relations within quantum mechanics.

\subsection{Symmetries and reduction} 
\label{sub:ser}

With respect to the foliation $\Phi^F$ introduced above: if  $L_{\Gamma}\varphi_a=0$ we see that the reduction procedure begins with a family of constants of the motion: $\Gamma$ turns to be tangent to each leaf $N_{m'}$, and this allows to reduce the problem of integrating the dynamics to the specific leaf $N_{m'}$ selected by the Cauchy data, entering with a role of parameters the  differential equations therefore  depending on a lower number of variables. In this case, one can analyse whether an equivalence relation $\sim$ in $N_{m'}$ compatible with the dynamics $\Gamma$ exists, i.e. such that $$n\sim n'\,\Leftrightarrow\,n(t)\sim n'(t)$$ where $n,n'$ are elements in $ N_{m'}$ and $n(t),n'(t)$ are the corresponding time evolved under the dynamics generated by $\Gamma$. Such equivalence relations  are commonly analysed by means of the action of a group, or by  
an involutive distribution $\mathcal D_{m'}$ on each leaf $N_{m'}$, which reads a further reduction for the restriction of the dynamics $\Gamma$ onto a suitable lower dimensional quotients\footnote{Not only do constants of the motions  for a given dynamics useful originate  reduction procedure, they are indeed useful to analyse global properties of the trajectories, especially when their level sets are compact submanifolds. Noether symmetries have been for instance exploited, within the Lagrangian setting,  to analyse    interesting classes of solutions in cosmology (see \cite{cosmol}).}. 

We notice that also the problem of restricting a vector field $\Gamma\in\mathcal X(M)$ to a submanifold $i_N:N\hookrightarrow M$ has an algebraic formulation. The set 
\beq
\label{intro1.4}
\mathcal I_N\,=\,\{f\,\in\,M\,:\,i_{N}^*f=0\}
\eeq
is an ideal in the algebra $\mathcal F(M)$. The short exact sequence of associative algebras 
\beq
\label{intro1.5}
0\quad\rightarrow\quad\mathcal I_N\quad\rightarrow\quad \mathcal F(M) \quad\rightarrow\quad \mathcal F(M)/\mathcal I_N \quad\rightarrow\quad 0
\eeq
allows to recover that $\mathcal F(N)\simeq\mathcal F(M)/\mathcal I_N$ and that a vector field $\Gamma$ on $M$ is tangent to $N$ if and only if $\Gamma(f)\in\mathcal I_N$ for any $f\in\mathcal I_N$. 

When $L_{\Gamma}\varphi_a=0$, each leaf of the foliation $\Phi$ is the invariant submanifold associated to the ideal $\mathcal I_{N_c}$ generated by $f_a=\varphi_a-c_a$ with $c_a$ giving the components\footnote{When the invariant submanifold is isolated, i.e. it is not a leaf  of an invariant foliation, we get what Levi-Civita called an \emph{invariant relation}.} of $c=(c_1,\dots, c_k)\in\R^k$.
An involutive distribution $\mathcal D_c$ on each leaf $N_c$ amounts to have a Lie algebra $\tilde{\mathcal D_c}$ of vector fields in $M$ such that $\tilde Y(f)\in\mathcal I_{N_c}$ for any $\tilde Y\in\tilde{\mathcal D}_c$: the dynamics can be reduced to the corresponding quotient if and only if $[\Gamma,\tilde Y]\in\tilde{\mathcal D}_c$ for any $\tilde Y\in\tilde{\mathcal D}_c$.   


The map $\tau$ introduced above allows to define the distribution $\mathcal D_X=\{\tau(\dd\varphi_a)\}$. One then analyses how to obtain a suitable involutive distribution $\tilde{\mathcal D_{c}}\subseteq\mathcal D_X$ on each invariant leaf $N_c$, and under which conditions it reads a  regular foliation $\Phi^X$ compatible with the foliations $\Phi^{F}$, so to reduce the dynamics $\Gamma$ from $N_c$ to a lower dimensional quotient. 

This observation is the compass we use to describe the Noether theorem and its role in reduction of dynamical system. 
Among the conditions one can consider, the notion of symmetry of $\Gamma$ under the action of a given Lie algebra plays, because of the Frobenius theorem,  a central role, and we shall analyse it in some detail. In particular, since a Lie algebra may be described in terms of 1-forms, or of Cartan's 1-forms, we shall consider the case of a family of invariant 1-forms suitably defining a Lie algebra.

\section{Noether theorem and reduction on Poisson manifolds}
\label{sec:poisson}

A contravariant 2-tensor is also a  bidifferential operator, therefore it may be used to define a bilinear bracket on functions. When additional conditions are met, i.e. it is skewsymmetric and satisfies an integrability condition,    we get a Poisson structure, that we now introduce. 
A Poisson bracket  on a $N$-dimensional manifold $M$ is a bilinear map $$\{~,~\}\,:\,\mathcal F(M)\times\mathcal F(M)\to\mathcal F(M)$$ that satisfies the following conditions, for any $f,g,h\,\in\,\mathcal F(M)$:
\begin{enumerate}[(1)] 
\item it is skewsymmetric, i.e. $\{f,g\}=-\{g,f\}$;
\item the Jacobi identity $\{f,\{g,h\}\}+\{g,\{h,f\}\}+\{h,\{f,g\}\}=0$ holds; 
\end{enumerate}
These two conditions define a Lie algebra structure on $\mathcal F(M)$. When, in addition to a Lie algebra structure, the bilinear map $\{~,~\}$ defines derivations with respect to the algebra structure on $\mathcal F(M)$, that is when the condition 
\begin{enumerate}[(3)] 
\item $\{fg, h\}=f\{g,h\}+\{f,h\}g$ (a Leibniz  rule) holds,
\end{enumerate}
then the set $(\mathcal F(M), \{~,~\})$ is called a Poisson algebra. A function $C\in\mathcal F(M)$ is defined to be a  Casimir function for the given Poisson structure if $\{C,f\}=0$ for any $f\in\mathcal F(M)$. 
The condition (3) implies that, for any $H\in\mathcal F(M)$,  the map 
\beq
\label{eq6}
X_H\,:\,f\,\mapsto\,\{f,H\}
\eeq
 is a derivation operator on $\mathcal F(M)$.  Such derivations are called \emph{inner} in $\mathcal F(M)$ with respect to the Poisson structure on $M$.
If, as implicitly assumed, the Poisson bracket is local\footnote{An interesting example comes upon considering the commutative algebra $(A, *)$ given as a suitable completion of the space of integrable functions on $\R^2$ with  the  abelian \emph{non local} convolution product  (here $\R^2\in x\,=\,(x_1,x_2)$) $$(f*g)(x)\,=\,\int\dd y f(x-y)g(y).$$ The pointwise multiplication operators $\hat x_a\,:\,f(x)\,\mapsto\,x_af(x)$ with $a=1,2$ is easily seen to be linear and to satisfy the Leibniz rule, i.e. it is a derivation of the commutative algebra $(A,*)$. It is now immediate to prove that 
$$
\{f,g\}\,=\,(\hat x_1f)*(\hat x_2g)\,-\,(\hat x_2f)*(\hat x_1g)
$$
is a non local  Poisson bracket in $(A,*)$.
} (which means that ${\rm Supp}\{f,g\}\subseteq {\rm Supp} \,f\cap {\rm Supp} \,g$), we have from the Willmore's theorem (see \cite{AM78, michor-book})  that the derivation $X_H$ is a vector field on $M$, referred to as the Hamiltonian vector field corresponding to the (Hamiltonian) function $f$ via the given Poisson bracket\footnote{Derivations in $\mathcal F(M)$ which are not Hamiltonian are called \emph{outer} with respect to the Poisson structure on $M$.}.  It is evident that $X_C=0$ if $C$ is a Casimir function.  From the Jacobi identity it is immediate to prove that
 \beq
 \label{eq8}
 [X_f,X_g]\,=\,X_{\{g,f\}},
\eeq
so one sees that the map \eqref{eq6} provides an (anti)homomorphism between the Poisson algebra on $\mathcal F(M)$ and the infinite dimensional 
 Lie algebra $(\mathfrak X(M), [~,~])$ of the vector fields on $M$ with respect to the commutator structure. 
 With respect to a local chart $\{x^a\}_{a=1,\dots,N}$ for $M$, the coordinate expression for a Hamiltonian vector field is given\footnote{We denote $\del_a\,=\,\del/\del x^a$.} by 
 \beq
 \label{eqa1}
 X_H\,=\,\{x^a, H\}\del_a,
 \eeq
 while the Poisson bracket reads
\beq
\label{eq1}
\{f,g\}\,=\,\frac{\del f}{\del x^a}\{x^a,x^b\}\frac{\del g}{\del x^b}.
\eeq
This expression can be translated into
\beq
\label{eq2}
\{f,g\}\,=\,\Lambda(\dd f,\dd g)
\eeq
where $\Lambda$ turns to be a  skewsymmetric $(2,0)$ tensor whose coordinate expression is $\Lambda\,=\,\Lambda^{ab}\del_a\wedge\del_b$  with 
$$
\Lambda^{ab}=\Lambda(\dd x^a, \dd x^b)=\{x^a,x^b\}.
$$
 The Jacobi identity reads that $[\Lambda, \Lambda]=0$ with respect to the Schouten bracket generalising the notion of commutator of vector fields to skewsymmetric contravariant tensors on $M$.  If $X\,=\,X_1\wedge\ldots\wedge X_a$ and $Y\,=\,Y_1\wedge\ldots Y_b$ with $X_i,Y_j\in\mathfrak X(M)$, then the Schouten bracket  can be defined as
\beq
\label{schoutend}
[X,Y]\,=\,\sum_{i,j}(-1)^{i+j}[X_i,Y_j]X_1\wedge\ldots\wedge X_{i-1}\wedge X_{i+1}\wedge\ldots X_a\wedge Y_1\wedge\ldots\wedge Y_{j-1}\wedge Y_{j+1}\wedge\ldots\wedge Y_b, 
\eeq 
with $[X_i,Y_j]$ the usual commutator of vector fields.  This definition is extended to the case $Y=f\in\mathcal F(M)$ upon setting 
$$ 
[X,f]\,=\,\sum_i-(-1)^{i}(L_{X_i}f)X_1\wedge\ldots\wedge X_{i-1}\wedge X_{i+1}\wedge\ldots X_a.
$$
The converse holds true: if $\Lambda$ is a skewsymmetric $(2,0)$ tensor on $M$ which satisfies the Schouten  condition $[\Lambda,\Lambda]=0$, then the expression \eqref{eq2} defines a Poisson structure on $\mathcal F(M)$ corresponding to the Poisson tensor $\Lambda$. It is interesting to notice that from this follows that, if $\Lambda, \Lambda'$ are Poisson tensors on $M$, then $\Lambda+\Lambda'$ is a Poisson tensor on $M$ if and only if $[\Lambda,\Lambda']=0$. When this condition is valid, any linear combination of $\Lambda,\Lambda'$ with real coefficients provide $M$ a Poisson structure.
 
A bilinear bracket on $\mathcal F(M)$ satisfying the conditions (1)-(2) written above defines a Lie algebra structure on $\mathcal F(M)$. An interesting example of Lie bracket on smooth functions on a  manifold is given by 
\beq
\label{jacobib}
\{f,g\}\,=\,\Lambda(f,g)+fD(g)-gD(f),
\eeq
where $\Lambda$ is a bivector field and $D$ a vector field which satisfy the conditions
\begin{align}
&[\Lambda, \Lambda]=2D\wedge\Lambda, \nn \\ &[D,\Lambda]=0\label{jacobic}
\end{align}
The bilinear \eqref{jacobib} is said a Jacobi bracket on $M$. It is clear that a Jacobi bracket with $D=0$ provides a Poisson tensor. Notice also that from \eqref{jacobib} follows that the restriction of a   Jacobi bracket on the subalgebra of functions in $M$ which are annihilated by $D$ provides a Poisson bracket. Conversely,  it  is possible to prove that if a  Lie bracket $\{~,~\}$ on $\mathcal F(M)$ is given by a bilinear differential operator, then there exist a bivector field $\Lambda$ and a vector field $D$ on $M$ such that the relations \eqref{jacobib} and \eqref{jacobic} hold. Jacobi structures on a manifold are often related to  contact structures, but we shall not consider these more general  structures in this paper and refer to \cite{jbrackets} for a more detailed review.

Since a Poisson structure on a manifold $M$ can be formulated in terms of a suitable contravariant tensor on it, it is natural to define $\alpha\in\Lambda^1(M)$ a Casimir 1-form if $\Lambda(\alpha, \beta)=0$ for any $\beta\in\Lambda^1(M)$, and to define a Poisson tensor \emph{locally degenerate}  if it has a (local) Casimir 1-form. Analogously, a Poisson structure is  called \emph{(globally) degenerate} if it has an exact non trivial Casimir 1-form, i.e. there exists a (non constant) function $C\in\mathcal F(M)$ (a Casimir function) such that  $\alpha=\dd C$  is a Casimir 1-form for $\Lambda$.  Global non degeneracy  is  equivalent to the condition that the Poisson tensor defines an isomorphism $T_mM\leftrightarrow T_m^*M$ for each $m\in M$.

 A diffeomorphism $\phi\,:\,M\,\to\,M$ is called a Poisson map, or a \emph{canonical} map, if 
 $$
 \phi_*\Lambda=\Lambda.
 $$ 
A vector field $X$ on the Poisson manifold $(M, \Lambda)$ is called \emph{canonical} if 
$$
L_X\Lambda=0;
$$
 it is clear that canonical vector fields on $(M,\Lambda)$ provide the infinitesimal version of one parameter groups of canonical diffeomorphisms.  This condition is proven to be equivalent to the condition 
\beq
\label{eq9}
L_X\Lambda\,=\,0\qquad\Leftrightarrow\qquad L_X\{f,g\}\,=\,\{L_Xf,g\}+\{f,L_Xg\},
\eeq  
that is a vector field $X$ is canonical if and only if it is a derivation for the Lie algebra structure on $\mathcal F(M)$ given by $\Lambda$. Notice that, as the relation \eqref{eq8} shows,  if both $X$ and $Y$ are canonical vector fields on $(M,\Lambda)$, their commutator is also canonical: this means that the set of canonical vector fields for  a Poisson structure on a  manifold is a Lie algebra.

Any Hamiltonian vector field is canonical, since  $L_{X_H}\Lambda=0$ for any $H\in\mathcal F(M)$: notice indeed that this implication can not be reversed, i.e. there are canonical derivations which are not inner.  Consider the Poisson tensor $\Lambda\,=\,\del_1\wedge\del_2$ on $\R^3$: the vector field $X=\del_3$ is canonical, i.e. $L_{\del_3}\Lambda=0$, but there is no element $f\in\mathcal F(\R^3)$ such that $\del_3=X_f$. This is related to the fact that $\Lambda$ is degenerate, with  $C=x^3$ providing  a global Casimir function for the Poisson tensor, i.e. a constant of the motion for any Hamiltonian vector field on $\R^3$ with respect to $\Lambda$. Another interesting example is given upon considering the Poisson tensor $\Lambda=xy\del_x\wedge\del_y$ on $\R^2$. The dilation vector field $\Delta=x\del_x+y\del_y$ is canonical, since one has $L_{\Delta}\Lambda=0$, but it is not Hamiltonian since one has $\Delta=X_H$ with $H=\log\left|\frac{y}{x}\right|$, which is defined only on $\R^2\backslash\{xy=0\}$.  

An important class of Poisson tensors comes from the theory of (finite dimensional) Lie algebras, namely $N$-dimensional vector spaces $\mathfrak g$ equipped with a commutator structure $[e_a,e_b]=f_{ab}^{\,\,\,\,\, c}e_c$ along  a basis $\{e_a\}_{a=1,\dots,N}$. The tensor 
\beq
\label{podu}
\Lambda_{\mathfrak g}\,=\,f^{ab}_{\,\,\,\,\,c}x^c\del_a\wedge\del_b
\eeq
 defines a Poisson structure\footnote{The coordinate chart $\{x^a\}_{a=1,\dots,N}$ is global and dual to the basis $\{e_a\}$ for $\mathfrak g$, where indices are raised and lowered via the Euclidean metric tensor on $\mathfrak g$ and its dual on $\mathfrak g^*$.} on $M\simeq\mathfrak g^*$, since the Jacobi identity for the Lie algebra $\mathfrak g$ turns to be equivalent to the Jacobi identity for $\Lambda$. 
Interesting examples come from 3-dimensional Lie algebras, following the analysis in \cite{gmp93}. 

If $\mathfrak g=\mathfrak{su}(2)$, then one has 
\beq
\label{podu1}
\Lambda_{\mathfrak{su(2)}}=\varepsilon^{ab}_{\,\,\,\,\,c}x^c\del_a\wedge\del_b
\eeq
 on $\mathfrak g^*\simeq\R^3$. Such a Poisson tensor has a  Casimir function, given by $C=\delta_{ab}x^ax^b$: the set 
$\ker(\Lambda_{\mathfrak{su(2)}})$ is the $\mathcal F(\R^3)$-bimodule generated by the exact Casimir 1-form $\dd C=\delta_{ab}x^a\dd x^b$.  

The Lie algebra $\mathfrak{sb}(2,\C)$ (also known as the \emph{book algebra}) reads the Poisson tensor on $\R^3$ given by 
\beq
\label{podu2}
\Lambda_{\mathfrak{sb}(2,\C)}=x^2\del_1\wedge\del_2+x^3\del_1\wedge\del_3=-(x^2\del_2+x^3\del_3)\wedge\del_1.
\eeq
One easily computes that $\ker(\Lambda_{\mathfrak{sb}(2,\C)})$ is generated by the 1-form $\alpha=x^2\dd x^3-x^3\dd x^2$, which is not closed (and then neither exact).  This shows that $\Lambda_{{\mathfrak sb}(2,\C)}$ is locally degenerate, but has no global Casimir function. 
Another interesting example of a Poisson tensor which is degenerate but has no global Casimir function comes upon considering the 3-dimensional torus $\mathbb{T}^3$, and the Poisson tensor given by $\Lambda=(a\del_1+b\del_2)\wedge\del_3$  (with $a/b\in\R\backslash\mathbb{Q}$) along the coordinate chart given by the angular   $\{\theta_a\}_{a=1,\dots,3}$. The orbits of the corresponding Hamiltonian vector fields span one dimensional submanifolds which are dense in the $\mathbb{T}^2$ defined upon setting $\dd\theta_3=0$
 One sees that the Casimir 1-form $\alpha=a\dd\theta_1-b\dd\theta_2\in\rm{ker}(\Lambda)$ is closed but not exact (unlike previous examples, coefficients are not functions but real numbers).

A dynamics $\Gamma$ on $M$ is said to have a Poisson description if $M$ is equipped with a Poisson tensor $\Lambda$ such that there exists an element $H\in\mathcal F(M)$ reading $\Gamma=X_H$. This condition is written, in local coordinates $\{x^a\}_{a=1,\dots,N}$ on $M$, as (see \eqref{eqa1}) $\Gamma=\Gamma^k\del_k$ with $\Gamma^k=\Lambda^{kj}\del_jH$.
The Poisson tensor $\Lambda$ is invariant along $X_H$, and therefore we realise  the  $\tau\,:\,\Lambda^1(M)\,\to\,\mathfrak X(M)$ as described in \eqref{intro00}, with 
$\tau\,:\,\alpha\,\mapsto\,\hat X_{\alpha} $ defined by
\beq
\label{eq5}
\hat X_{\alpha}(f)\,=\,i_{\hat X_{\alpha}}\dd f\,=\,\Lambda(\dd f,\alpha)
\eeq 
for any $f\in\mathcal F(M)$. 
Notice that the map $\tau$ is not injective, since $\hat X_{\alpha}=\hat X_{\alpha+\xi}$ if $\xi$ is a Casimir 1-form for the Poisson tensor $\Lambda$. As we already discussed in the introduction, the Hamiltonian function $H$ is a constant of the motion for the dynamics $X_H$, since the Poisson tensor is skewsymmetric, while it is the Jacobi identity which provides the set of constants of the motion for $X_H$ a Lie (sub)algebra structure in $\mathcal F(M)$  with respect to the Poisson bracket.
 The Poisson tensor also allows to define a Lie algebra structure on $\Lambda^1(M)$, given by 
\begin{align}
[\alpha, \beta]_{\Lambda}&=\,L_{\hat X_{\beta}}\alpha\,-\,L_{\hat X_{\alpha}}\beta\,-\,\dd(\Lambda(\alpha, \beta)) \nn \\ &=\,i_{\hat X_{\beta}}\dd\alpha\,-\,i_{\hat X_{\alpha}}\dd\beta\,+\,\dd(\Lambda(\alpha, \beta)),
\label{eq11}
\end{align}
which, for exact 1-forms, reads 
\beq
\label{eq11b}
[\dd f,\dd g]_{\Lambda}\,=\,\dd\{f,g\}.
\eeq


The comparison between \eqref{eq6} and \eqref{eq5} shows that $\hat X_{\dd f}\,=\,X_f$. Directly from \eqref{eq5} one has 
$$
\hat X_{\alpha}(f)\,=\,-\alpha(X_f)\,=\,-i_{X_f}\alpha:
$$
 together with \eqref{eq8}, it gives
\begin{align}
&(L_{\hat X_{\alpha}}\Lambda)(\dd f, \dd g)\,=\,-\dd\alpha(X_f,X_g) \nn ; \\
\label{eq10}
&[X_H,\hat X_{\alpha}](f)\,=\,-(L_{X_H}\alpha)(X_f)
\end{align}
for any 1-form $\alpha$ and any triple $f,g,H$ of functions on $M$. 

\subsection{Symmetries and reduction within the Poisson formalism}
\label{susec:poisson}
The above  relations show that the conditions under which the vector field $\hat X_{\alpha}$ provide an infinitesimal canonical symmetry for the dynamics $X_H$ on the Poisson manifold $(M,\Lambda)$ can be studied in terms of the properties of the 1-form $\alpha$ along the family of all Hamiltonian vector fields on $M$. 

In particular, the first relation shows that $\hat X_{\alpha}$ is canonical if and only if $\alpha$ is \emph{relatively closed}, i.e. closed along all Hamiltonian vector fields on $M$, while $\hat X_{\alpha}$ commutes with the dynamics $X_H$ if and only if $i_{X_f}(L_{X_H}\alpha)=0$ for any Hamiltonian vector field $X_f$. 

In order to study under which conditions a class of 1-forms which are invariant along a dynamics $\Gamma=X_H$ provide (via the map $\tau$) vector fields which are infinitesimal canonical symmetries for $\Gamma$, we consider the following cases, emphasizing the associated reduction procedure.
\subsubsection{\blu{Constants of the motions for a Poisson dynamics and the associated reduction procedure}}
\label{subsecP1}
 Let $u\in\mathcal F(M)$ be a constant of the motion for $\Gamma=X_H$, that is $$L_{X_H}u=\{u,H\}=0.$$ This implies that $$\dd (L_{X_H}u)=L_{X_H}\dd u=0,$$  that is  the exact 1-form $\alpha=\dd u$ is invariant along the integral curves of $X_H$. From \eqref{eq10} it is immediate to see that  $\hat X_{\dd u}=X_u$  is a canonical infinitesimal symmetry for the dynamics. This is the standard form of the Noether theorem within the Poisson formalism, that we report as follows.
\begin{prop}
\label{noepoisson1}
Given a Hamiltonian dynamics $\Gamma=X_H$ on a Poisson manifold $(M,\Lambda)$, if $u\in\mathcal F(M)$ is a constant of the motion, then the corresponding Hamiltonian vector field $X_u$ is an infinitesimal canonical symmetry for $\Gamma$.
\end{prop} 
We describe now how this allows for an interesting reduction of the dynamics, when $u$ and $H$ are functionally independent, i.e. $\dd u\wedge\dd H\neq0$.  
Let  $c\,\in\R$ be  a regular value for the function $u$, i.e. $\dd u\neq 0$ so that $u$ is a (local) submersion (see the appendix \ref{app2}). With respect to \eqref{equazione2}, we have $F=u$.
Assume that the  sets 
$$
N_{c}\,=\,\{m\in M\,:\, u(m)=c\}
$$
 give the leaves of a regular foliation $\Phi^{F}$ and that the quotient $M/\Phi^{F}$ has a manifold structure. The vector fields which are tangent to the leaves of this foliation provide the (integrable) distribution (see \eqref{equazione2p})
\beq
\label{eqpoi1}
\mathcal D_{F}\,=\,\{Y\,\in\,\mathfrak X(M)\,:\,i_Y\dd u\,=\,L_Yu\,=\,0\}.
\eeq
 From $L_{X_H}u=0$ we immediately see that $X_H\in\mathcal D_{F}$; moreover, we also trivially have that $X_u\in\mathcal D_{F}$. Since $[X_u, X_H]=0$, the dynamics $X_H$ turns to be  projectable also with respect to the foliation (the condition that $u$ and $H$ are functionally independent reads  that the vector fields $X_H, X_u$ span a two dimensional distribution at each point of interest) on each $N_c$ generated by $X_u$. This shows that the presence of a \emph{single} constant of the motion (provided it is functionally independent from the Hamiltonian $H$) for a dynamics which has a Poisson description allows for a \emph{two dimensional} reduction, i.e. the flow of the vector field  $X_H$ projects onto the $(N-2)$-dimensional quotient (when it turns out to have a manifold structure)  given upon identifying the points in $N_c$ which belong to the same orbit of $X_u$, and that we denote by $M_c\simeq N_c/X_u$. 
In the following lines we shall outline  how to see that each leaf $M_c$ has a Poisson structure coming from the one on $M$, and that the reduced vector field, which we still denote by $X_H\in\mathfrak X(M_c)$ has a Hamiltonian description. 

The reduction we described is developed through two steps: the first is the restriction of the dynamics to an invariant submanifold, the second is the projection of such restriction onto a suitable quotient submanifold of the invariant one. Both these steps can be illustrated within an algebraic formalism, as we outlined in sections \ref{ss:red} and \ref{sub:ser} in terms of the short exact sequences \eqref{intro1.3} and \eqref{intro1.5}.   
Being a bidifferential operator (i.e. a contravariant tensor) on $M$ which defines a  Lie algebra structure on $\mathcal F(M)$ compatible with the (pointwise commutative) product of the algebra $\mathcal F(M)$,  a Poisson tensor $\Lambda$ provides a relation between the two above steps of the reduction procedure for  a dynamics compatible with it. In this sense, a single constant of the motion may allow for a double reduction. 


\subsubsection{\blu{Reduction of a Poisson algebra}}\label{sususe:Po1}
Consider then  the distribution $\mathcal D_X$ on $M$ spanned by the Hamiltonian vector field $X_u$, and assume it generates a regular foliation $\Phi^{X}$, with the quotient space $M/\Phi^{X}$ having  a manifold structure. The set 
\beq
\label{eq10d}
\mathcal F_{X}=\{f\,\in\,\mathcal F(M)\,:\,X_u(f)=\{f,u\}=0\}
\eeq 
gives a  subalgebra corresponding to  the distribution $\mathcal D_X$, and one immediately sees that $\mathcal F_{X}\simeq\mathcal F(M/\Phi^X)$. 
The relation (an immediate consequence of the Jacobi identity) $X_u\{f,f'\}=0$ for any pair $f,f'\in\mathcal F_{X}$ proves that $\mathcal F_X$ is a Poisson algebra, or equivalently that the Poisson tensor $\Lambda$ is \emph{projectable} onto the quotient $M/\Phi^{X}$.

Denote by $\mathcal I_{a}\subset \mathcal F(M)$ the ideal generated by the function $\varphi_{(a)}=u-a$ with $\dd a=0$ and its \emph{normalizer} with respect to the Poisson tensor, that is $$\mathcal N_a\,=\,\{f\,\in\,\mathcal F(M)\,:\,\{f, f_I\}\in\mathcal I_a\,\forall\,\,f_I\in\mathcal I_a\}.$$  After noticing that one has the  inclusion  
$\mathcal F_{X}\subset \mathcal N_a$, it is easy to prove also  that $$\mathcal I'_a=\mathcal I_a\cap\mathcal F_{X}$$ is a Poisson ideal, so that the Poisson bracket is meaningfully defined on the quotient algebra $\mathcal F_{X}/\mathcal I'_a$. Notice that, with respect to the reduction described above,  the equivalence $$\mathcal F_X/\mathcal I'_a\simeq\mathcal F(N_c/X_u)$$ holds. 
The elements of this quotient can be written, for any element $f\in\mathcal F_{X}$,  as 
\beq
\mathcal F_{X}/\mathcal I'_a\,\ni\,[f]\,=\,f\,+\,\varphi_{(a)} f'\quad{\rm with}\quad X_u(f)=X_u(f')=0,
\label{eq10a}
\eeq
while the action of the Poisson bracket is easily proven to be 
\beq
\{[f],[g]\}\,=\,[\{f,g\}].
\label{eq10e}
\eeq
Since  $[X_H,X_u]=0$, we see that the vector field $X_H$ is a derivation of the algebra $\mathcal F_{X}$, i.e. the dynamics $X_H$ is projectable onto $M/\Phi^X$. Moreover, from  \eqref{eq10a} we have
\begin{align} 
X_H\,:\,[f]\quad&\mapsto\quad X_H(f)\,+\,X_H(\varphi_{(a)} f')\nn \\
&=\quad \{f,H\}\,+\,\varphi_{(a)} X_H(f')\,=\,[\{f,H\}],
\label{eq10b}
\end{align}
where the second line comes from $u$ (and then $\varphi_{(a)}$) being a constant of the motion for the dynamics.  This means that we can write 
\beq
\label{eq10c}
X_H\,:\,[f]\quad\mapsto\quad \{[f],[H]\}:
\eeq 
a (suitably regular) constant of the motion for the dynamics $X_H$ allows for a reduction, and the reduced dynamics is Hamiltonian.  

\noindent How to generalise this construction when the dynamics has a higher number of  canonical symmetries? Assume in general that $F\,:\,M\,\to\,(u_1,\dots,u_{\delta})$ is a (local) submersion,  
with the corresponding involutive distribution 
\beq
\label{eqpoi2}
\mathcal D_{F}\,=\,\{Y\in\mathfrak X(M):i_Y\dd u_a=0\}\,=\,\cap_{j=1}^{\delta}{\rm ker}\,\dd u_a,
\eeq
generalising \eqref{eqpoi1}. We assume that the rank of the distribution $\mathcal D_{F}$ is maximal at each point of interest, which means that the elements $u_a$ are functionally independent. Each leaf of the corresponding foliation $\Phi_{F}$ is given by the $(N-\delta)$-dimensional manifold $$N_c\,=\,\{m\in M\,:\,u_j(m)=c_j\}$$ with $c\in\R^{\delta}$ a regular value for $F$. The algebra $\mathcal F(N_c)$ can be identified with the quotient $\mathcal Q_c=\mathcal F(M)/\mathcal I_c$ where $\mathcal I_c$ is the ideal generated by  the elements $\varphi_j=(u_j-c_j)$ with $\dd c_j=0$, so that we can write 
\beq
\label{trans1}
\mathcal I_c\,=\,\{f\,=\,\varphi_jg_j, \quad g_j\in\mathcal F(M)\}.
\eeq
Since the Poisson bracket on $M$ can not be consistently defined  onto $\mathcal Q_c$, the path   \cite{marared} starts from the  subalgebra 
\beq
\label{eqco1}
\mathcal F_X\,=\,\{f\,\in\,\mathcal F(M)\,:\,\{f,u_j\}=0\quad\forall \,j=1,\dots,\delta\}\,\simeq\,\mathcal F(M/\Phi^X)
\eeq
generated by the distribution $\mathcal D_X$ spanned by the Hamiltonian vector fields $\{X_{u_j}\}$, with $\Phi^X$ denoting the corresponding regular foliation. As already showed above, the Poisson structure on $M$ can be consistently defined  onto the subalgebra $\mathcal F_X$, since $X_{u_j}\{f,g\}=0$  for $f,g\in\mathcal F_X$. The vector field $X_H$ defines a derivation on $\mathcal F_X$ if and only if 
\beq
\label{introcop}
L_{X_H}u_a\,=\,f_a(u)
\eeq
with $f_a$ depending  only on the $u_j$. In order to reduce the dynamics further, consider the ideal $\mathcal I_c$ defined in \eqref{trans1} and its normaliser\footnote{Notice that the elements in $\mathcal N_c$ correspond to functions on $M$ whose corresponding Hamiltonian vector fields are tangent to $N_c$.}
\beq
\label{trans2}
\mathcal N_c\,=\,\{f\,\in\,\mathcal F(M)\,:\,\{f, \varphi_jg_j\}\,\in\,\mathcal I_c\}.
\eeq
 It is immediate to see that $\mathcal F_X\subset\mathcal N_c$. The intersection
\beq
\label{trans3}
{\mathcal I}'_c\,=\,\mathcal I_c\cap\mathcal F_X\,=\,\{f\,\in\,\mathcal F(M)\,:\,f=\varphi_jg_j\quad{\rm and}\quad\{\varphi_b, \varphi_ag_a\}\,=\,0, \,\,g_a\in\,\mathcal F(M)\}
\eeq 
is (by a direct proof) a Poisson ideal, so the Poisson bracket on $\mathcal F_X$ consistently defines a Poisson bracket onto the quotient $\mathcal F_X/{\mathcal I}'_c$.  Moreover, it is the condition \eqref{introcop} which allows to prove that $X_H\,:\,{\mathcal I}'_c\to{\mathcal I}'_c$, so  $X_H\in{\rm Der}(\mathcal F_X/ {\mathcal I}'_c))$. 

\noindent When is such a derivation \emph{inner} with respect to the Poisson structure? Only if $\{H,u_j\}=0$, i.e. only if $H\in\mathcal F_X$. 
We therefore assume that 
the elements $\{u_j\}_{j=1,\dots,\delta}$ are functionally independent  constants of the motion for the dynamics, i.e. $L_{X_H}u_j\,=\,\{u_j, H\}\,=\,0$, with  $[X_H,X_{u_j}]=0$.  The analysis described in the previous lines shows that 
 $X_H$ can be reduced to the quotient algebra $\mathcal F_X/{\mathcal I}'_a$, where its action turns out to be Hamiltonian, generalising \eqref{eq10c}. 
That this procedure is equivalent to the Poisson dynamics reduction  developed in \cite{marared} comes from \cite{gralamavi}. The quotient $\mathcal F_X/\mathcal I'_a$ can be identified with the algebra $\mathcal F(N_c/\mathfrak g_c)$ where $\g_c$ is the  Lie subalgebra of Hamiltonian vector fields $X_{u_j}$ which are tangent to $N_c$.

\subsubsection{\blu{Invariant closed 1-forms along a Poisson dynamics and the associated reduction procedure}}
\label{subsecP2}
 Assume that $\alpha$ is an invariant 1-form for the dynamics, i.e. $L_{X_{H}}\alpha=0$, and that $\dd \alpha=0$, i.e. $\alpha$ is closed. Under these two conditions, one still has from \eqref{eq10} that $\hat X_{\alpha}$ is an infinitesimal canonical symmetry for the dynamics. If the de Rham first  cohomology class for the manifold $M$ is trivial, then one can write $\alpha=\dd u_{\alpha}$ with $u_{\alpha}\in\mathcal F(M)$. In such a case, from the identity $L_{X_H}\dd u=\dd L_{X_H}u$ follows that the function 
\beq
\label{eq23}
L_{X_H}u_{\alpha}=\{u_{\alpha},H\}=-\{H,u_{\alpha}\}=-L_{\hat X_{\alpha}}H
\eeq
 is constant on each connected component of $M$.  We have sketched the proof of the following result
 \begin{prop}
 \label{noepoisson2}
Given a Hamiltonian dynamics $\Gamma=X_H$ on a Poisson manifold $(M,\Lambda)$, if  
 $u\in\mathcal F(M)$ satisfies the condition  $\dd L_{X_H}u=0$, then the corresponding Hamiltonian vector field $\hat X_{\dd u}=X_u$ gives a canonical infinitesimal symmetry for the dynamics $X_H$.
 \end{prop}
\noindent Notice that this result reduces to the previous proposition \ref{noepoisson1} when $\{u,H\}=0$. If $\dd L_{X_H}u=0$ and $\{u,H\}\neq 0$, 
it is immediate to see that the  one parameter group of canonical diffeomorphisms generated by $X_u$ maps integral curves of $X_H$ into integral curves of $X_H$ lying on different level sets for $H$. 
  Dynamics sharing these properties have been considered, also within the quantum setting, in \cite{dothan, hermann1,hermann2}, and are referred to as \emph{spectrum generating algebras}. 

\bigskip

\noindent Following what we outlined above in the previous section \ref{subsecP1} about the reduction procedure, it is clear that, if we have a set of closed invariant 1-forms $\alpha_j=\dd u_j$ for the dynamics $X_H$, and 
$\mathcal I_{c}$ denotes the ideal in $\mathcal F(M)$ generated by the elements $\varphi_j=u_j-c_j$ with $\dd c_j=0$, the quotient algebra $\mathcal F_X/(\mathcal I_c\cap\mathcal F_X)$ has a Poisson structure coming from the one on $M$. The operator $X_H$ is projectable onto $\mathcal F_X/(\mathcal I_c\cap\mathcal F_X)$, its action can be written as an \emph{outer} derivation.

\subsubsection{\blu{Lie algebras of infinitesimal canonical symmetries and the momentum map}}
\label{subsub:momp}
 Assume now   there exists a finite set $\{\alpha_j\}_{j=1\dots,\delta}$ of closed 1-forms which are invariant along $X_H$. From \eqref{eq11} we have that 
$$
[\alpha_j,\alpha_k]_{\Lambda}=\dd(\Lambda(\alpha_j,\alpha_k)),
$$
and this shows that the 1-form  $[\alpha_j,\alpha_k]_{\Lambda}$ is not only invariant along $X_H$, but also  exact,  its primitive $\Lambda(\alpha_j,\alpha_k)$ giving a constant of the motion for the dynamics. Together with  the identity 
\beq
\hat X_{\dd(\Lambda(\alpha_j,\alpha_k))}\,=\,X_{\Lambda(\alpha_j,\alpha_k)}\,=\,[\hat X_{\alpha_j},\hat X_{\alpha_k}],
\label{eq23b}
\eeq
we can prove that, if such closed invariant 1-forms close a  $\delta$ dimensional Lie algebra $\g$ 
\beq
\label{eq23c}
[\alpha_j,\alpha_k]_{\Lambda}\,=\,c_{jk}^{\,\,\,\,s}\alpha_s,
\eeq
 then the corresponding vector fields give 
 \beq
 \label{eq23d}
 [\hat X_{\alpha_j},\hat X_{\alpha_k}]=c_{kj}^{\,\,\,\,s}\hat X_{\alpha_s},
 \eeq
  i.e. a Lie algebra of  canonical infinitesimal symmetries for the dynamics, or equivalently a representation of $\g$ in terms of canonical infinitesimal symmetries for the dynamics.  
  
 Assume further that such 1-forms are exact, i.e.  $\alpha_j=\dd u_{j}$: under the condition \eqref{eq23c} we can  prove that 
\beq
\label{eq23e}
\{u_j,u_k\}\,=\,c_{jk}^{\,\,\,\,s}u_s\,+\,\sigma_{jk}(C)
\eeq
where $\sigma_{jk}(C)$ are (skewsymmetric in $(jk)$) functions of the Casimirs $C$ for the Poisson structure. If the Poisson structure is non degenerate, then $\dd\sigma_{jk}=0$.  If the Lie algebra is perfect, i.e. it coincides with its derived algebra, then it is possible to add an integration constant $u_j\,\mapsto\,u_j+c_j$ such that $\sigma_{jk}=0$.

\noindent Paralleling what we described above, we know from  the Frobenius theorem that both the distribution $\mathcal D_X$ spanned by the canonical vector fields $\{\hat X_{\alpha_j}\}_{j=1,\dots,\delta}$ (from \eqref{eq23d} we see that the vector fields $\hat X_{\alpha_j}$  close a Lie algebra) and the distribution $\mathcal D_{\alpha}\,=\,\cap_{j=1}^{\delta}{\rm ker}\,\alpha_j$ are integrable (since $\dd\alpha_j=0$, see the proposition \ref{prop.frofor}). Under the natural hypothesis of regularity for the quotients $M/\Phi^X$ and $M/\Phi^{\alpha}$ we can prove that $X_H$ is projectable with respect to both foliations. The Poisson structure is not  in general projectable onto the subalgebra 
\beq
\label{eqp2}
\mathcal F_{\alpha}\,=\,\{f\,\in\,\mathcal F(M)\,:\,i_Y\dd f=0\quad\forall\,\, Y\,\in\,\mathcal D_{\alpha}\}\,\simeq\,\mathcal F(M/\Phi^{\alpha}). 
\eeq
A sufficient condition for that is that the distribution $\mathcal D_{\alpha}$ is spanned by the kernel of a set $\{\beta_{j}=\dd u'_j\}_{j=1,\dots,\delta}$ of independent exact 1-forms on $M$ such that
\beq
\label{eqp3}
\{u'_j,u'_k\}\,=\,c_{jk}^{\,\,\,\,s}u'_s, 
\eeq
i.e. the Hamiltonian vector fields $X_{u'_j}$ give a so called Hamiltonian action of the Lie algebra $\mathfrak g$ on the Poisson manifold\footnote{Notice that a slightly more general sufficient condition for the Poisson structure to be defined on $\mathcal F_{\alpha}$ is that $$\{u'_j,u'_k\}=c_{jk}^{\,\,\,s}u'_s\,+\,\sigma_{jk}$$ with $\sigma_{jk}=-\sigma_{kj}$ and $\dd \sigma_{jk}=0$.} $M$. Correspondingly to this action, one can define the map $\mu\,:\,M\to\mathfrak g^*$ given by 
$$
\mu(m)\,=\,u_j'(m)\epsilon^j,
$$
where $\{\epsilon^j\}_{j=1,\dots,\delta}$ is the basis of $\mathfrak g^*$ dual to the basis of $\mathfrak g$ mapped under the Hamiltonian action into the vector fields $X_{u'_j}$. The map $\mu$ is called the \emph{momentum map} associated to the given  Hamiltonian action of $\mathfrak g$ on the Poisson manifold $M$\footnote{A generalisation of the notion of momentum map for a given Hamiltonian action of a Lie algebra $\mathfrak g$ on a Poisson manifold $M$ is provided  by the notion of function groups, which is a set $\{f_j\}_{j=1,\dots,k}$ of independent elements in $\mathcal F(M)$ such that $\dd\{f_j,f_s\}=c_{js}^{\,\,\,m}\dd f_m$ with $c_{js}^{\,\,\,m}\in\mathcal F(M)$ depending only on the $f_j$, satisfying the antisymmetry condition $c_{js}^{\,\,\,m}=-c_{sj}^{\,\,\,m}$  and the Jacobi identity. One can say that a function group is a momentum map  when the Hamiltonian vector fields represent an infinite dimensional Lie algebra.  
Following the general reduction scheme outlined in the previous sections, the elements $f_j$ of a function group define a map $F\,:\,M\,\to\,(f_1,\dots,f_k)\,\in\,M'$ from the Poisson manifold to another Poisson manifold (target manifold) on which the canonical derivations of the subalgebra project.  }. With respect to the general reduction scheme sketched in the section \ref{ss:red}, the momentum map $\mu$ provides an example of the map $F$, with $M'=\mathfrak g^*$, the algebra $\mathcal F_{\alpha}$ defined in \eqref{eqp2} being the corresponding subalgebra (see \eqref{equa2pp}) $$\mathcal A_{\mu}=\{\mu^*(w)\,:\,w\,\in\,\mathcal F(\mathfrak g^*)\}.$$ 
 The dynamics $X_H$ is projectable onto $M/\Phi^{\alpha}$, i.e. well defined on $\mathcal F_{\alpha}$, if and only if $[X_H,Y]\in\mathcal D_{\alpha}$ for any $Y\in\,\mathcal D_{\alpha}$, which is equivalent to the \eqref{introcop}, 
\beq
\label{intrococo}
L_{X_H}u'_a\,=\,f_a(u')
\eeq
with $f_a$ depending only on $u'_j$.  The elements $u'_j$ provide a local coordinate chart on the quotient manifold $M/\Phi^{\alpha}$, which turns to be  equipped with  a Poisson structure. The relation \eqref{intrococo} allows to write the dynamics (\cite{brunov, svz67}) on $M/\Phi^{\alpha}$ as
\beq
\label{poissont}
X_H\,=\,f_a\,\frac{\del}{\del u'_a}
\eeq
We close this description upon noticing that the constants of the motion  $\Gamma_{ij}=\{u_j,u_k\}$ which close the commutation relations $\{\Gamma_{jk}, \Gamma_{ab}\}\,=\,c_{jk}^{\,\,\,s}c_{ab}^{\,\,\,m}\Gamma_{sm}$ provide a realisation of a derived algebra $\g'$ from $\g$, and allow for a reduction procedure.

\noindent An interesting generalisation of the formalism sketched above appears when there exists a finite set $\{\alpha_j\}_{j=1,\dots,\delta}$ of 1-forms which are  invariant along the orbits of the dynamics $X_H$, with (as in \eqref{eq23c})
\beq
\label{eqE1}
[\alpha_j,\alpha_k]_{\Lambda}\,=\,c_{jk}^{\,\,\,\,s}\alpha_s,
\eeq
but such that they are no longer closed, satisfy instead a Maurer-Cartan relation
\beq
\label{eqE2}
\dd\alpha_a\,+\,\frac{1}{2}\,\phi_{a}^{\,\,bc}\alpha_b\wedge\alpha_c\,=\,0
\eeq
with $c_{jk}^{\,\,\,s}$ and $\phi_{a}^{\,\,bc}$ respectively giving the structure constants of the Lie algebras $\g$ and $\tilde\g$. From \eqref{eq10} we see that the corresponding vector fields $\hat X_{\alpha_j}$ commute with the dynamics $X_H$ but need not to be canonical. A direct computation shows that
\beq
\label{eqE3}
[\hat X_{\alpha_k},\hat X_{\alpha_j}]\,=\,\left(c_{jk}^{\,\,\,\,s}\,+\,(\phi_k^{\,\,sa}\Lambda(\alpha_a,\alpha_j)\,-\,\phi_j^{\,\,sa}\Lambda(\alpha_a,\alpha_k))\right)\hat X_{\alpha_s}.
\eeq
When the elements in round bracket provide the structure constants of a Lie algebra, then one can see that $\g$ and $\tilde\g$ are the building blocks for a double Lie group (see \cite{AGMM1,AGMM2}). In this case, one also has that there exists a Lie group $\tilde G$ and a map $\tilde\Phi\,:\,M\,\to\,\tilde G$ such that $\alpha_j\,=\,\tilde\Phi^*(\omega_j)$, where $\omega_j$ are the left-invariant 1-forms associated to the basis $e_j$ of the Lie algebra of $\tilde{G}$. Moreover, the dynamics $X_H$ is in this case related to a one parameter subgroup on $\tilde{G}$. This is the starting point for  what is known as the formalism of \emph{group valued} constants of the motion.

\subsubsection{\blu{Relatively closed 1-forms invariant along a Poisson dynamics}}
\label{subsecP3}
 Assume that $\dd\alpha(X_f,X_g)=0$  for any $f,g\in\mathcal F(M)$. This condition is usually referred to as the 1-form $\alpha$ being \emph{relatively} (i.e. along Hamiltonian vector fields) closed. From the first relation out of \eqref{eq10}, we have then that $\hat X_{\alpha}$ is canonical. From the further identity
\beq
\label{eq12}
[X_H,\hat X_{\alpha}]f\,=\,(L_{\hat X_{\alpha}}\Lambda)(\dd H, \dd f)\,-\,X_f(\alpha(X_H))
\eeq
for any $f,H\in\mathcal F(M)$, it follows that the vector field $\hat X_{\alpha}$, with $\alpha$ relatively closed, provide an infinitesimal symmetry for the dynamics $X_H$ if and only if 
\beq
\label{eq13}
X_f(\alpha(X_H))\,=\,\{\alpha(X_{H}),f\}\,=\,\{i_{X_H}\alpha, f\}\,=\,0,
\eeq
i.e. if and only if the element $i_{X_H}\alpha\in\mathcal F(M)$
is a Casimir for the Poisson tensor $\Lambda$. From the identity $L_{\hat X_{\alpha}}H=-i_{X_H}\alpha$,  we see that the condition \eqref{eq13} is fulfilled if and only if $L_{\hat{X}_{\alpha}}H$ is a Casimir for the Poisson tensor.   We write this result as
\begin{prop}
\label{noepoisson3}
Given the Hamiltonian dynamics $\Gamma=X_H$ on a Poisson manifold $(M,\Lambda)$ and a relatively closed 1-form $\alpha$, the corresponding vector field $\hat X_{\alpha}$ is an infinitesimal canonical symmetry if and only if the element $L_{\hat X_{\alpha}}H\in\mathcal F(M)$ is a Casimir function for the Poisson tensor $\Lambda$.
\end{prop}
\noindent As in the case studied in the previous section \ref{subsecP2} above, there exist infinitesimal canonical symmetries $\hat X_{\alpha}$ for the  dynamics $X_{H}$ which do not preserve the level sets of $H$.

\bigskip

\noindent Assume that the 1-form ${\alpha}$ is invariant along the integral curves of the dynamics $X_H$, that is 
$L_{X_H}\alpha=0$. From \eqref{eq10} we see  then that  $\hat X_{\alpha}$ gives an infinitesimal symmetry for $X_H$. We notice that such a vector field $\hat X_{\alpha}$ is  not necessarily canonical, even under the further assumption (considered in the proposition \ref{noepoisson3} above) of $i_{X_H}\alpha$ being a Casimir for the Poisson tensor $\Lambda$: the two conditions $L_{X_{H}}\alpha=0$ and $\{i_{X_H}\alpha,f\}=0$ for any $f\in\mathcal F(M)$ imply that $\hat X_{\alpha}\{H,f\}\,=\,\{\hat X_{\alpha}H,f\}+\{H,\hat X_{\alpha}f\}$ for any $f$, and this is not equivalent to \eqref{eq9}.  It is indeed immediate to prove (using the relations \eqref{eq12}-\eqref{eq13})  that, if $L_{X_H}\alpha=0$ and $L_{\hat X_{\alpha}}\Lambda=0$, then $L_{\hat X_{\alpha}}H$ is a Casimir element for the Poisson structure.

\subsection{Symmetries  for linear Poisson dynamics}
\label{susec:linPoisson}
Given the linearity of quantum mechanics, classical linear systems provide a first  class of examples where the theory described so far give interesting results. 

On a $N$-dimensional vector space $E$, the one parameter group of diffeomorphisms $\phi_t\,:\,u\mapsto e^tu$ realises a dilation, for any vector $u\in E$. Its infinitesimal generator is the vector field $\Delta=x^a\del_a$ with respect to a  global coordinate system $\{x^a\}_{a=1,\dots,N}$, usually referred to as the Liouville vector field on $E$. Such a vector field is complete, has only one critical point $x^a=0$, and introduces a grading (see the well known Euler's theorem) within the ring $\mathcal F(E)$ of smooth functions on $E$ given by 
\beq
\label{Fri1}
\mathcal F^{(k)}(E)=\{f\in\mathcal F(E)\,:\,L_{\Delta}f=kf\}
\eeq
 for not negative integer $k$. Moreover, one has that $\mathcal F^{(0)}(E)$ is given by constant functions, while the elements in  $\mathcal F^{(1)}(E)$ separate derivations. 
 
 It is possible to prove that a Liouville vector field allows to tensorially characterise a linear finite dimensional space \cite{lie-scheffers}. 
If $M$ is a smooth finite dimensional manifold, and a complete vector field $\Delta\in\mathfrak{X}(M)$ exists, such that $\Delta$ has  only one non-degenerate
critical point,  with $\mathcal F^{(0)}(M)\simeq\R$  and  $\mathcal F^{(1)}(M)$ separating  the derivations on $\mathcal F(M)$, then $M$ can be given a unique vector space structure such that $\Delta$ coincides with the infinitesimal generators of the dilations.  Under such hypothesis, the elements in $\mathcal F^{(1)}(M)$ provide a global coordinate chart for $M$.

On such a linear space, a vector field $X\in\mathfrak{X}(M)$ is linear if and only $[X,\Delta]=0$. This condition is easily seen to be equivalent to the existence of a matrix $A$ with constant entries such that $X=A^a_{\,b}x^b\del_a$. 
In general we shall say that a dynamical system $X$ on a manifold $M$ is compatible with a linear structure on $M$ if a Liouville vector field $\Delta$ on $M$ exists and it commutes with $X$. Such a formulation has been used to analyse also the linearisation problem for non linear dynamics within a non  perturbative approach (see \cite{bambusi-etal,gaeta-m})

Once outlined a geometric (i.e. intrinsic) characterization for a linear dynamics, we consider 
$\R^N$, and  we associate to any matrix $A$ with constant entries $A^a_{\,b}$ the vector field $X_A$, to any skewsymmetric matrix $\Lambda=-\Lambda^T$ with constant entries $\Lambda^{ab}$ the Poisson structure on $\R^N$ given by $\{x^a,x^b\}=\Lambda^{ab}$, to any symmetric matrix $F=F^T$ with constant entries $F_{ab}$ the quadratic element $$f_F(x)\,=\,\frac{1}{2}F_{ab}x^ax^b,$$ to any matrix $\phi$ with constant entries $\phi_{ab}$ the 1-form $$\alpha_{\phi}=\phi_{ab}x^a\dd x^b.$$ 
It is 
\beq
\label{eq17}
\{f_A,f_B\}\,=\,f_{A\Lambda B-B\Lambda A}
\eeq
with $A=A^T$ and $B=B^T$, while the action of the Lie derivative $L_{X_M}$ is easily seen to give the maps
\begin{align}
B=B^T, \,f_B&\qquad\mapsto\qquad f_{BM+M^TB} \nn \\
X_B&\qquad\mapsto\qquad[X_M,X_B]=X_{[B,M]} \nn \\
\alpha_{\phi}&\qquad\mapsto\qquad \alpha_{M^T\phi+\phi M} \nn \\
\Lambda=-\Lambda^T&\qquad\mapsto\qquad-(M\Lambda+\Lambda M^T)
\label{eq15}
\end{align}
thus reading the differential calculus in terms of matrix calculus. The dynamics 
$\Gamma=X_A\,=\,A^a_{\,b}x^b\del_a$,  has a Poisson description on $(\R^N, \Lambda=-\Lambda^T)$ if and only if the matrix equation 
\beq
\label{eq14}
\Lambda H=A
\eeq
 is satisfied, with $H=H^T$.  This factorization problem for linear dynamics has been analysed in \cite{gio-ma-ru93} within the Hamiltonian formalism and reviewed\footnote{We notice that, if the matrix $A$ can be factorised as in \eqref{eq14}, it follows that ${\rm Tr}(A^{2k+1})=0$ for  any $k\in\N$, and $A^{2k+1}=(-1)^k\Lambda((A^T)^kHA^k)$. This means that, if a meaningful factorization $\Lambda H=A$ exists, then all odd powers of $A$ give raise to Hamiltonian vector fields with respect to the same Poisson structure: all of them commute, but they are not independent, as the Cayley theorem proves. The functions $f_k=(-1)^k((A^T)^kHA^k)_{ij}x^ix^j$ are constants of the motion for $X_A$ in involution. Even powers of $A$  on the other hand provide symmetries  which are not canonical. It is possible to show that they are related to \emph{complexified} Hamiltonian vector fields, very much as it happens for quantum dynamics on K\"ahler manifolds.  } in chapter 4 in \cite{gfd} within the Poisson formalism.  
 Assume then that $X_A$ has such a Poisson description\footnote{Notice that the isotropic harmonic oscillator dynamics corresponds to $H=1_{N}$ for $N=2k$.} . Using \eqref{eq15} it is immediate to prove that $$\hat X_{\alpha_{\phi}}=X_{\Lambda\phi^T},$$ and that the relations \eqref{eq10} can be read as the following (with constant entries) matrix relations 
\begin{align}
&L_{\hat X_{\alpha_{\phi}}}\Lambda\,\sim\,\Lambda(\phi-\phi^T)\Lambda, \nn \\
&[X_A,\hat X_{\alpha_{\phi}}]\,\sim\,\Lambda\phi^TA-A\Lambda\phi^T\,=\,\Lambda\{\phi^T\Lambda H-H\Lambda\phi^T\}, \nn \\
&L_{X_A}\alpha_{\phi}\,\sim\,A^T\phi+\phi A\,=\,\phi\Lambda H-H\Lambda\phi
\label{eq16}
\end{align} 
Recall the case described in the proposition \ref{noepoisson3}. The vector field $\hat X_{\alpha_{\phi}}$ is canonical, i.e. from the first relation in \eqref{eq16} the 1-form $\alpha_{\phi}$ corresponds to a matrix $\phi$ reading 
\beq
\label{eq16punto}
\Lambda\phi\Lambda=\Lambda\phi^T\Lambda.
\eeq The condition under which such canonical vector field is an infinitesimal symmetry for the Poisson dynamics $X_A$ comes upon considering the second relation in \eqref{eq16} together with \eqref{eq16punto}, and gives $$\Lambda\{\phi\Lambda H-H\Lambda\phi^T\}=0.$$ Since the matrix in curly bracket is symmetric, such condition amounts to state, recalling \eqref{eq17}, that the function  $$f_{\phi\Lambda H-H\Lambda\phi^T}=L_{X_{\alpha_{\phi}}}f_H$$ is a Casimir for the Poisson structure $\Lambda$.

\noindent Furthermore, the condition that $\alpha_{\phi}$ is invariant along the dynamics  $X_A$ is written as $$\phi H\Lambda=H\Lambda\phi.$$ This condition implies that $\hat X_{\alpha_{\phi}}$ is an infinitesimal symmetry for $X_A$. It is evident that such infinitesimal symmetry is also canonical if and only if the skewsymmetric matrix $\Sigma=\phi-\phi^T$ satisfies the condition $\Lambda\Sigma\Lambda=0$. Notice that $\Sigma=0$ amounts to a symmetric $\phi$, which gives $\alpha_{\phi}=\dd f_{\phi}$ and then the case studied in section \ref{subsecP1} above.  

We also  notice that the most general conditions under which the  vector field $\hat X_{\alpha_{\phi}}$ gives an infinitesimal canonical symmetry for the linear dynamics $X_A=X_{\Lambda H}$ can be written as  $$\Lambda\phi\Lambda=\Lambda\phi^T\Lambda$$ and $$\Lambda\{\phi^T\Lambda H-H\Lambda\phi^T\}=0.$$ Similar equations were found and used in \cite{be-go} by analysing constraints in general relativity.

We close this section upon providing an explicit example of the role that an invariant Poisson tensor for a dynamics has in relating symmetries with constants of the motion. 
\begin{example}
\label{expoil}

Consider the dynamics described by  the Liouville vector field itself $\Delta=x\del_x+y\del_y$ on $\R^2$, which is clearly linear and associated to the identity matrix in $\R^2$. Any linear vector field $X_A$ is an infinitesimal symmetry for $\Delta$, with the 1-parameter group of linear diffeomorphisms $\Phi(t)=e^{tA}$ mapping each integral curve of $\Delta$ into an integral curve of $\Delta$, since (by linearity)  it maps  any  vector subspace into a vector subspace in $\R^2$. At the same time, a function $f\in\mathcal F(\R^2)$ is invariant along $\Delta$ (i.e. a constant of the motion) if and only if $\Delta f =0$, which means that the set of constants of the motion can be identified with the set of elements $f\in\mathcal F({\rm S^1})$ with clearly ${\rm S}^1\simeq\R^2\backslash\{0\}/\R^+$. 

As we discussed at length  in section \ref{ss:syco},  a skewsymmetric bivector field invariant under $\Delta$ allows to connect functions invariant along $\Delta$ with infinitesimal symmetries. We begin by noticing that the equation \eqref{eq14} has no solution for a skewsymmetric matrix $\Lambda$ and a symmetric matrix $H$. 

We considered indeed  this example already when introducing Poisson structures: the Poisson tensor $\Lambda=xy\del_x\wedge\del_y$ is invariant along $\Delta$, although $\Delta$ is not Hamiltonian with respect to such $\Lambda$, since one would have $\Delta=X_H$ with an Hamiltonian function  $H=\log|y/x|$ which is defined only on $\R^2\backslash\{xy=0\}$. The tensor $\Lambda$ nonetheless allows to map an invariant function to an infinitesimal symmetry for the dynamics:  if $\Delta f=0$, then one proves  $[X_f,\Delta]=0$. Such infinitesimal symmetry might be non Hamiltonian (due to eventual singularities of $f$ as in the case of $X_H$ itself), but turns nevertheless out to be canonical, i.e. $L_{X_f}\Lambda=0$. 

Concerning the infinitesimal symmetries $X_A$, one proves that they are canonical if and only if  $A={\rm diag}(\alpha, \beta)$, and $f\,=\,\alpha\log|x|-\beta\log|y|$ defined on $\R^2\backslash\{xy=0\}$ gives $X_f=X_A$. Such $f$ is invariant along $\Delta$ if and only if $\alpha=\beta$.

\end{example}

\section{Noether theorem and  symplectic structures}
\label{sec:hamilton}

We have described at length how a Poisson structure $\Lambda$ on a manifold $M$ allows to define, at each point $m\in M$, the map  $\tau: T_m^*M\to T_mM$ (see \eqref{intro00}). The range  of the map $\tau$ defines a distribution $D^{\Lambda}$ with $D^{\Lambda}_{(m)}\subseteq T_mM$ whose rank defines the rank of the Poisson structures at $m$. Points where the rank of $\Lambda$ is maximal (and hence constant) are said to give the regular part of the Poisson tensor (such a subset is proven to be open and dense in $M$), points where such a rank is not maximal are naturally called singular. Hamiltonian vector fields on $M$ span the distribution $D^{\Lambda}$, while from the  Jacobi identity and the relation \eqref{eq8} it is clear that the set of Hamiltonian vector fields on $M$ give an involutive distribution. It can indeed be proven (this result is known as the \emph{Kirillov symplectic foliation theorem}) that there exists a singular foliation of the Poisson manifold $M$ whose leaves are such that their tangent spaces are spanned by the Hamiltonian vector fields, and that  the level sets of the Casimir functions for the Poisson tensor $\Lambda$ provide the regular leaves for the foliation. Moreover, each leaf is invariant under the flow of any Hamiltonian vector field, and this allows to define, on each leaf, a 2-form $\omega$ via 
\beq
\label{kyri}
\Lambda(\dd f,\dd g)\,=\,\omega(X_f,X_g)
\eeq
for any $f,g\in\mathcal F(M)$. This 2-form turns to be non degenerate and closed, that is \emph{symplectic}. 

Coherently with our approach to the analysis of classical dynamical systems from the perspective of the principle of analogy, we also notice that symplectic manifolds are  the classical counterparts of the set of states for  irreducible representations of  $C^*$-algebras in the quantum setting, arising when \emph{neutral elements} or Casimir operators on the Hilbert space of a quantum system  act as multiples of the identity.
This is the reason why, after we analysed the geometry of a Poisson bivector on a manifold,  we devote this section to  the study of the geometry of non degenerate closed skewsymmetric tensors.

A \emph{symplectic} structure on an orientable manifold $M$ is defined by a 2-form $\omega$ which is non degenerate (i.e. the dimension of $M$ is then necessarily even, say $2N$, and the matrix representing $\omega$ is invertible) and closed (i.e. $\dd\omega=0$). The reduction to normal form of a non degenerate bilinear antisymmetric form on a finite dimensional  vector space generalises to the Darboux theorem: if $(M,\omega)$ is a symplectic manifold, for each point $m_0\in M$ there exists an open subset $U\subset M$  and a local coordinate system on $U$ given by $(x^a, y^a)_{a=1,\dots,N}$ such that $m_0\in U$ and  $\omega_{\mid U}=\dd x^a\wedge\dd y^a$.
A diffeomorphism $\phi$ on the symplectic manifold $(M,\omega)$ is said \emph{symplectic}, or \emph{canonical}, if 
\beq
\label{ds00}
\phi^*\omega=\omega.
\eeq
A 2-form $\omega$ which is closed and has constant (non maximal) rank on a (possibly odd dimensional) manifold $M$ is called a \emph{pre}-symplectic structure.  Dynamics on a pre-symplectic manifold will be described in sections  \ref{sec:pre}.
 
The relation between symplectic structures and non degenerate Poisson structures on a manifold can be described further.
A symplectic structure $\omega$ allows to define, using the contraction operator,  the  map $\tilde{\tau}\,:\,\mathfrak X(M)\,\to\,\Lambda^1(M)$ (see \eqref{intro01}) with 
\beq
\label{eq33}
\tilde{\tau}\,:\,X\,\mapsto\,\hat{\alpha}_X=i_X\omega.
\eeq
  The relation (for $X,Y\in\mathfrak X(M)$), (see \eqref{kyri}) 
\beq
\label{eq19}
\Lambda(i_X\omega,i_Y\omega)\,=\,\omega(X,Y)
\eeq
defines, given the non degeneracy of $\omega$, a bivector field (i.e. a skewsymmetric (2,0) tensor) $\Lambda$ on $M$, which turns  to be a non degenerate Poisson tensor since the condition $\dd\omega=0$ is easily seen to be equivalent to the Jacobi identity for $\Lambda$ (see \cite{pauli-nc}). With respect to a local chart $\{x^a\}_{a=1,\dots,2N}$ on $M$, one has $$\omega=\omega_{ab}\dd x^a\wedge\dd x^b$$ with $$\hat \alpha_{X}=\omega_{ab}X^a\dd x^b$$ for $X=X^i\del_i\in\mathfrak X(M)$, 
and
\beq
\label{eq18}
\Lambda^{ab}\,=\,\{x^a,x^b\}\,=\,(\omega^{-1})^{ba},
\eeq
with
$$
\omega_{ab}(\omega^{-1})^{bc}=\delta_{a}^c,
$$
so that one immediately proves, recalling \eqref{eq5},  that $\tilde{\tau}=\tau^{-1}$.

What we have described shows that any symplectic manifold $(M,\omega)$ \emph{has} a non degenerate Poisson structure $\Lambda$ defined by \eqref{eq19}. This can be reversed. If a manifold $M$ has a non degenerate Poisson tensor $\Lambda$, written as $\Lambda=\Lambda^{ab}\del_a\wedge\del_b$ with respect to a local chart $\{x^a\}_{a=1,\dots,2N}$, the tensor $$\omega=-(\Lambda^{-1})_{ab}\dd x^a\wedge\dd x^b$$ is a symplectic structure on $M$ and the relation \eqref{eq19} holds.  

A vector field $X$ on the symplectic manifold $(M,\omega)$ is said \emph{locally Hamiltonian} if 
\beq
\label{loch}
L_X\omega=0.
\eeq
 Since $\dd\omega=0$, this condition is equivalent to the condition $\dd(i_X\omega)=0$, i.e. a vector field is locally Hamiltonian 
 if and only if the 1-form $\hat\alpha_X=i_X\omega$ is closed. We denote by $\mathfrak{X}_{LH}(M)\subset\mathfrak X(M)$ the set of locally Hamiltonian vector fields on $(M,\omega)$. 

 A vector field $X$ is said \emph{globally Hamiltonian}, or simply Hamiltonian, if the 1-form $i_X\omega$ is exact, that is if a function $f\,\in\mathcal F(M)$ exists such that \beq
 \label{eq20}
 i_X\omega=\dd f.
 \eeq
  In such a case the function $f$, which is determined by $X$ up to a constant, is said the Hamiltonian function of $X$. We denote by $\mathfrak{X}_H(M)\subset\mathfrak{X}_{LH}(M)$  the set of globally Hamiltonian vector fields on $(M, \omega)$.  It is immediate to recover (recalling the non degeneracy of the  2-form $\omega$) that, if $i_X\omega=\dd f$, then $X=X_f$ as defined in \eqref{eq6}, so one can write 
  \beq
  \label{eq21}
  \{f,g\}\,=\,\Lambda(\dd f, \dd g)\,=\,\omega(X_f,X_g),
  \eeq 
which is \eqref{kyri}. The following chain of identities based on the properties of the exterior Cartan calculus proves  that the commutator of two locally Hamiltonian vector fields $X,Y$ is globally Hamiltonian, since 
\beq
\label{eq22-1}
i_{[X,Y]}\omega\,=\,(L_Xi_Y-i_YL_X)\omega\,=\,L_Xi_Y\omega\,=\,i_X\dd(i_Y\omega)+\dd(i_Xi_Y\omega)\,=\,\dd 
(i_Xi_Y\omega).
\eeq 
This shows that $\mathfrak{X}_{LH}(M)$ is  a Lie subalgebra in $\mathfrak{X}(M)$, while the relation \eqref{eq8} shows that $\mathfrak{X}_{H}(M)$ is 
a Lie subalgebra of $\mathfrak{X}_{LH}(M)$, indeed a Lie algebra ideal\footnote{Notice also that, due to the non degeneracy of the 2-form $\omega$, it is possible to prove that on a symplectic manifold $(M,\omega)$ the $\mathcal F(M)$-module of vector fields $\mathfrak X(M)$ has a basis of Hamiltonian vector fields, namely those corresponding to the coordinate functions.}, as the identity  (for any $L_A\omega=0$)
\beq
\label{eq22}
i_{[A,X_f]}\omega\,=\,\dd(L_Af\,+\,i_Ai_{X_f}\omega),
\eeq
together with \eqref{eq8} proves.  Thus the derived Lie algebra is always made of Hamiltonian vector fields  and the quotient of locally Hamiltonian with respect to Hamiltonian ones is related to the first cohomology group of the manifold $M$.

A dynamical system $\Gamma$ on $M$ is said to have a Hamiltonian description if a symplectic structure $\omega$ on $M$ exists and if a (then so called) Hamiltonian function $H\,\in\,\mathcal F(M)$ exists, such that $\Gamma=X_H$, or equivalently such that $$i_{\Gamma}\omega=\dd H.$$  It is clear that, if $\Gamma=X_H$, then $H$ \emph{is} a constant of the motion\footnote{We pointed out already in the introduction that this comes from the skewsymmetry of the tensor $\omega$.}. Moreover, it is from the closedness condition $\dd\omega=0$ that we can prove that, as within the Poisson formalism,  the constants of the motion close a Lie subalgebra of $\mathcal F(M)$ with respect to the Poisson bracket.

\subsection{Submanifolds of a symplectic manifold}
\label{sub:dirac}
The symplectic 2-form $\omega$ allows to define, on the vector space $T_mM$ tangent in the point $m\in M$ to the manifold $M$, a notion of symplectic orthogonality: for any linear subspace $V_m\subset T_mM$, one defines 
\beq
\label{orthosymp}
V_m^{\perp}\,=\,\{u\in T_mM\,:\,\omega(u,v)=0\,\,\forall\, v\in V_m\}.
\eeq
From the definition it is easy  to directly prove that $$V_m\subset W_m\,\Rightarrow W^{\perp}_m\subset V^{\perp}_m$$ and $(V_m^{\perp})^{\perp}=V_m$. One defines a vector subspace $V_m$ \emph{isotropic} if $$V_m\subset V_m^{\perp}$$ (so that ${\rm dim}\,V_m\leq N$); as \emph{coisotropic} if $$V_m^{\perp}\subset V_m$$ (so that ${\rm dim}\,V_m\geq N$); as \emph{Lagrangian} if $$V_m=V_m^{\perp}$$ (so that ${\rm dim}\,V_m=N$), as \emph{symplectic} if $$V_m\cap V_m^{\perp}=0.$$ A distribution $\mathcal D$ on a symplectic manifold $(M,\omega)$ will be said isotropic (resp. coisotropic, Lagrangian, symplectic) if the vector subspaces in $\mathcal D_{(m)}\subseteq T_mM$ satisfy such conditions at each point $m$.  
A submanifold $M'\hookrightarrow M$ will be called isotropic (resp. coisotropic, Lagrangian, symplectic) if its tangent space corresponds to an isotropic (resp. coisotropic, Lagrangian, symplectic) distribution in $TM$. 

It is useful for what we shall describe in the following sections to relate the definitions above to a description of the submanifold in terms of a system of constraints \cite{dirac50, dirac-lec, me-tu78, sm-book}. Let $M'$ be the $\delta$-dimensional manifold defined by  
\beq
\label{mprimo}
M'\,=\,\{m\in \,M\,:\,f_j(m)=0\}
\eeq
with $\{f_j\}_{j=1,\dots,2N-\delta}$ a set of independent smooth functions, i.e. $\dd f_1\wedge\dots\wedge\dd f_{2N-\delta}{}_{\mid M'}\neq0$. It is immediate to see that the tangent distribution $\mathcal D'$ to $M'$ is spanned at each point $m'\in M'$ by the vector fields $X$ on $M$ which give $i_X\dd f_j=0$: upon denoting by $\mathcal D_{H}'$ the distribution spanned at each point $m'\in M'$ by the Hamiltonian vector fields $X_{f_j}$, one proves that 
\beq
\label{dco1}
\mathcal D'=\mathcal D_{H}'^{\perp}.
\eeq 
Consider the antisymmetric matrix ${\rm C}_{js}(m)=\{f_j,f_s\}$, and assume its rank is constant on $M'$, with ${\rm rk}({\rm C}_{ab})_{\mid M'}=2N-\delta-k$. This amounts to say that the null eingenspace $\ker({\rm C}_{js})$ is spanned by a set of independent (on $M'$)  elements $\phi_{aj}\in\mathcal F(M)$ solving   
\beq
\label{dco2}
\{f_j,f_s\}\phi_{as}{}_{\mid M'}=0
\eeq
with ${a=1,\dots,k}$; $j,s=1,\dots,2N-\delta$. The functions $\phi_a=\phi_{aj}f_j\in\mathcal F(M)$  are then, following Dirac, a maximal set of independent \emph{first class} constraints with the properties 
\begin{align}
&\{\phi_a, f_j\}_{\mid M'}=0, \nn \\
&\{\phi_a,\phi_b\}_{\mid M'}=0
\label{dco3}
\end{align}
for any $a,b=1,\dots,k$ and $j=1,\dots,2N-\delta$. We see from the first relation out of \eqref{dco3} that each $X_{\phi_a}\in\mathcal D'$, while from \eqref{dco1} (since $\mathcal D_{H}'$ is spanned by the Hamiltonian vector field associated to \emph{any} constraint function) we see that $X_{\phi_a}\in\mathcal D'^{\perp}$, i.e. 
\beq
\label{dco4}
X_{\phi}\in\mathcal D'\cap\mathcal D'^{\perp}
\eeq
for any first class combination $\phi\in\mathcal F(M)$.
 Moreover, from the second relation out of \eqref{dco3} we see that, given the independence of the $X_{\phi_a}$, the number of first class independent constraints $k$ cannot exceed $N$. 

If $M'$ is isotropic, then $\mathcal D'\subset\mathcal D'^{\perp}$ and one can prove this happens (from \eqref{dco1} and \eqref{dco4})  if and only if the number of independent first class constraints equals the dimension of $M'$. If $M'$ is coisotropic, then $\mathcal D'^{\perp}\subset\mathcal D'$ and from the same relations one sees that it happens if and only if all the $f_j$ are first class. The submanifold $M'$ is Lagrangian when both conditions are simultaneously satisfied, that is when $M'$ is defined via $\delta=N$ independent first class constraints. Finally, the submanifold $M'$
is symplectic if and only if there are no first-class constraints among the $f_j$ elements. 

We close this presentation by noticing that, if we compare the above description with that of the reduction for a Poisson algebra (section \ref{sususe:Po1}), then it is easy to see that, given the submanifold $M'$ in terms of the defining constraints \eqref{mprimo}, we can define the ideal $$\mathcal I_{M'}\,=\,\{f\in\mathcal F(M): f_{\mid M'}=0\}$$ and its normaliser with respect to the Poisson structure, i.e. $$\mathcal N_{M'}\,=\,\{u\in\mathcal F(M):\{u,f\}\in\mathcal I_{M'}\,\forall\,f\in\mathcal I_{M'}\}.$$ Elements in $\mathcal N_{M'}$ give the Hamiltonian vector fields which are tangent to $M'$, while elements in the intersection $\mathcal I_{M'}\cap\mathcal N_{M'}$ give the set of first class constraints.  

\subsection{Exact symplectic manifolds}\label{sub:exasy}
Since $\omega$ is closed, a symplectic manifold $(M,\omega)$  is called \emph{exact} if a 1-form $\theta\in\Lambda^1(M)$ is globally defined and reads 
\beq
\label{symfo1}
-\dd\theta=\omega
\eeq
 (notice that the minus sign is a matter of historical convention). An example of a symplectic manifold which is not exact symplectic is given by the 2-dimensional torus ${\rm T}^2$, with local angular coordinates $\{\phi_a\}_{a=1,2}$. The volume form $\omega=\dd\phi_1\wedge\dd \phi_2$ is symplectic, but not exact, because if we had a globally defined potential 1-form $\theta$ on ${\rm T}^2$ then by the Stokes theorem it would be 
 $$
 4\pi^2\,=\,\int_{{\rm T}^2}-\dd\theta\,=\,\int_{\del{\rm T}^2}-\theta\,=\,0
 $$ since $\del{\rm T}^2=\emptyset$.
 If $Q$ is a smooth $N$-dimensional orientable manifold, its cotangent bundle $\pi_Q\,:\,T^*Q\to Q$ gives a natural example of an exact symplectic manifold, with $\omega_Q=-\dd\theta_Q$ where the 1-form $\theta_Q\in\Lambda^1(T^*Q)$ is globally defined via $$\sigma^*(\theta_Q)=\sigma,$$ where $\Lambda^1(Q)\ni\sigma:Q\to T^*Q$ is  a smooth section of $T^*Q$  and $\sigma^*(\theta_Q)$ denotes the pullback of $\theta_Q$ via $\sigma$. With respect to the natural local bundle coordinates $(q^a, p_a)_{a=1,\dots, N}$ (with $q^a$ providing local coordinates on the base manifold  $Q$ and $p_a$ global fiber coordinates) one has 
 \begin{align}
 &\theta_Q=p_a\dd q^a, \nn \\ &\omega_Q=-\dd\theta_Q=\dd q^a\wedge\dd p_a.
 \label{symfo2}
 \end{align}
  This shows that on a cotangent bundle manifold the base manifold is given as the quotient of a Lagrangian foliation, and each fiber of this foliation is a vector space, so that a suitable Liouville-type  vector field $\Delta=p_a\del_{p_a}$ is defined, with 
  $$L_{\Delta}\omega_Q=\omega_Q.
  $$
It is possible to prove \cite{gfd} that these structures locally\footnote{There exist examples, within the category of Stein manifolds in complex geometry, of exact symplectic manifolds which are not globally canonically diffeomorphic to a cotangent bundle manifold.} characterise a cotangent bundle manifold: there exists a local symplectic diffeomorphism between $(M,\omega)$  (with ${\rm dim}\, M=2N$) and $(T^*Q, \omega_Q)$ if and only if there exists a partial Liouville structure $\Delta\in\mathfrak X(M)$ such that the zero manifold (i.e. the manifold of critical points) for $\Delta$ is $N$-dimensional and $L_{\Delta}\omega=\omega$. By a partial Liouville structure on $M$ we mean, generalising what we described in section \ref{susec:linPoisson}, a complete vector field $\Delta\in\mathfrak X(M)$ 
such that the subalgebra 
$\mathcal F^{(0)}$ (defined as in \eqref{Fri1}) and the 
$\mathcal F^{(0)}$-bimodule $\mathcal F^{(1)}$  (again from  \eqref{Fri1}) are finitely generated and together with 
the 1-forms $\dd \mathcal F^{(0)},\, \dd \mathcal F^{(1)}$ give a $\mathcal F(M)$-bimodule basis for the whole $\Lambda^1(M)$; and moreover such that the critical set for $\Delta$ is a submanifold whose algebra of functions is isomorphic to 
$\mathcal F^{(0)}$. The distribution
 $$
 \mathcal D=\{Y\in\mathfrak X(M)\,:\,L_Yf=0\,\,\forall \,f\in\mathcal F^{(0)}\}
 $$ 
 is integrable; the condition given by $L_{\Delta}\omega=\omega$ allows to recover that the corresponding foliation is Lagrangian, so that  one can denote by $\{q^a\}_{a=1,\dots N}$ a local coordinate chart on the quotient manifold (the base $Q$  for the bundle) corresponding to the foliation generated by the distribution $\mathcal D$ and by $\{p_a\}_{a=1,\dots,N}$ the fiber coordinates on each leaf, so that $\Delta=p_a\del_{p_a}$. 
We notice that  a given symplectic manifold can have alternative cotangent bundle structures. A trivial example for that is given by the vector space $M=\R^{2N}$, where $\theta=p_a\dd q^a$ and $\theta'=-q^a\dd p_a$ result in the same symplectic 2-form, but give base manifolds for the associated cotangent bundle which are related by a Fourier transform. 

We close this introduction to the geometry of exact symplectic manifolds by noticing that, if $M=T^*Q$, then we can lift  any diffeomorphism $\phi:Q\to Q$ to the canonical diffeomorphism $\tilde\phi:T^*Q\to T^*Q$ whose action on the fibres of $T^*Q$ is given by the condition $\tilde \phi^*\theta_Q=\theta_Q$. With respect to the Darboux chart on $T^*Q$, this reads
\beq
\label{calif}
\tilde\phi\,:\,(q^a,p_a)\quad\mapsto\quad(\phi^a(q), \left(\frac{\del\phi^a}{\del q^b}\right)^{-1}p_b).
\eeq
The map $\tilde\phi$ is called the cotangent lift to $T^*Q$ of $\phi$ on $Q$.  If $X\in\mathfrak X(M)$ is the infinitesimal generator of a one parameter group of diffeomorpisms $\Phi_s$ on $M$, with a local expression $X=X^a\del_{q^a}$, then the cotangent lift $\tilde\phi_s$ gives a one parameter group of canonical diffeomorphisms whose infinitesimal generator $\tilde{X}$ on $T^*M$ has the coordinate expression   
\beq
\label{califf}
\tilde X\,=\,X^a\frac{\del}{\del q^a}\,-\,p_b(\frac{\del X^b}{\del q^a})\frac{\del}{\del p_a}.
\eeq


\subsection{Symmetries and reduction within the symplectic formalism}
\label{susec:symplectic}

Before entering a systematic analysis of the Noether theorem within the Hamiltonian formalism, we notice  that, if $[A,X_H]=0$, then the function $$\phi_A=i_{X_H}i_A\omega=-i_A\dd H=-L_AH$$ is a constant of the motion, since $L_{X_H}(L_AH)=L_A(L_{X_H}H)+L_{[X_H,A]}H=0$. It is nonetheless crucial to stress that there is no control on how effective this procedure may be, and there are several example where it turns to be empty (see \cite{mssv85}).

Following the path outlined in the introduction, we devote our attention to the relations between invariant 1-forms and infinitesimal symmetries for a given Hamiltonian dynamics $X_H$ on a symplectic manifold $(M,\omega)$. 
\subsubsection{\blu{Constants of the motions and symplectic reduction}}
\label{subsub:H1}
It is immediate to see from \eqref{eq8},\eqref{eq9} and \eqref{eq23} that, if $L_{X_H}u=0$, i.e. $u$ is a constant of the motion, then the vector field $X_u$ is a canonical (i.e. it is compatible with the symplectic structure) infinitesimal symmetry for the dynamics $X_H$. This claim is obviously equivalent to the case studied in section \ref{subsecP1}  within the Poisson formalism, see the proposition \ref{noepoisson1}.
This is the well known  form of the Noether theorem within the Hamiltonian formalism, that we write as
\begin{prop}
\label{noesymp1}
Given a Hamiltonian dynamics $\Gamma=X_H$ on a symplectic manifold $(M,\omega)$, if $u\in\mathcal F(M)$ is a constant of the motion, then the corresponding Hamiltonian vector field $X_u$ is an infinitesimal canonical symmetry for $\Gamma$. 
\end{prop}

\noindent 
We now focus  on a reduction procedure related to the existence of a constant of the motion for a given Hamiltonian dynamics (see \cite{mss79, mssv85}). The example we are going to describe is intended to  clarify what one means by symplectic reduction and to prepare for a more general analysis of the problem. 

\begin{example}
\label{exemplum1} Consider the dynamics $\Gamma$ of a point particle moving in an external radial force field. The configuration space for such system is $Q=\R^2$ with global coordinates $(x,y)$ the corresponding  phase space $M=T^*\R^2$ has the canonical symplectic structure $$\omega=\dd x\wedge\dd p_x+\dd y\wedge\dd p_y,$$ so $\Gamma=X_H$ with 
$$
H=\frac{1}{2}(p_x^2+p_y^2)+V(r)
$$
where $r^2=x^2+y^2$. The function 
\beq
\label{Lfun}
L=xp_y-yp_x
\eeq
 (the angular momentum of the given particle) is invariant along the integral curves of $X_H$, $$\{H,L\}=0.$$ Consider $L=l\neq 0$. The set 
$$
N_l\,=\,\{m\in M\,:\,L(m)=l\}
$$
is   a three dimensional embedded $i_l\,:\,N_l\,\hookrightarrow\,M$ submanifold. The integral curves of $X_H$ lies at any time $t$ on $N_l$ for any choice of Cauchy data giving $L=l$. 
In order to elaborate such example, we consider the set of two dimensional spherical coordinates $(\theta, r>0)$ on $Q_0=Q\backslash\{0\}$ (notice that the choice $L=l\neq0$ amounts to consider integral curves which do not pass through the origin in $Q$) with 
\begin{align*}
&x=r\cos\theta, \\ & y=r\sin\theta,
\end{align*} and 
\begin{align}
&rp_r\,=\,xp_x\,+\,yp_y, \nn \\
&p_{\theta}\,=\,xp_y\,-\,yp_x\,=\,L
\label{trasfopp}
\end{align}
for the corresponding conjugate variables. On $T^*Q_0$ it is $\Gamma=X_H$ with 
$$
H\,=\,\frac{1}{2}(p_r^2\,+\,\frac{p^2_{\theta}}{r^2})+V(r),
$$ 
reading the o.d.e. system
\begin{align}
\dot\theta\,=\,p_{\theta}/r^2, &\qquad \dot p_{\theta}\,=\,0, \nn \\
\dot r\,=\,p_r, &\qquad 
\dot p_r\,=\,p_{\theta}^2/r^3\,-\,\del_rV
\label{sODE1}
\end{align}
The embedding $i_l$ that comes from the Cauchy data corresponds  to $N_l\,=\,\{m\in T^*Q_0\,:\,p_{\theta}=l\}$. The equations above are written on $N_l$ as 
\begin{align}
\dot\theta\,=\,l/r^2, &\qquad p_{\theta}\,=\,l\neq0, \nn \\
\dot r\,=\,p_r, &\qquad 
\dot p_r\,=\,l^2/r^3\,-\,\del_rV
\label{sODE2}
\end{align}
The radial motion is described by the second line equations \eqref{sODE2}, and is decoupled from the angular motion: once a solution $r(t)$ is obtained, the angular motion (given by the first line equation in \eqref{sODE2}) can be determined, thus giving a complete solution to the dynamical problem. 

 Notice that  if the Cauchy data select $l=0$ it is easy to see that there exists a straight line $S=\R$ through the origin in $Q$ such that the dynamics reduces to the Newtonian equation $$\ddot s=-\del_sV$$ with $V(s)=V(-s)$ for the global coordinate $s$ on $S$.

As we already commented on, when discussing about the reduction procedure within the Poisson formalism in the previous section, the presence of the \emph{single}  constant of the motion $L=p_{\theta}$ for the Hamiltonian dynamics $X_H$ on a phase space can give a  \emph{two} dimensional reduction of the dynamics. Moreover, the equations for the radial motion turn to have a Hamiltonian formulation, with 
\beq
\label{tildeHex}
\tilde H\,=\,\frac{1}{2}\left(p_r^2\,+\,\frac{l^2}{r^2}\right)+V(r)
\eeq
 on the phase space $T^*\R_+$ with the symplectic form $\tilde\omega=\dd r\wedge\dd p_r$. 

We cast this example within the general reduction scheme outlined in section \ref{sec:uno}. The constant of the motion  $L:T^*\R^2\,\to\,\R_0$ gives $L_*(X_H)=0$ (see \eqref{equazione2}), 
the involutive distribution 
$$
\mathcal D_L\,=\,\{Y\,\in\,\mathfrak X(M)\,:\,i_Y\dd L=0\}
$$ 
gives the set of vector fields in $M$ which are tangent to $N_l$ (see \eqref{equazione2p}): one clearly has that both  $X_H$ and $X_L$ are in $ \mathcal D_L$.  
Each integral curve of $X_L$ gives a leaf of a (regular, for $l\neq0$) foliation in $N_l$, whose quotient turns to be the  2 dimensional manifold $T^*\R_+$, so we have the fibration 
$$
\pi\,:\,N_l\,\to\,T^*\R_+.
$$ The basis of such a fibration has the symplectic structure $\tilde\omega$, and from the condition $$[X_H,X_L]=0$$ we see  that the vector field $X_H$ is projected onto  a vector field on $T^*\R_+$, and such projected vector field  turns to be  Hamiltonian with respect to the Hamiltonian function $\tilde H$ and the symplectic form $\tilde\omega$.   

\noindent Such a reduction procedure, which seems quite \emph{ad hoc}, can be analysed under a more general approach. On the embedded submanifold $i_{l\neq0}:N_l\hookrightarrow M$ the 2-form 
$$
\omega_l=i_l^*\omega=\dd r\wedge\dd p_r
$$ is closed and degenerate. Its kernel is spanned by the Hamiltonian vector field   $X_L=\del_{\theta}$ in $\mathcal D_{L}$  corresponding to the constant of the motion $L$. This  kernel gives an involutive regular distribution. One proves that the corresponding fibration is $\pi\,:\,N_l\,\to\,T^*\R_+$, and that $\omega_l$ induces the symplectic structure $\tilde\omega$ on the basis. 

\noindent The \emph{restriction} of the vector field $X_H$ to $N_l$ is projected  onto $T^*\R_+$, where it turns to be Hamiltonian with respect to $\tilde\omega$ and Hamiltonian function $\tilde H$ such that 
\beq
\label{riduH}
H_{\mid N_l}=\pi^*\tilde H.
\eeq 
Such a restriction clarifies the reason why the corresponding Hamiltonian function $\tilde H$ \eqref{tildeHex} depends on the value of $l$. Notice that not the whole dynamics is projected on the quotient $T^*\R_+$, since a piece of the problem, namely the  angular equation \eqref{sODE2} $\dot\theta=l/r^2$, has to be analysed after the reduced motion on $T^*\R_+$ had been solved.  
\end{example}

\noindent This approach to the reduction procedure based on the existence of constants of the motion for a Hamiltonian dynamics is what is generalised in the following pages.

\bigskip
\noindent Assume that a symplectic dynamics $X_H$ on $(M,\omega)$ has a set  $\{u_j\}_{j=1,\dots,\delta}$ of constants of the motion which are functionally independent and provide a (local) submersion $$F\,:\,M\,\to\,(u_1, \dots, u_{\delta})\subset\R^{\delta}.$$ Such a submersion defines the involutive distribution 
\beq
\label{eqsy6}
\mathcal D_{F}\,=\,\{Y\,\in\,\mathfrak X(M)\,:\,L_Yu_j\,=\,0\}
\eeq
that  gives the space of vector fields which are tangent to the leaves of the corresponding foliations $\Phi^{F}$. We denote each leaf by 
\beq
\label{eqsy7}
N_c\,=\,\{m\,\in\,M\,:\,F(m)=c\}
\eeq
with $c\,\in\,\R^{\delta}$ and restrict our attention to regular values for $\dd F$, so that the leaf $N_c$ is a manifold embedded into $M$, with $i\,:\,N_c\hookrightarrow M$ and ${\rm dim}\,N_c\,=\,2N-\delta$. From the condition  $L_{X_H}u_j=0$ we have  $X_H\in\mathcal D_{F}$, which means that the dynamical flow lies on a (given by the initial conditions)  leaf $N_c$. The tensor $$\omega_c=i^*\omega$$ is a (possibly degenerate) closed 2-form on $N_c$. Its kernel is seen  to give an involutive distribution which can be written, recalling the notion of orthosymplectic subspace \eqref{orthosymp}, as the intersection
$$
{\rm ker}\,\omega_c\,=\,\mathcal D_{F}\cap\mathcal D_{F}^{\perp}.
$$
The identity $\mathcal D_{F}=\mathcal D_X^{\perp}$  (the analogue of the relation \eqref{dco1}, where  
the distribution $\mathcal D_X$ is the span of the Hamiltonian vector fields $\{X_{u_j}\}_{j=1,\dots,\delta}$  in analogy to what we considered in the previous section\footnote{Within the classical approach developped by Whittaker and Levi-Civita, a reduction procedure associated to a set of constants of the motion for a symplectic dynamics $\Gamma=X_H$ was based on the fact that a suitable parametrization for the integral curves of the vector fields $X_{u_j}$ gave the set of variables $\varphi_j$ \emph{conjugate} to the $u_j$. Along a local coordinate system given  upon completing the $u,\varphi$, the Hamiltonian would indeed depend on the $\varphi_j$ \emph{ignorable} coordinates. })
allows to prove that  the intersection above is  
\beq
\label{eqsy1}
{\rm ker}\,\omega_c\,=\,\mathcal D_{F}\cap\mathcal D_X,
\eeq
(at each point in $N_c$) or equivalently 
\beq
\label{eqsy1poi}
{\rm ker}\,\omega_c\,=\,\mathcal D_{X}\cap\mathcal D_X^{\perp}.
\eeq
Notice  that, since we have not assumed that the functions $\{u_j\}$ close a function group, the distribution $\mathcal D_X$ is not necessarily involutive, and the Hamiltonian vector fields $X_{u_j}$ are indeed not necessarily tangent to  $N_c$, since 
$$
L_{X_{u_j}}f_k=\{f_k,f_j\}.
$$ 
If we assume that the rank of the distribution \eqref{eqsy1} is constant on each point of $N_c$ and for values of $c$ in a suitable open subset in $\R^\delta$,  then the  corresponding foliation $\Phi^{\omega_c}$ reads the quotient $N_c/\Phi^{\omega_c}$ which has a manifold structure\footnote{In analogy to what we described in section \ref{sub:dirac}, it is possible to prove that the dimension of the quotient is given by $(2N-2\delta+\lambda)$, where $\lambda$ is the (functional) rank of the matrix $F_{ij}=\{u_i,u_j\}$.}. On such a quotient manifold the 2-form $\omega_c$ turns to be closed and  non degenerate, i.e. symplectic $\tilde{\omega}$. 
It is indeed easy to see that 
\begin{align*}
&i_Y\omega_c=0, \\
&L_Y\omega_c=0
\end{align*}
for any $Y\in\ker\,\omega_c$, and this reads that there exists a 2-form $\tilde\omega$ on $N_c/\Phi^{\omega_c}$, which is also non degenerate, such that 
$$
\omega_c=\pi^*\tilde\omega
$$
where 
$$\pi\,:\,N_c\to N_c/\Phi^{\omega_c}$$ denotes  the canonical projection defined by the foliation $\Phi^{\omega_c}$.  
The dynamics $X_H$ is projectable onto $(N_c/\Phi^{\omega_c}, \tilde{\omega})$, with 
\beq
\label{eqsy2}
i_{X_H}\tilde{\omega}\,=\,\dd \tilde H
\eeq
where one has for the restriction $H|_{N_c}\,=\,\pi^*(\tilde H)$ with $\tilde H\in\mathcal F(N_c/\Phi^{\omega_c})$. This means that a set of constants of the motion allows to reduce a symplectic dynamics $X_H$ to a vector field on a suitable quotient, and such a reduced vector field is Hamiltonian with respect to a symplectic structure induced on the quotient. Notice that, since the symplectic structure can be formulated in terms of  a 2-form, i.e. a covariant tensor, we have described the reduction process not in terms of algebra quotients (as we did within the Poisson formalism), but in terms of manifolds and maps between them. 

The motion of a point particle in a central force field in $\R^3$ gives an interesting generalization of the analysis performed in the two dimensional example \ref{exemplum1}. 
\begin{example}
\label{exemplum2}
Consider the Hamiltonian dynamics $\Gamma=X_H$ given by 
$$
H\,=\,\frac{1}{2}p^2+V(r), \qquad\qquad \omega=\dd x_j\wedge\dd p_j
$$
on $M=\R^6=T^*\R^3$ with the usual global symplectic coordinates $(x_j,p_j)_{j=1,\dots,3}$ and $p^2=p_jp_j, \,r^2=x_jx_j$. The three functions 
\beq
\label{angum}
L_j\,=\,\varepsilon_{jab}x_ap_b
\eeq
give a set of constants of the motion. Fixing their values amounts to select  $$N_L\,=\,\{m\in M\,:\,L_j(m)=l_j\},$$ which for $l_jl_j\neq 0$ gives a three dimensional submanifold with the embedding $i_L\,:\,N_L\hookrightarrow M$. Notice that, since it is $$\{L_j, L_k\}=\varepsilon_{jks}L_s,$$ the Hamiltonian vector fields $X_j$ given by $$i_{X_j}\omega=\dd L_j$$ are not tangent to $N_L$. It is indeed 
easy to see that the kernel of the closed 2-form $$\omega_L=i_{L}^*\omega$$ is spanned (see \eqref{eqsy1}) by $$X\,=\,L_jX_j.$$ Such a vector field $X\in\mathfrak X(N_L)$ is the vertical vector field for the fibration $\pi:N_L\to M'$, where $M'$ turns to be symplectic.

\noindent In order to understand the topology of the symplectic quotient $M'$ we proceed as in the example \ref{exemplum1}, i.e. we study the problem in $T^*(\R^3\backslash\{0\})$ using spherical coordinates
\begin{align*} 
&q_1=r\sin\theta\cos\phi, \\ &q_2=r\sin\theta\sin\phi, \\ & q_3=r\cos\theta
\end{align*}
with $\theta\in(0,\pi), \,\phi\in(0,2\pi), r>0$, (notice that this amounts to consider only integral curves for $\Gamma$ in $M$ which do not intersect the origin in the configuration space $Q=\R^3$, i.e. to fix $l_jl_j\neq 0$). The dynamics can be written as
\begin{align}
\dot\theta\,=\,p_{\theta}/r^2,&\qquad\qquad\dot p_{\theta}\,=\,0, \nn \\
\dot\phi\,=\,p_{\phi}/(r^2\sin\theta),&\qquad\qquad\dot p_{\phi}\,=\,0, \nn \\
\dot r\,=\,p_r,&\qquad\qquad\dot p_r\,=\,\{p_r,H\}
\label{sODE3}
\end{align}
with the Poisson bracket corresponding to $\omega=\dd r\wedge\dd p_r+\dd\theta\wedge\dd p_{\theta}+\dd\phi\wedge\dd p_{\phi}$ and 
$$
H\,=\,\frac{1}{2}\left(p_r^2\,+\,\frac{1}{r^2}\left(p_{\theta}^2+\frac{p_{\phi}^2}{\sin\theta}\right)\right)+V(r).
$$
Fixing the values $p_{\theta}=c_{\theta}, \,p_{\phi}=c_{\phi}$ for the constants of the motion results in the embedding $$i_c\,:\,N_c\,\hookrightarrow\,M,$$ with the degenerate closed 2-form $$\omega_c=i_{c}^*\omega=\dd r\wedge\dd p_r.$$ Its kernel is two dimensional and spanned by the Hamiltonian vector fields $\{\del_{\theta}=X_{p_{\theta}}, \del_{\phi}=X_{p_{\phi}}\}$. These vector fields provide the vertical fields for the fibration $\pi:N_c\to T^*\R_0\simeq M'$. The restriction $X_{H}\mid_{N_c}$ is projected onto the Hamiltonian vector field $X_{\tilde H}$ on $T^*\R_0$ with 
$$
\tilde H\,=\,\frac{1}{2}\left(p_r^2\,+\,\frac{1}{r^2}\left(c_{\theta}^2+\frac{c_{\phi}^2}{\sin\theta}\right)\right)+V(r). 
$$
Once the radial problem is solved, the angular motion given by
$$
\dot\theta\,=\,c_{\theta}/r^2, \qquad\qquad 
\dot\phi\,=\,c_{\phi}/(r^2\sin\theta),
$$
can be analysed. Alternatively, we can notice that, since it is $q_jL_j=0$, the motion in the configuration space is planar, and up to a rotation in $Q_0=\R^3\backslash\{0\}$, it is no loss of generality to fix the values of the angular momentum functions \eqref{angum} as 
$$
L_1=L_2=0, \qquad L_3=l,
$$
which selects $N_l$ as 
$$
\theta=\pi/2,\qquad p_{\theta}=0, \qquad p_{\phi}=l,
$$ 
and then its tangent space $\D_L$ as the distribution spanned by the vector fields $\{\del_{\phi}, \del_r, \del_{p_{r}}\}$. It is 
$$
\omega_l=i_{l}^*\omega=\dd r\wedge\dd p_r;
$$
It is evident that $\ker\omega_l=\{\del_\phi\}$, and it is an easy calculation to prove it coincides with the restriction of $X=L_jX_j$ to $N_l$. We close the analysis of this example by noticing that the conditions $\theta=0, p_{\theta}=0$ map this three dimensional dynamics into the two dimensional considered in the example \ref{exemplum1}.

\end{example}

\subsubsection{\blu{Symmetry action of a Lie algebra and the symplectic reduction}}
\label{subsub:liesym}
The reduction procedure presented above does not require any Lie algebra structure: the constants of the motion $u_j$ need not be in involution nor close a Lie algebra with respect to the Poisson tensor corresponding to the symplectic form. We describe now how the reduction procedure can be modified, when this occurs (see \cite{mawe74}).

\noindent Assume that  a $\delta$-dimensional Lie algebra $\g$ with commutator structure given, along a basis $\{e_a\}_{a=1,\dots,\delta}$, by $$[e_a,e_b]=c_{ab}^{\,\,\,s}e_s,$$ has a symplectic action on $(M,\omega)$. In analogy to what we described in section \ref{subsub:momp}, under the name of symplectic action of a Lie algebra we mean that there exists a Lie algebra homomorphism $\psi\,:\,e_a\,\mapsto\,\hat X_{a}$, with $\hat X_a\in\mathfrak X(M)$,  such that $$L_{\hat X_a}\omega=0:$$ this reads that locally $$i_{\hat X_a}\omega=\dd u_a$$ with $u_a\in\mathcal F(M)$. As already pointed out within the Poisson formalism (see \eqref{eq23e}), this implies that  $$\{u_a,u_b\}=c_{ba}^{\,\,\,s}u_s+\sigma_{ab}$$ with $\sigma_{ab}=-\sigma_{ba}$ and $\dd\sigma_{ab}=0$. If there exists a basis for $\g$ which reads $\sigma_{ab}=0$, then the action $\psi$ is said \emph{strongly Hamiltonian}\footnote{This is general not possible, notice indeed that any action of a perfect Lie algebra is strongly Hamiltonian. Notice also that in case of interest it could be preferable to have  a symplectic action which is not strongly Hamiltonian. Consider the symplectic vector space $(\R^2, \omega=\dd x\wedge\dd y)$, with the Hamiltonian vector fields $X=\del_x=\{~~, y\}$ and $Y=\del_y=\{~~, -x\}$; one clearly has $[X,Y]=0$, and $\{x,y\}=1$. The vector fields $X,Y$ are seen to realise a symplectic action of the commutative Lie algebra in 2 dimensions, or equivalently (and this is one of the cornestones of the geometric formulation of the correspondence principle in quantum mechanics) a strongly Hamiltonian action of the 3 dimensional Lie algebra corresponding to the Heisenberg group.}. In this case, the set $\{u_a\}_{a=1,\dots,\delta}$ closes a function group.

\noindent Assume that $\psi$ is a strongly Hamiltonian action of $\g$ upon $(M, \omega)$, with $\dim M=2N$. We analyse how the assumptions on the (Lie) algebraic properties of the elements $\{u_{a}\}_{a=1,\dots,\delta}$ modify the reduction procedure we outlined in the previous pages. 
We define the \emph{momentum map} $$\mu\,:\,M\,\to\,\g^*$$ corresponding to the strongly symplectic action $\psi$ via (here $\g^*$ is the dual vector space to $\g$)
\beq
\label{eqsy3}
\mu\,:\,m\,\mapsto\,(u_a(m))\epsilon^a
\eeq
where $\{\epsilon^a\}_{a=1,\dots,\delta}$ is the dual basis to the $\{e_a\}$ (which reads $\epsilon^a(e_b)=\delta^a_b$), with $\hat X_a\,=\,X_{u_a}$. The momentum map $\mu$ is in general not \emph{onto}, i.e. it is not surjective on all $\g^*$. 
Upon restricting the map $\mu$ so to have that the functions $\{u_a\}$ are functionally independent, we see that  $\mu$ is  a (local) submersion and gives a foliation $\Phi^{\mu}$ depending\footnote{In such a case one usually says that $\psi$ \emph{foliates}.} on the action $\psi$. In this case it is possible to prove that the momentum  map $\mu$ transforms the action $\psi$ into (a \emph{restriction} of) the co-adjoint action of $\g$ on $\g^*$, which means that (locally, again, since the vector fields $\hat X_a$ are not necessarily complete) 
\beq
\label{eqsy4}
\mu_*\,:\,\hat X_a\,\mapsto\,\tilde X_a\,=\,c_{ba}^{\,\,\,s}x_s\frac{\del}{\del x_b},
\eeq
where $\{x_a\}_{a=1,\dots,\delta}$ is the global coordinate system for $\g^*\ni x=x_a\epsilon^a$. This amounts to say\footnote{It is also possible to prove that, if $\psi$ is a symplectic action of $\g$ upon $(M,\omega)$ with $\{u_a,u_b\}=c_{ba}^{\,\,\,s}u_s+\sigma_{ab}$, then for the momentum map defined as in \eqref{eqsy3} it is $\mu_*(\hat X_a)\,=\,(c_{ba}^{\,\,\,s}x_a\,+\,\sigma_{ba})\del_b$.}  that the Hamiltonian vector fields $\hat X_a$ are $\mu$-\emph{related} to the vector fields $\tilde X_a$. Since the vector fields $\hat X_a$ close a Lie algebra, the distribution $\mathcal D_X$ they generate is integrable. If $W_a\subset M$ is the $\delta$-dimensional leaf of the corresponding foliation $\Phi^X$ (with $a\,\in\,M$) from \eqref{eqsy4} we see that $\mu(W_a)\,\subseteq\,S_{\mu(a)}$, where $S_{\mu(a)}$ is the orbit of the coadjoint action of $\g$ upon $\g^*$ which passes through $\mu(a)$.  

\noindent The reduction procedure runs parallel to what we have already described. If $\alpha\in{\rm Range}(\mu)\subseteq\g^*$, we denote by  
\beq
\label{eqsy5}
N_{\alpha}\,=\,\{m\,\in\,M\,:\,\mu(m)=\alpha\}
\eeq
the leaf (corresponding to the value $\alpha\in\g^*$) of the foliation generated by the momentum map. Each leaf $N_{\alpha}$ is a $(2N-\delta)$-dimensional manifold embedded in $M$, with $i_{\alpha}:N_{\alpha}\hookrightarrow M$.
One proves that 
\beq
\label{eqsyin1}
{\rm ker}\,i_{\alpha}^*\omega\,=\,\mathcal D_X\,\cap\,\mathcal D_{\mu},
\eeq
(see \eqref{eqsy1}) with 
\beq
\label{eqsy10}
\mathcal D_{\mu}\,=\,\{Y\,\in\,\mathfrak X(M)\,:\,i_Y\dd u_a=0\}
\eeq
 (as in \eqref{eqsy6}), so mimicking \eqref{eqsy1}. The distribution
 spanned by ${\rm ker}\,i_{\alpha}^*\omega$ on $N_{\alpha}$ is regular, and generates a foliation. Each leaf of such a foliation turns out to be the $(\delta-{\rm dim}\,S_{\alpha})$-dimensional manifold $N_{\alpha}\cap W_a$.  On the quotient manifold (which we write as $N_{\alpha}/(N_{\alpha}\cap W_a)$) the tensor $i_{\alpha}^*\omega$ induces a symplectic form $\tilde\omega$, which can be written as (see \cite{mawe74})
\beq
\label{eqsy13} 
i_{\alpha}^*\omega\,=\,\pi_{\alpha, a}\tilde\omega
\eeq
where the projection is  $\pi_{\alpha,a}\,:\,N_{\alpha}\,\to\,N_{\alpha}/(N_{\alpha}\cap W_a)$. It is interesting to notice that the Poisson tensor on \eqref{podu} defined in $\g^*$ induces a non degenerate Poisson tensor, i.e. equivalently a symplectic form $\tilde\omega_{\alpha}$ on each coadjoint orbit $S_{\alpha}\subset\g^*$ (this is the well known Kirillov - Kostant - Souriau structure on coadjoint orbits of finite dimensional Lie algebras, see \cite{k1,k2,s1}). The pullback form $\mu^*\tilde\omega_{\alpha}$ is seen to coincide with the restriction of $\omega$ to each leaf $W_a$ when $\mu(a)=\alpha$.  

\noindent When a Hamiltonian   dynamics $X_H$ is invariant under a strongly simplectic action of a Lie algebra $\g$, i.e. $L_{X_H}u_a=0$ for the corresponding function group, one proves that $X_H$ is projectable onto $(N_{\alpha}/(N_{\alpha}\cap W_r))$ and it is Hamiltonian with respect to the symplectic form $\tilde\omega$, with a relation analogue to \eqref{eqsy2} valid.

On the symplectic vector space $(M=\R^4, \,\omega=\dd q_a\wedge\dd p_a,\,\,a=1,2)$, with a point $m\in \R^4$ identified by the global coordinates $m=(q_a,p_a)_{a=1,2}$ we consider  the following examples for a symplectic reduction, which comes from \eqref{podu1} and \eqref{podu2}. 

\begin{example}
\label{exemplum-mom1}
Such example may be understood as related to the classical dynamics of a two-dimensional harmonic oscillator or to the geometrical description of the quantum dynamics of a q-bit.
Let $\g=\mathfrak{su}(2)$ with commutator structure $[e_a,e_b]=\varepsilon_{ab}^{\,\,\,c}e_c$. The functions 
\begin{align}
&u_1\,=\,\frac{1}{2}\,(q_1q_2+p_1p_2), \nn \\ 
&u_2\,=\,\frac{1}{2}\,(q_1p_2-q_2p_1), \nn \\
&u_3\,=\,\frac{1}{2}\,(q_1^2+p_1^2-q_2^2-p_2^2)
\label{eqsy8}
\end{align}
give the realization of the Lie algebra\footnote{\label{notina} Notice that the elements $\{u_j\}$ in $\mathcal F(M)$ are more than a function group, since a function group is a Poisson subalgebra of $\mathcal F(M)$ and need not be related to any finite dimensional Lie algebra.} $\{u_a,u_b\}=\varepsilon_{ab}^{\,\,\,c}u_c$, the corresponding Hamiltonian vector fields 
\begin{align}
&\hat X_1\,=\,\frac{1}{2}\,(p_2\del_{q_1}\,+\,p_1\del_{q_2}\,-\,q_2\del_{p_1}\,-\,q_1\del_{p_2}), \nn \\
&\hat X_2\,=\,\frac{1}{2}\,(-q_2\del_{q_1}\,+\,q_1\del_{q_2}\,-\,p_2\del_{p_1}\,+\,p_1\del_{p_2}), \nn \\
&\hat X_3\,=\,\frac{1}{2}\,(p_1\del_{q_1}\,-\,p_2\del_{q_2}\,-\,q_1\del_{p_1}\,+\,q_2\del_{p_2})
\label{eqsy9}
\end{align}
give  a strongly Hamiltonian action $\psi$ of the Lie algebra $\mathfrak{su}(2)$ on $(\R^4, \omega)$ with corresponding momentum map $\mu:\,m\,\mapsto\,u_a(m)\epsilon^a$.  

\noindent This momentum map is a regular submersion for any $m\neq m_0=(0,0,0,0)$, the leaves $W_r$ corresponding to the Hamiltonian distribution $\mathcal D_X$ spanned by the vector fields $\hat X_a$ in \eqref{eqsy9} at any regular point $m$ with $$u_4(m)=r^2/4$$ are easily seen to be diffeomorphic to a 3-dimensional sphere of radius $r>0$, having defined $$u_4\,=\,\frac{1}{4}(q_1^2+p_1^2+q_2^2+p_2^2),$$  reading $\{u_4, u_a\}=0$, with $u_4^2=u_1^2+u_2^2+u_3^2$. The momentum map $\mu$ gives $$\mu_*(\hat X_a)\,=\,\varepsilon_{a}^{\,\,bc}u_b\del_{u_c},$$ which are tangent to the orbit $$S_{\mu(m)}\simeq{\rm S}^2$$ of the coadjoint action of $\g=\mathfrak{su}(2)$.  
The distribution $\mathcal D_{\mu}$ (see \eqref{eqsy10}) turns to be the span of the vector field $$\Gamma\,=\,(p_1\del_{q_1}-q_1\del_{p_1})+(p_2\del_{q_2}-q_2\del_{p_2}),$$ which is the Hamiltonian vector field $\Gamma=\hat X_4$ corresponding to the Hamiltonian function  
$u_4$. This means that the leaf $$N_{\alpha}\,=\,\{M\ni m\neq m_0\,:\,\mu(m)=\alpha\in\g^*\}$$ (assumed that $\alpha\neq 0\in\g^*$) is given by the orbit of $\Gamma$ through $m$, which is diffeomorphic to a circle $\mathrm S^1$. Aiming at a symplectic reduction driven by the action $\psi$, we notice that ${\rm dim}\, W_r>{\rm dim }\,N_{\alpha}$, so we  consider the embedding $i_{r}\,:\,W_r\hookrightarrow\R^4$ and the 2-form $i_r^*\omega$, whose kernel is the span  by $\mathcal D^{\mu}$, since $$\Gamma\,=\,u_4^{-1}u_a\hat X_a.$$ This means that, when $N_{\alpha}\cap W_r\neq\{\emptyset\}$, it is $N_{\alpha}\subset W_r$, thus reading the  quotient $W_r/(W_r\cap N_{\alpha})\simeq W_r/N_{\alpha}$ which is  the well known Hopf fibration ${\rm S}^2\,\simeq\,{\rm S}^3/{\rm S}^1$.   In this case, the symplectic structure $\tilde\omega$ induced on the topological quotient ${\rm S}^2$ can be written in terms of the symplectic form on the $S_{\alpha}$ coadjoint orbit as  $$\tilde\omega\,=\,\varepsilon^{abc}u_a\dd u_b\wedge\dd u_c$$ with $u_4^2\,=\,u_1^2+u_2^2+u_3^2=r^2$.

\noindent In this specific example, it is clear that the subalgebra  introduced in \eqref{eqco1} is given by $$\mathcal F_X=\{f\in\R^4\,:\,f=f(u_4)\},$$ while the subalgebra introduced in \eqref{eqp2} is given by $$\mathcal F_{\alpha}=\mathcal F(\g^*\backslash\{0\}).$$ When, as an example of  \eqref{poissont} we have the case studied in section \ref{subsub:momp}, we see that the dynamics can be projected as $\mu_*(X_H)=\gamma\del_{u_a}$, which is not tangent to any coadjoint orbit and therefore not canonical. 
\end{example}
\begin{example}
\label{exemplum-mom2}
 Consider now a strong symplectic action on the same manifold $(\R^4, \omega)$ of the  3-dimensional Lie algebra $\g=\mathfrak{sb}(2,\C)$ whose Poisson tensor $\Lambda_{\g}$ on $\mathcal F(\g^*)$ has no global Casimir, as we showed after  \eqref{podu2}. The functions
\begin{align}
&u_1\,=\,-(q_1p_1+q_2p_2), \nn \\
&u_2\,=\,q_1, \nn \\
&u_3\,=\,q_2
\label{eqsy11}
\end{align}
give a representation (see the footnote \ref{notina}) for the Lie algebra $\g=\mathfrak{sb}(2,\C)$ with 
$$
\{u_1, u_2\}=u_2, \qquad\{u_1,u_3\}=u_3, \qquad \{u_2,u_3\}=0,
$$
the corresponding Hamiltonian vector fields 
\begin{align}
&\hat X_1\,=\,-q_1\del_{q_1}-q_2\del_{q_2}+p_1\del_{p_1}+p_2\del_{p_2}, \nn \\
&\hat X_2\,=\,-\del_{p_1}, \nn \\
&\hat X_3\,=\,-\del_{p_2}
\label{eqsy12}
\end{align}
give a strongly Hamiltonian action $\psi$ with momentum map $\mu$. Such Hamiltonian vector fields span the integrable  distribution $\mathcal D_X$. The analysis of the corresponding leaves begins upon noticing that the derivations $\hat X_2, \hat X_3$ generate the translation group onto the Lagrangian subspaces given by the $(p_1,p_2)$ planes. Moreover, the action of the vector field $\hat X_1$ amounts to independent dilations on the $(q_1,q_2)$ and on the $(p_1,p_2)$ planes. It is now clear that the 2-dimensional manifold  $W_0=(0,0,p_1, p_2)$ gives the leaf of the foliation through each point $m$ with $q_1^2+q_2^2=0$, while the 3-dimensional manifold $$W_{\theta}=(e^sq_1, e^sq_2, p_1,p_2)$$ gives the leaf of the foliation through the point $m$ with $q_1^2+q_2^2\neq0$, which can be locally labelled by an  angle $\theta$ on the $(q_1,q_2)$ plane. This shows that $W_{\theta}\simeq \R^2\times\R^+$.

\noindent The momentum map associated to the function group \eqref{eqsy11} is surjective onto $\g^*$, but provides a submersion only if $m\notin W_0$. The distribution $$\mathcal D_{\mu}\,=\,\{Y\,\in\,\mathfrak X(\R^4)\,:\,i_Y\dd u_a=0\}$$ which generates the corresponding foliation is span by $$Y\,=\,q_2\del_{p_1}-q_1\del_{p_2}.$$ For $\alpha\neq0\in\g^*$ this reads that the leaf through $m$ with $\mu(m)=\alpha$ is $$N_{\alpha}=\{m(s)\,=\,(q_1,q_2, p_1+sq_2, p_2-sq_1)\},$$ which is a line. 

\noindent We proceed as in the previous example.  With respect to the embedding  $i_{\theta}\,:\,W_{\theta}\hookrightarrow \R^4$, we see that ${\rm ker}\,i_{\theta}^*\omega$ is spanned by  $Y$, which is indeed not Hamiltonian. 
For the quotient we have $W_{\theta}/N_{\alpha}\simeq \R\times\R^+$, which is diffeomorphic to each of the orbits which do not pass through the zero of the coadjoint action of $\g=\mathfrak{su}(2)$ on $\g^*$, whose generators are 
\begin{align*}
&\mu_*(\hat X_1)\,=\,u_2\del_{u_2}+u_3\del_{u_3},\\ &\mu_*(\hat X_2)\,=\,-u_2\del_{u_1}, \\ &\mu_*(\hat X_3)\,=\,-u_3\del_{u_1},
\end{align*}
while $Y\,=\,u_2\hat X_3-u_3\hat X_2$ and then $\mu_*(Y)=0$. These examples have been considered in relation with  a Lie-Poisson dynamics and in T-duality field theory, see \cite{lizzi-etal, pezz-vita}, while the properties of the associated momentum map have been used to study non commutative differential calculi in \cite{mvz06,mvz19}.
\end{example}

\subsubsection{\blu{Invariant closed 1-forms and constants of the motions}}
\label{subsecH2}

Let us assume that $\alpha\in\Lambda^1(M)$ is invariant along the integral curves of $X_{H}$, i.e. $L_{X_H}\alpha=0$. Since the relations \eqref{eq10} are valid on any Poisson manifold, and then on $(M,\omega)$ with respect to the Poisson tensor $\Lambda$ dual to $\omega$, we immediately conclude that $\hat X_{\alpha}$ is an infinitesimal symmetry for $X_H$. Such an infinitesimal symmetry is canonical (by definition)   if $L_{\hat X_{\alpha}}\omega=0$. This condition is equivalent to the condition $\dd(i_{\hat X_{\alpha}}\omega)=0$ which results in $\dd\alpha=0$, i.e. $\alpha$ closed\footnote{Notice that we are clearly considering the case (P4) from the previous section, which becomes simpler due to the non degeneracy of the Poisson tensor on $M$.}. From $\dd\alpha=0$ we  have that (locally, i.e. depending on the topology of $M$) $\alpha=\dd f$, so that $\hat X_{\alpha}=X_f$ and the straightforward computation  $\dd(L_{X_H}f)=\dd\{f,H\}=0$ shows that $\phi=\{f,H\}$ is therefore a constant function on (each connected component) of $M$.

The analogy with the theory described within the Poisson formalism  in section \ref{subsecP2}  is natural,  so we write, recalling the maps defined in \eqref{intro00} and \eqref{intro01},  the analogue of the proposition \ref{noepoisson2} as 
\begin{prop}
\label{noesymp2}
Given a Hamiltonian dynamics $\Gamma=X_H$ on a symplectic manifold $(M,\omega)$, a 1-form $\alpha$ on $M$ is invariant along the dynamics if and only if the vector field $\hat X_{\alpha}\,=\,\tau(\alpha)$ is an infinitesimal symmetry for $\Gamma$. Equivalently, a vector field $X$ on $M$ is an infinitesimal symmetry for $\Gamma$ if and only if the 1-form $\alpha\,=\,\tilde{\tau}(X)\,=\,i_X\omega$ is invariant along the dynamics. 
An infinitesimal symmetry $X$ for $\Gamma$ is canonical if and only if the corresponding invariant 1-form $\alpha$ is closed. 
\end{prop} 

\begin{example}
\label{exemplum3}
An example that may clarify the difference between the meaning of the propositions \ref{noesymp1} and \ref{noesymp2}  is given by considering the vector field describing the dynamics of a charged particle moving in the external stationary and homogeneous electric field among two plates of a plane condenser,
$$
\Gamma=p_x\frac{\del}{\del x}\,+\,p_y\frac{\del}{\del y}\,-\,k\frac{\del}{\del p_y}
$$
which is Hamiltonian on $\R^4$ with respect to the canonical symplectic form $\omega=\dd x\wedge\dd p_x+\dd y\wedge\dd p_y$, with Hamiltonian function $H\,=\,1/2(p_x^2+p_y^2)+ky$. The vector field $T_1=\del_{x}$ 
is an infinitesimal canonical symmetry for the dynamics, i.e. $[T_1,\Gamma]=0$, and the corresponding Hamiltonian function $f_1=p_x$ is a constant of the motion, since $L_{\Gamma}p_x=0$. The vector field $T_2=\del_{y}$ is an infinitesimal canonical symmetry for the dynamics whose  corresponding Hamiltonian function $f_2=p_y$ is \emph{not} a constant of the motion, since $L_{\Gamma}p_y=k$, while $\dd f_2$ is an invariant exact 1-form for the dynamics.
\end{example}

We close our analysis upon assuming  that the vector field $A$ is an infinitesimal symmetry for the dynamics $X_H$, i.e. $[A,X_H]=0$. We consider the corresponding (under the map $\tilde{\tau}$) 1-form $\hat\alpha_A=i_A\omega$. It is immediate to prove that 
$$
L_{X_{H}}i_A\omega\,=\,i_{[X_H,A]}\omega\,+\,i_AL_{X_H}\omega\,=\,0.
$$
This means that (as expected) $i_A\omega$ is invariant along the integral curves of $X_H$.  Again, we see that $A$ is canonical if and only if $\dd( i_A\omega)=0$, so we fall within the case  above.

\subsubsection{\blu{A symplectic description of the free particle dynamics and of the isotropic harmonic oscillator with their  symmetries}}
\label{subfree}
We analyse the examples of the free particle dynamics and of the isotropic harmonic oscillator in detail, since they allow for an interesting observation, which naturally comes along with the principle of analogy.  

Consider an associative algebra $A$ on $\C$ polynomially generated by elements $q,p$ and the identity $1$; let $A$ have a Hermitian structure  (compatible with $\C$) such that $q^*=q$ and $p=p^*$. 
Let a dynamics on it be  given by a derivation operator $D$ on $A$ such that $D(q)=p$ and $D(p)=0$. We can call $(A,D)$ the \emph{free particle dynamics algebra}:  it describes  both the quantum and the classical evolution of a free particle system. If the associative algebra $A$ is realised with a non commutative product $*$, such that $[q,p]=q*p-p*q=1$, then the derivation $D$ is inner, with $D(q)=[q,H]$ and $D(p)=[p,H]$ for $H=p^2/2$, and we recover the quantum dynamics of a free particle within the so called Heisenberg picture. If the associative algebra $A$ is realised with a commutative product, then the derivation $D$ is inner with respect to the Poisson structure defined by $\{q,p\}=1$, and we recover the classical dynamics of a free particle in terms of a non degenerate Poisson structure, thus providing a symplectic form. 

Analogously, let a dynamics $D'$ on $A$ be given by $D(q)=p$ and $D(q)=-p$. We can call $(A,D')$ the \emph{isotropic harmonic oscillator dynamics algebra}. It encodes both the classical and the quantum description of the isotropic harmonic oscillator, depending whether the algebra $A$ is realised as commutative or non commutative, with $D$ an inner derivation for $H=(p^2+q^2)/2$ with respect to the commutator or to the non degenerate Poisson bracket structure $\{q,p\}=1$.

On $M=\R^2$ with respect to the global coordinate system $(q,p)$, both the vector fields $\Gamma_F=p\del_q$ and $\Gamma_O=p\del_q-q\del_p$ are linear, that is they commute with the Liouville vector field $\Delta=q\del_q+p\del_p$ encoding (as we described in section \ref{susec:linPoisson}) the linear structure on $\R^2$. We shall analyse general linear dynamics on symplectic vector spaces in the next section: in the following examples we describe these dynamics per se, our aim being to show how a symplectic structure connects symmetries to  constants of the motion and viceversa.

\begin{example}
\label{efree}
Consider the vector field $$\Gamma=p\frac{\del}{\del_q}$$ on $M=\R^2$ with global coordinates $(q,p)$. 
A 2-form $\omega=a(q,p)\dd q\wedge\dd p$ is symplectic if and only if $a$ is never vanishing. Since $\R^2$ is simply connected, $\Gamma$ has a symplectic description if and only if $i_{\Gamma}\omega=pa(q,p)\dd p$ is closed, that is if and only if $\del_qa=0\,\Leftrightarrow a=a(p)$. In such a case, the Hamiltonian $H$ is defined by $\dd H=pa\dd p$.  Notice that, since we suppose $a$ to be smooth, its sign remains constant, so one has  $$H=\int_0^p\dd k\, ka(k)$$ (setting $H(q=0,p=0)=0$). This also reads that the  Darboux chart given  by the map $\Phi:(q,p)\mapsto(q,P)$ with $\dd P=a(p)\dd p$ is globally defined. 

It is easy to directly compute that a vector field $X$ on $\R^2$ is an infinitesimal symmetry for $\Gamma$  if and only if there exist functions $\alpha=\alpha(p)$ and $\beta=\beta(p)$ such that 
\beq
\label{hm1}
X\,=\,(\alpha+q\beta)\frac{\del}{\del q}+p\beta\frac{\del}{\del p};
\eeq
the corresponding 1-form $\tilde{\tau}(X)=i_X\omega$ is invariant along $\Gamma$, i.e. $L_\Gamma i_X\omega=0$. 
A function $f\in\mathcal F(\R^2)$ is a constant of the motion if and only if $f=f(p)$, the corresponding Hamiltonian vector field is 
$$
X_f\,=\,\frac{1}{a}(\frac{\dd f}{\dd p})\frac{\del}{\del q}.
$$
The vector field $X$ defined in \eqref{hm1} is Hamiltonian with respect to the symplectic form $\omega=a(p)\dd q\wedge\dd p$ if and only if 
\beq
\label{hm2}
a\beta+\del_p(pa\beta)=0.
\eeq This condition shows that there exist infinitesimal symmetries for $\Gamma$ which do not correspond to any constant of the motion, even considering the most general symplectic structure compatible with the free particle dynamics. If we set $\beta=1$, for example, so that from \eqref{hm1} it is $X=\alpha\del_q+\beta\Delta$,  the condition \eqref{hm2} reads $a=c/p^2$ (with $c$ a constant), which clearly does not define a  meaningful symplectic structure on $\R^2$. 
\end{example}

\begin{example}
\label{eiso}
With respect to the global coordinate system $(q,p)$ adopted in the previous example, the dynamics of the isotropic harmonic oscillator on $M=\R^2$ is given by the vector field
$$
\Gamma\,=\,p\frac{\del}{\del q}-q\frac{\del}{\del p}.
$$
As we did in the previous example, we do not focus on the aspects of the problem related to the linearity of $\Gamma$, so we consider the problem on $M'=\R^2\backslash\{0\}$, removing the origin $(q=0,p=0)$ which is the only fixed point solution for $\Gamma$. With respect to the radial coordinate system $(r>0, \theta=[0,2\pi))$ given by $$q=r\cos\theta,\qquad p=r\sin\theta$$
we have that the 2-form $\omega=a(r,\theta)\dd\theta\wedge\dd r$ is symplectic if and only if $a$ is never vanishing. The dynamical vector field is written as $\Gamma=-\del_{\theta}$, and it has a symplectic description with a Hamiltonian function $H=H(r)$  provided $a=a(r)$ and $\dd H=-a(r)\dd r$, which gives 
$$
H\,=\,-\int_{0}^{r}\dd k\,a(k).
$$
The Darboux chart given by the map $\Phi\,:\,(r,\theta)\,\mapsto\,(R,\theta)$ with $\dd R=a(r)\dd r$ is globally defined. 
This also proves that any globally injective function $H=H(r)$ for $r>0$ is an admissible Hamiltonian for $\Gamma$ with corresponding symplectic form $\omega=-(\dd H/\dd r)\dd\theta\wedge\dd r$.

A vector field $X\in\mathfrak X(M')$ is an infinitesimal symmetry for $\Gamma$ if and only if 
\beq
\label{hm3}
X\,=\,\alpha(r)\frac{\del}{\del \theta}+\beta(r)\frac{\del}{\del r}, 
\end{equation}
the corresponding 1-form $\tilde\tau(X)=i_X\omega=a\alpha\dd r-a\beta\dd\theta$ on $M'$ is invariant along $\Gamma$. An element $f\in\mathcal F(M')$ is a constant of the motion for $\Gamma$ if and only if $f=f(r)$, its corresponding Hamiltonian vector field 
$$
X_f\,=\,\frac{1}{a}(\frac{\dd f}{\dd r})\frac{\del}{\del \theta}
$$
is an infinitesimal symmetry for $\Gamma$. Unlike the previous example, since $M'$ is not simply connected, canonical vector fields are not necessarily Hamiltonian. It is easy to compute that the vector field $X$ defined in \eqref{hm3} is canonical, i.e. $L_X\omega=0$, if and only if $\del_r(a\beta)=0$. It turns out to be Hamiltonian if and only if $\beta=0$.
This result once more shows that there exist infinitesimal symmetries for a symplectic dynamics which do not correspond to any constant of the motion. The restriction to $M'$ of the Liouville vector field $\Delta=r\del_r$ gives the 1-form $i_\Delta=-ra(r)\dd\theta$ which is not exact.

We close this example by noticing that the vector fields $\Gamma, \Delta$ provide a basis for the left $\mathcal F(M')$-module $\mathfrak X(M')$, with dual basis for the $\mathcal F(M')$-bimodule $\Lambda^1(M')$ given by $(\alpha_\Gamma=\dd\theta, \alpha_{\Delta}=r^{-1}\dd r=\dd(\log r))$. A vector field $X$ in \eqref{hm3} can be written as 
$$
X\,=\,\alpha\Gamma+\beta\Delta
$$
with $\alpha, \beta\in\ker \Gamma$. We limit ourselves to mention that the vector fields $\Delta$ and $\Gamma$ giving the Liouville vector field and the isotropic harmonic oscillator vector field on $\R^{2N}$ allow for the analysis of the reduction procedure from $\C^N$ to its projective space $\C^N/\C_0$, which describes the set of states for a finite level quantum mechanical system (see \cite{gfd, mz19}). 

\end{example}

\subsection{Symmetries for linear symplectic dynamics}\label{susec:linearsimplectic}
As within the context of the Poisson formalism, we focus on linear Hamiltonian dynamics on a symplectic vector space. If $(E,\omega)$ is a $2N$-dimensional vector space, with $$\omega=\frac{1}{2}\omega_{ab}\dd x^a\wedge\dd x^b, \quad \omega_{ab}\in\R$$ a linear symplectic structure, then the closedness condition $\dd\omega=0$ implies, since a vector space is simply connected, that a so called potential 1-form $\theta$ exists, such that $\dd\theta=-\omega$ (notice that the minus sign is a matter of historical convention). If $\Delta$ is the Liouville vector field describing the linear structure on $E$, then one proves that $\dd(i_\Delta\omega)=2\omega$, so that one has 
$2\theta=-i_{\Delta}\omega+\dd f$
for an arbitrary function $f$. Upon redefining $\theta$, one has\footnote{As we commented upon in section \ref{sub:exasy}, different $f$ allow to identify $E$ with alternative cotangent bundle structures.} 
$$
2\theta=-i_{\Delta}\omega.
$$
It is possible to prove (see \cite{gfd}) that these conditions tensorially characterise a linear symplectic manifold, so let $(M,\omega)$ be a symplectic manifold: there exists a linear structure on $M$ such that $\omega$ is a linear symplectic structure if and only if there exists a Liouville vector field $\Delta$ on $M$ such that the 1-form $\theta=-\frac{1}{2}i_{\Delta}\omega$ is a symplectic potential, i.e. $\dd i_{\Delta}\omega=2\omega$.

If we consider the case of a linear Hamiltonian dynamics, with $M=\R^{2N}$, the relations \eqref{eq17}-\eqref{eq15} are clearly valid, while the existence of a symplectic structure with representing matrix $\omega$ reads (given the relation \eqref{eq18})  $\omega=-\Lambda^{-1}$. The dynamics 
$$\Gamma=X_A\,=\,A^a_{\,b}x^b\del_a,$$  has therefore a symplectic description on  $(\R^{2N}, \omega=-\omega^T)$ with ${\rm det}\,\omega\neq0$ if and only if the matrix equation 
\beq
\label{eq23bi}
H=-\omega A
\eeq
 is satisfied, with $H=H^T$.  Notice that, if such a relation is verified, then from $\omega A=(\omega A)^T$ one proves that 
 \beq
 \label{eq23pi}
 {\rm Tr}(A)=0.
 \eeq
 Since the relations \eqref{eq16} are valid within the symplectic formalism, provided $\omega=-\Lambda^{-1}$, we focus on a different class of constants of the motion related to a linear dynamics. Assume that the  \eqref{eq23bi} is valid, then $L_{X_{A^{2k+1}}}\omega=0$ and this is equivalent (since $\R^{2N}$ is simply connected, then a vector field is globally Hamiltonian if and only if it is locally Hamiltonian) to have that  that there exists a symmetric matrix $F_{(k)}=F_{(k)}^T$ such that $i_{A_{X^{2k+1}}}\omega=\dd f_{F_{(k)}}$. As recalled in section \ref{subsecP2}, we directly have then that 
 $L_{A_{X^{2s+1}}}f_{F_{(k)}}=c_{(k)}$ with $c_{(k)}$ a real constant, for each $s\in\N$. As $f_{F_{(k)}}$ is quadratic and $X_{A^{2s+1}}$ is linear, the constant $c_{(k)}$ must vanish. This shows that, for linear vector fields $X_A$ which preserve a constant symplectic 2-form $\omega$ there is a natural set of constants of the motions given by $f_{F_{(k)}}$, with $\{f_{F_{(k)}}, f_{F_{(j)}}\}=0$. Notice that, if $X_A$ denotes the linear dynamics of the isotropic harmonic oscillator, then $A=-\omega$, and one immediately sees that $f_{F_{(k)}}=f_H$. 

If the constant matrices $\{B_{a}\}_{a=1,\dots,k}$ span a $k$ dimensional Lie algebra $\mathfrak g$ of infinitesimal linear symmetries $X_{B_a}$ for the symplectic linear dynamics $X_A$,  i.e. $[A,B_a]=0$, and they preserve the symplectic structure, i.e. $L_{X_{B_a}}\omega=0$, then we have as before that $i_{X_{B_a}}\omega=\dd f_{(B_a)}$, where $f_{(B_a)}$ is quadratic function on $\R^{2N}$ such that $L_{X_A}f_{(B_a)}=0$. It is easy to see that such quadratic functions $\{f_{(B_a)}\}_{a=1,\dots,k}$, with respect to the Poisson structure corresponding to $\omega$, realise the Lie algebra $\mathfrak g$, that is $\{f_{(B_a)}, f_{(B_b)}\}=f_{([B_a,B_b])}$. With respect to the Poisson bracket, linear functions on $\R^{2N}$ close the Heisenberg-Weyl algebra in $2N+1$ dimensions, while quadratic functions (corresponding to symmetric matrices) close the Lie algebra of the symplectic group in $\R^{2N}$. The quadratic functions generate canonical derivations for the Heisenberg-Weyl algebra. Linear and quadratic functions generate the Lie algebra corresponding to the inhomogeneous symplectic group in $\R^{2N}$. Such Lie algebra contains as Lie subalgebras both the free particle and the harmonic oscillator Lie algebras introduced in section \ref{subfree}.

\subsubsection{\blu{Reduction for linear Hamiltonian dynamics}}
\label{susu:Hl}
The first natural example of a reduction of a linear dynamics comes by considering a vector field $X_A$ associated to a matrix $A$ having an eigenspace $W$. If $W'$ denotes one of the (infinitely many) subspaces of $\R^{2N}$ such that $\R^{2N}=W\oplus W'$, then it is immediate to see that $X_A$ can be reduced to $W'$, where it is linear. Indeed, if the global coordinate systems $\{x^a, y^b\}_{a=1,\dots,\dim W; b=1,\dots,\dim W'}$ is adapted to the direct sum, with $\{x^a\}$ identifying an element in $W$ and $\{y^b\}$ an element in $W'$, 
one sees that the subalgebra $\mathcal A=\{f\in\mathcal F(W)\,:\,\del_{x^a}f=0\}$  is invariant under $X_A$. Each leaf of the corresponding foliation is isomorphic to $W$, the quotient space (that is, the space of leaves) turns to be isomorphic to $W'$. 

 If in addition also $W'$ is an eigenspace for $A$, then the dynamics is \emph{completeley} separated in two non interacting linear (sub)systems: the vector field $X_A$ can be split into the sum of two commuting vector fields on $W$ and $W'$. The flow generated by $X_A$ can then be recovered by \emph{composing} two independent motions.
 
  If, in particular, the dynamics $X_A$ is Hamiltonian, then the requirement the reduced dynamics to be Hamiltonian  forces the subspaces $W,W'$ to be even dimensional, and such that the restrictions of $\omega$ and $H$ on them have suitable properties. 

Interesting examples of reductions of linear Hamiltonian dynamics resulting in a linear dynamics on the quotient manifold come by considering the reduction associated to a function group.  
Consider $M=T^*\R^N$ with the canonical symplectic form and the elements 
\beq
\label{2g1}
u_1\,=\,\frac{1}{2}p^2, \qquad u_2\,=\,q^ap_a,\qquad u_3\,=\,\frac{1}{2}q^2
\eeq
with respect to the global Darboux coordinate system $(q^a,p_a)$, where we wrote $p^2=p_ap_b\delta^{ab}$ and $q^2=q^aq^b\delta_{ab}$. Such elements close a function group modelled on the Lie algebra $\mathfrak g=\mathfrak{sl}(2,\R)$, since 
$$
\{u_1,u_2\}=-2u_1, \qquad \{u_1,u_3\}\,=\,-u_2, \qquad \{u_2,u_3\}=-2u_3.
$$
Following the general description of a reduction procedure in section \ref{ss:red}, one defines the momentum map  $\mu\,:\,M\,\to\,\mathfrak g^*$ given by $\mu(m)\,=\,u_a\epsilon^{a}$ with $\{\epsilon^a\}_{a=1,\dots,3}$ a suitable basis for $\mathfrak g^*$. Without discussing the details of the associated symplectic reduction, which closely follows the examples  \ref{exemplum-mom1} and \ref{exemplum-mom2}, we limit ourselves to notice that, after removing the points where the foliation $\Phi^{\mu}$ associated to $\mu$ is not regular, if $\Gamma=p_a\del_{q^a}$ is the free particle dynamics, then 
$$
\Gamma(u_1)\,=\,0, \qquad \Gamma(u_2)=2u_1, \qquad \Gamma(u_3)=u_2.
$$
This shows that $\Gamma$ reduces to the quotient manifold $M/\Phi^{\mu}$, its $\mu$-related vector field $$\Gamma_{\mu}\,=\,2u_1\frac{\del}{\del u_2}+u_2\frac{\del}{\del{u_3}}$$ describes the dynamics among the leaves of the foliation. 
The same momentum map allows for a reduction also of the isotropic harmonic oscillator $\Gamma=p_a\del_{q^a}-q^a\del_{p^a}$, with $\mu$-related vector field
$$
\Gamma_{\mu}\,=\,-u_2\frac{\del}{\del u_1}+u_2\frac{\del}{\del u_3}.
$$

It is interesting to notice that the Hamiltonian reduction of a linear dynamics may result to be non linear. 

Consider the example \ref{exemplum1} which describes, for $V=0$, the linear dynamics of a free particle on a plane. The angular momentum $L$ defined in \eqref{Lfun} gives a quadratic constant of the motion, the reduced dynamics $\tilde\Gamma_L$ on $T^*\R_+$ (given by the equations for $(\dot r=0,\dot p_r=l^2/r^3)$ in \eqref{sODE2}) is Hamiltonian with \eqref{tildeHex}, but no longer linear.

An interesting example of a linear dynamics whose reduction is non linear, the space of solutions of such reduced equation nonetheless presenting a peculiar composition law, is given by considering  $\Gamma=X_A$ on $\R^2$ with  
$$
\begin{pmatrix} \dot x_1 \\ \dot x_2 \end{pmatrix}\,=\,\begin{pmatrix} a_{11}  & a_{12} \\ a_{21} & a_{22} \end{pmatrix}\,\begin{pmatrix} x_1 \\ x_2\end{pmatrix}.
$$
Since $[X_A, \Delta]=0$ with $\Delta=x_1\del_{x_1}+x_2\del_{x_2}$ the infinitesimal generator of the dilation group on $\R^2$, the dynamics projects onto the quotient given upon identifying points in $\R^2\backslash\{0\}$ lying on the same integral curve of $\Delta$, i.e. on the same ray on the plane. If one introduces  $y=x_1/x_2$ locally on the quotient manifold, the projected dynamics reads $\dot y\,=\,a_{12}+(a_{11}-a_{22})y-a_{21}y^2$, which is an equation of a Riccati type. Although such equation is non linear,  there exists a superposition  rule within the space of its solutions. If $\{y_1,y_2,y_3\}$ are independent solutions, then any other solution $y$ is given\footnote{For a more general analysis of equations presenting a non linear superposition rules we refer to \cite{lie-scheffers}.}  by the ratio
$K\,=\,(y-y_1)(y_2-y_3)/(y-y_2)(y_1-y_3)$.

\subsubsection{\blu{Nonlinear reductions for the free dynamics}}
\label{sss:free}
The example of the free dynamics is interesting also because it provides examples of reduction of the same dynamics with respect to alternative symplectic structures. We know from \cite{gralamavi} that the vector field $\Gamma=p_1\del_{x_1}+p_2\del_{x_2}$ is Hamiltonian with respect to the class of symplectic 2-forms $\omega_s=\dd x^a\wedge\dd p_a+s(\dd p^2)\wedge\dd(x^ap_a)$ (with $s\in\R$ and $p_ap_b\delta^{ab}=p^2$) and $H_s=p^2(1+sp^2)/2$. By considering the Hamiltonian reduction driven by the  same invariant function $L$, one sees that the reduced dynamics $\tilde\Gamma_L$ is Hamiltonian on $T^*\R_+$ with 
$$
\tilde H_s\,=\,\frac{1}{2}(p^2+\frac{l^2}{r^2})(1+s(p^2+\frac{l^2}{r^2})).
$$
once fixed $L=l\neq0$.

We conclude this presentation of examples of Hamiltonian  reduction of a free dynamics by considering (see \cite{gralamavi}) the problem on the symplectic manifold $T^*\R^3$ with the canonical symplectic form $\omega=\dd q^a\wedge\dd p_a$. Identifying the position coordinates with the symmetric matrix 
$$
Q\,=\,\begin{pmatrix} q_1 & q_2/\sqrt{2} \\ q_2/\sqrt{2} & q_3\end{pmatrix}
$$
there is an action of the group ${\rm SO}(2)$ via $Q\,\to\, G^{-1}QG$ with 
$$
G\,=\,\begin{pmatrix} \cos\varphi & \sin\varphi \\ -\sin\varphi & \cos\varphi\end{pmatrix}.
$$  
This action is seen to be Hamiltonian, with infinitesimal generator $X=\del_\varphi$ with respect to the set of local coordinates $(x^1,x^2,\varphi)$ on $\R^3$ given by
$$
q^1\,=\,x^1\cos^2\varphi+x^2\sin^2\varphi, \qquad q^3\,=\,x^1\sin^2\varphi+\cos^2\varphi, \qquad q^2\,=\,(x^2-x^1)(\sin 2\varphi)/\sqrt{2}
$$
and corresponding symplectic conjugate fiber variables $(k_1,k_2,k_{\varphi})$ given by 
\begin{align*}
&k_{\varphi}\,=\,(x^2-x^1)((p_1-p_3)\sin 2\varphi+\sqrt{2}p_2\cos2\varphi, \\ &k_1\,=\,p_1\cos^2\varphi+p_3\sin^2\varphi-p_2(\sin2\varphi)/\sqrt{2}, \\ 
&k_2\,=\,p_1\sin^2\varphi+p_3\cos^2\varphi+p_2(\sin2\varphi)/\sqrt{2}.
\end{align*}   
The free particle Hamiltonian $H=p^2/2$ (with $p^2=p_ap_b\delta^{ab}$) is invariant under the vector field $X_{\varphi}$, and the Hamiltonian reduction associated to a regular value $\alpha\in\R$ of  the momentum map $k_{\varphi}$ gives a reduced dynamics $\tilde X$ on $T^*\R^2$ with the symplectic structure given by $\omega_{\alpha}=\dd x^a\wedge\dd k_a$ which is Hamiltonian with 
$$
H_{\alpha}\,=\,\frac{1}{2}(k_1^2+k_2^2)+\frac{\alpha^2}{4(x^1-x^2)^2}.
$$   
This dynamics is a Calogero-Moser system.

\subsubsection{\blu{Gram dynamics}}
\label{sss:gram}
Let $M=T^*\R^2$ be equipped with the canonical symplectic form $\omega=\dd q^a\wedge\dd p_a$, and consider   the Hamiltonian $H=L^2$ where $L=(q^1p_2-q^2p_1)$ is the usual angular momentum function. The Hamiltonian dynamics $X_H$ is clearly non linear, since $H$ is a polynomial of degree 4. The corresponding equations of motions can be written as
\begin{align}
\dot q^1=-2Lq^2, \qquad\qquad &\dot q^2=2Lq^1, \nn \\
\dot p_1=-2Lp_2, \qquad\qquad &\dot p^2=2Lp_1.
\label{2g2}
\end{align}
Since $\{H,L\}=0$, we see that $X_H$ is tangent to the  leaves defined by the level sets of $L$, and for $L=l\neq0$ fixed by the Cauchy data the flow generated by $X_H$ comes by exponentiating a matrix with constant entries. 

Consider an analogous system on $M=T^*\R^3$. Let the Hamiltonian be again the square of the angular momentum, i.e.  $H=L^2=\delta^{ab}L_aL_b$ with $L_a=\varepsilon_{aj}^{\,\,\,\,\,s}q^jp_s$. The corresponding Hamiltonian dynamics is given by 
\begin{align}
\dot q^a\,=\,-2u_2\,q^a\,+\,4u_3\,p_a, \nn \\ 
\dot p_a\,=\,-4u_1\,q^a\,+\,2u_2\,p_a
\label{2g3}
\end{align}
with respect to the elements defined in \eqref{2g1}. Since $\{L,u_a\}=0$ for any $a=1,\dots,3$, then we see that the non linear dynamics $X_H$ is tangent to the leaves given by the level sets of the momentum map $\mu$ described in the previous section \ref{susu:Hl}. As for the equations \eqref{2g2},  for each initial condition the coefficients of the differential equations \eqref{2g3} acquire a constant value, so the flow comes by exponentiating a matrix with constant entries. 

These two examples can be generalised. Recall that, as we mentioned in section \ref{susec:linearsimplectic}, linear and quadratic functions on $M=\R^{4}$ generate, with respect to the canonical Poisson bracket, the Lie algebra of the inhomogeneous symplectic group in four dimensions. Such a  Lie algebra contains as Lie subalgebras all three dimensional Lie algebras (see \cite{gmp93}). Let $\mathfrak g$ be one of the three dimensional Lie algebras having a global Casimir function: denote by $\{u_a\}_{a=1,\dots,3}$ the quadratic functions closing $\mathfrak g$ with respect to the Poisson bracket, and denote by $C_{\mathfrak g}$ (which is a quadratic function on $\R^4$)  its Casimir. For example, the elements \eqref{eqsy8} give such elements for $\mathfrak g=\mathfrak{su}(2)$, those in \eqref{2g1} for $\mathfrak g=\mathfrak{sl}(2,\R)$. The Hamiltonian vector field  given by $X_H$ with $H(C_{\mathfrak g})$ gives in general a non linear dynamics on $M$. Nonetheless, since $C_{\mathfrak g}$ is quadratic on $M$ and  commutes with  the $u_a$, one sees  that the vector field $X_H$ is tangent to the leaves given by the level sets of the corresponding momentum map 
$\mu:\R^4\to\mathfrak g^*$. The differential equations corresponding to  $X_H$ have the same structure of the above \eqref{2g2}-\eqref{2g3}, with coefficients given by constants of the motion. On each leaf selected by the Cauchy data, the flow comes again by exponentiating a matrix with constant entries.

\section{Symmetries and conservation laws for presymplectic dynamics}
\label{sec:pre}

Consider the problem of determining  a vector field $\Gamma\in\mathfrak X(M)$ which fullfills the relation
\beq
\label{ps1}
i_{\Gamma}\omega=\alpha
\eeq
where $M$ is a smooth $N$-dimensional manifold, $\omega$ is a closed 2-form on $M$ with (constant, we assume) rank ${\rm rk}\,\omega=k<N$ and $\alpha$ is a closed 1-form on $M$. Such relation generalises to the case of a (constant rank)  degenerate closed 2-form  $\omega$  the problem  represented by the relation \eqref{eq20}. Along what we already introduced in the previous section, we call $(M,\omega)$ a pre-symplectic manifold and $(M,\omega, \alpha)$ a pre-symplectic system. 

Problems of this kind arise  naturally in mathematical physics. As a first example consider the linear dynamics on $M=\R^4$ given by 
\beq
\label{presuno}
X_A=\,x^1\del_1+x^2\del_2+x^4\del_3-x^3\del_4, \qquad\qquad A=\begin{pmatrix} 1 & 0 & 0 & 0 \\ 0 & 1 & 0 & 0 \\ 0 & 0 & 0 & 1 \\ 0 & 0 & -1 & 0 \end{pmatrix} 
\eeq
with respect to the coordinate chart $\{x^a\}_{a=1,\dots,4}$. Since ${\rm Tr}\,A\neq0$, from the analysis on linear symplectic dynamics described in the previous section (see \eqref{eq23bi} and \eqref{eq23pi}) we know that there is no globally defined  skewsymmetric invertible  matrix $\omega$ on $\R^4$ with constant entries such that $i_\Gamma\omega=\dd H$ for a quadratic $H$. Moreover, it is 
clear that $\Gamma$ can be completely decomposed (reduced) as the sum $\Gamma=\Delta_{(1,2)}+\Gamma_{(3,4)}$ with $\Delta_{(1,2)}$ giving the Liouville dilation field on the $x^1x^2$-subspace and $\Gamma_{(3,4)}$ giving an harmonic oscillator dynamics on the $x^3x^4$-subspace. Generalising the analysis performed in the example \ref{expoil}, it is possible to see  that there is no invertible Poisson tensor on the $x^1x^2$-subspace such that $\Delta_{(1,2)}$ is Hamiltonian with a Hamiltonian function $H$ globally defined on the $x^1x^2$-subspace. If we indeed consider $\omega=\dd x^3\wedge\dd x^4$ and $H=\frac{1}{2}((x^3)^2+(x^4)^2)$, then we directly compute
\beq
\label{presdue}
i_{X_A}\omega=\dd H:
\eeq
the vector field $X_A$ is described in terms of the pre-symplectic system $(\R^4, \omega, \dd H)$ given above. Moreover, any vector field $X=X_A+K$ with $\mathfrak X(\R^4)\,\ni\,K=K^1\del_1+K^2\del_2$ fulfils the relation $i_X\omega=\dd H$.

Another interesting class of examples where pre-symplectic systems arise is when a classical system  has a set of non holonomic constraints (see \cite{AM78, AMP82, mssv85, michor-book}). Among those examples, one of the most interesting is given upon considering the dynamics of a free relativistic particle, i.e. the equations of the motion 
$$
\dot q^{\alpha}=p^\alpha, \qquad \dot p_\alpha=0
$$ 
with respect to the coordinate system $\{q^\alpha\}_{\alpha=0,\dots,3}$ on $M=T^*\R^4$.  Given the Minkowski metric tensor in its covariant form   $${\rm g}=g^{\alpha\beta}\dd q^\alpha\otimes\dd q^\beta=\dd q^0\otimes\dd q^0-\dd q^1\otimes\dd q^1-\dd q^2\otimes\dd q^2-\dd q^3\otimes\dd q^3$$
on $\R^4$, the \emph{rest mass } of the particle is given by $$m^2=g^{\alpha\beta}p_\alpha p_\beta$$ with $g^{\alpha\beta}g_{\beta\gamma}=\delta_{\gamma}^{\alpha}$. Fixing the value of the rest mass, which is a relativistic invariant, amounts to fix a constraint. Such constraint selects the 
odd dimensional submanifold $\Sigma_m$. Since the dynamical flow restricts to $\Sigma_m$, a Hamiltonian description of such dynamics is more naturally studied  in terms of a pre-symplectic geometry on $\Sigma_m$.

Before providing a deeper analysis of pre-symplectic geometry, we start by noticing that the  degeneracy of $\omega$ reads the map (see \eqref{eq33}) $\tilde{\tau}\,:\,X\,\mapsto\,i_X\omega$ to be no longer a bijection between the set of vector fields $\mathfrak X(M)$ and the set of 1-forms $\Lambda^1(M)$. A solution $\Gamma$ to the problem \eqref{ps1} for a given closed $\alpha$ may  exist only on a suitable subset $M'\subseteq M$, and indeed be not unique, since any $\Gamma'=\Gamma+K'$ with $K'\in\mathfrak{X}(M')\cap\ker\omega$ gives\footnote{We recall that  $\ker\omega\,=\,\{X\in\mathfrak X(M)\,:\,\omega(X,Y)=0\,\,\forall\,\,Y\in\,\mathfrak X(M)\}$.}  a  solution. 

In order to determine the subset $M'\subseteq M$ where the relation \eqref{ps1} has solutions, one notices that a necessary condition for that is  $$i_K\alpha=0$$ for any $K\in\ker\omega$. This condition selects $M_1\subseteq M$, so that there exists a  restriction $\Gamma_1=\Gamma\mid_{M_1}$ to $M_1$ of an element $\Gamma\in\mathfrak X(M)$ which solves \eqref{ps1}. Assume that $M_1$ is a submanifold in $M$: 
the further natural requirement that a Cauchy datum on $M_1$ evolves under the o.d.e. system without leaving $M_1$ is satisfied only if   
the solution $\Gamma_1$ is tangent to $M_1$, i.e. $\Gamma_1\in\mathfrak X(M_1)$. Such a requirement restricts the set of points in $M$ on which a meaningful solution to \eqref{ps1} exists to a suitable $M_2\subseteq M_1$, and the procedure just outlined goes along.  
This sequence of nested subsets (see \cite{GoNe79}), which we assume to be submanifolds, can be equivalently defined by (set $M_0=M$)
\beq
\label{ps1.1}
M_{s+1}=\{m\in\,M_s\,:\,i_X\alpha=0\,\quad\forall\,\,X\,\in\,\mathfrak X^{\perp}(M_s)\}
\eeq
 with 
 \beq
 \mathfrak X^{\perp}(M_s)\,=\,\{X\,\in\,\mathfrak X(M)\,:\,(i_Xi_Y\omega)_{\mid_{m\in M}}=0\,\,\forall\,\,Y\,\in\,\mathfrak X(M_s)\}.
 \label{ps1.2}
 \eeq
When such a sequence has a non trivial fixed point, i.e. there exists an integer $s$ such that $M_{s+1}=M_s$ with $\dim M_s\neq 0$, we denote it  by $M'$ and consider it as the final constraint submanifold corresponding to the pre-symplectic system $(M,\omega, \alpha)$. 
It is proven in \cite{GoNe79,hgn78} that the conditions above that select $M'$ are also sufficient to solve the equation \eqref{ps1}, and that the submanifold $M'$ obtained upon this recursion is maximal, i.e. if $N$ is a submanifold in $M$ on which \eqref{ps1} is solved, then $N\subseteq M'$.
We consider then the vector field $\Gamma\in\mathfrak X(M')$ a solution to \eqref{ps1} if, on any $m\in M'$ 
\beq
\label{ps2}
\omega(\Gamma, Y)\,=\,\alpha(Y)
\eeq
for any $Y\in\mathfrak X(M)$. The set of solutions for \eqref{ps2}, that we denote by $\mathfrak X^{\omega}(M')$,  is an affine space modelled on $\ker\omega\cap\mathfrak X(M')$.

\subsection{A Noether theorem for global dynamics on pre-symplectic systems}
\label{ss:glob}
Our analysis on the relations between symmetries and constants of the motions for the dynamics described by the vector field defined by \eqref{ps1} begins upon considering first a pre-symplectic system such that 
\beq
\label{glody0}
i_K\alpha=0
\eeq for any $K\in\ker\omega$, so that $$M'=M.$$ In such a case, we say the pre-symplectic system $(M,\omega, \alpha)$  admits  a \emph{global} dynamics.  A vector field $X\in\mathfrak X(M)$ defines an infinitesimal symmetry for $\Gamma\in\mathfrak X^{\omega}(M)$ if $[X,\Gamma]\in\ker\omega$ for any $\Gamma\in\mathfrak X^{\omega}(M)$; a function $f\in\mathcal F(M)$ is a constant of the motion for the given pre-symplectic system if $L_{\Gamma}f=0$ for any $\Gamma\in\mathfrak X^{\omega}(M)$. It is immediate to see that, if $f$ is a constant of the motion, it is  $L_Kf=0$ for any $K\in\ker\omega$.
Since $[X,\ker\omega]\subset\ker\omega$, it is also immediate to see that the set of infinitesimal symmetries is a Lie subalgebra in $\mathfrak X(M)$. 

Assume that $(M,\omega,\alpha)$ is a global dynamics on a pre-symplectic manifold. Given a vector field $X\in\mathfrak X(M)$ it is easy from \eqref{ps1} and \eqref{glody0} to prove the relation
\beq
\label{glody1}
i_{[X,\Gamma]}\omega\,=\,-L_{\Gamma}i_X\omega,
\eeq
which shows that $X$ is an infinitesimal symmetry for the given global dynamics if and only if the 1-form $i_X\omega$ is invariant along \emph{any} $\Gamma\in\mathfrak X^{\omega}(M)$. As the relation \eqref{glody1} shows, there can exist infinitesimal symmetries $X\in\mathfrak X(M)$ for a given global pre-symplectic dynamics which do not provide a function $f$ on $M$ invariant along $\Gamma$. 
Examples of infinitesimal symmetries and invariant functions for a global dynamics on a pre-symplectic system are given as follows.

Consider a constant of the motion $f\in\mathcal F(M)$, i.e. $L_{\Gamma}f=0$ for any $\Gamma\in\mathfrak X^{\omega}(M)$. From the  condition $L_Kf=0$ for any $K\in\ker\omega$ we see that a vector field $X$ on $M$ which satisfies the relation $i_X\omega=\dd f$ exists, and that the 1-form $i_X\omega$ is invariant along the integral curves of any $\Gamma$, so that $X$ (see \eqref{glody1}) is an infinitesimal symmetry for the dynamics. Moreover, one easily computes that the function given upon contracting $i_X\alpha$ vanishes, since 
$$
i_X\alpha\,=\,i_Xi_{\Gamma}\omega\,=\,-i_{\Gamma}i_X\omega\,=\,-i_{\Gamma}\dd f\,=\,-L_{\Gamma}f\,=\,0.
$$
On the contrary, assume that a vector field $X\in\mathfrak X(M)$ satisfies the condition $i_X\alpha=0$ and that a function $f$ on $M$ exists such that $i_X\omega=\dd f$. Such a vector field is called a \emph{Cartan} infinitesimal symmetry for the given global pre-symplectic dynamics. Notice that such a vector field  $X$ \emph{deserves} the name of  infinitesimal symmetry, since, from \eqref{glody1}, it is clearly $[X,\Gamma]\in\ker\omega$. 
One immediately proves further that  
\begin{align*}
&L_{\Gamma}f\,=\,i_{\Gamma}\dd f\,=\,i_{\Gamma}i_X\omega\,=\,-i_Xi_{\Gamma}\omega\,=\,-i_X\alpha\,=\,0, \\
&L_Kf\,=\,i_K\dd f\,=\,i_Ki_X\omega=0
\end{align*}
for any $K\in\ker\omega$. These relations show that $f$ is a constant of the motion for the given pre-symplectic dynamics\footnote{If $\{X_a\}_{a=1,\dots,r}$ is a set  of infinitesimal Cartan symmetries for the global dynamics $(M,\omega,\alpha)$, with $i_{X_a}\omega=\dd f_a$ and $i_{X_a}\alpha=0$, then one sees that $i_{[X_a,X_b]}\omega=\dd(L_{X_a}f_b)$ and $i_{[X_a,X_b]}\omega=0$. This reads that  infinitesimal Cartan symmetries give a Lie subalgebra in $\mathfrak X(M)$.}. 

\noindent The above  result is usually referred to (see \cite{CaRa88, dLdD96}) as the Noether theorem (and its inverse) for a pre-symplectic system having a global dynamics, and we write it as
\begin{prop}
\label{psgd} Given a pre-symplectic system $(M,\omega,\alpha)$ with a global dynamics $\Gamma$ on it, for any infinitesimal Cartan symmetry for $\Gamma$ there exists an invariant function $f$, and for any  invariant function $f$ on $M$ there exists an infinitesimal Cartan symmetry for $X\in\mathfrak X(M)$ for $\Gamma$.  
\end{prop}

Assume now  that $X$ is an infinitesimal symmetry for $\Gamma$ and that the 1-form $i_X\omega$ is exact, namely that there exists an element $f\in\mathcal F(M)$ such that $i_X\omega=\dd f$. It is immediate to see that one has in this case the relations $L_Kf=0$ for any $K\in\ker\omega$ and  $\dd(L_{\Gamma}f)=0$, i.e. $L_{\Gamma}f$ is constant on $M$. This mimicks the result of the example \ref{exemplum3}  studied within the Hamiltonian setting, and shows that not every infinitesimal symmetry corresponds to a constant of the motion.

\subsection{Symmetries and constants of the motion for pre-symplectic systems}
\label{ss:genps}

When the sequence defined by \eqref{ps1.1} and \eqref{ps1.2} has a non trivial limit given by the manifold $M'$ with $$i_{M'}\,:\,M'\,\hookrightarrow\,M,$$ one has that $$(M', \omega'=i_{M'}^*\omega, \alpha'=i_{M'}^*\alpha)$$ is (assuming that  both  ${\rm rk}\,\omega'$ and ${\rm rk}\,\omega\mid_{M'}$ are constant) is a pre-symplectic system  which clearly has a global dynamics. Together with the equation \eqref{ps2}, whose set of solutions is given by $\mathfrak X^{\omega}(M')$, one can introduce the set $\mathfrak X^{\omega'}(M')$ of elements $\Gamma\in\mathfrak X(M')$ which solve the relation
\beq
\label{ps3}
i_{\Gamma}\omega'=\alpha',
\eeq
which we equivalently write as
\beq
\label{ps3.a}
\omega'(\Gamma, Y)\,=\,\alpha(Y)
\eeq
for any $Y\in\mathfrak X(M')$. Since $\mathfrak X(M')\subseteq\mathfrak X(M)$, the comparison between the relations  \eqref{ps2} and \eqref{ps3.a} makes it evident that   
\beq
\label{ciocio}
\mathfrak X^{\omega}(M')\subseteq\mathfrak X^{\omega'}(M').
\eeq
 The notion of infinitesimal Cartan symmetry for $(M',\omega',\alpha')$ comes directly from the one introduced in the previous subsection, as a vector field $X\in\mathfrak X(M')$ such that $i_X\omega'=\dd f$ and $i_X\alpha'=0$ for a given $f\in\mathcal F(M')$; the result proven in the subsection \ref{ss:glob} above can be easily translated within this setting. The proposition \ref{psgd} can be restated as: if $X$ is an infinitesimal Cartan symmetry for $(M',\omega', \alpha')$, then $f$ is invariant along the integral curves of any element in $\mathfrak X^{\omega'}(M')$. Conversely, if an element $f\in\mathcal F(M')$ gives $L_{\Gamma}f=0$ for any $\Gamma\in\mathfrak X^{\omega'}(M')$, then there exists a vector field $X\in\mathfrak X(M')$ such that $i_X\omega'=\dd f$ and $i_X\alpha'=0$: the affine space $X+\ker\omega'$ gives the set of infinitesimal Cartan symmetries for $\mathfrak X^{\omega'}(M')$.  
 
 \begin{example}
 \label{exemplum:ps}
 We elaborate an example  from \cite{dLdD96}. Let $M=\R^6$ with $$\omega=\dd  x_1\wedge\dd x_4+\dd x_3\wedge\dd x_2$$ and $$\alpha=x_4\dd x_4-x_3\dd x_5-x_5\dd x_3$$ a pre-symplectic system. With respect to the submanifolds introduced in \eqref{ps1.1} and \eqref{ps1.2} it is immediate to compute that the vector field $\Gamma$  defined by  $i_{\Gamma}\omega=\alpha$ can be defined on the embedded $i_{M'}:M'\hookrightarrow M$ given by 
\begin{align*}
&M_1\,=\,\{m\in M\,:\,x_3=0\}, \\
&M'=M_2=M_1
\end{align*} 
with $\ker\omega\,=\,\{\del_5,\del_6\}$. It is then 
\beq
\label{chiomep}
\mathfrak X^{\omega}(M')\,=\,\{x_4\del_1+x_5\del_2+\ker\omega\}.
\eeq
One further defines  $\omega'=i_{M'}^*\omega=\dd x_1\wedge\dd x_4$  and $\alpha'=i_{M'}^*\alpha=x_4\dd x_4$ and computes  $$\ker\omega'\,=\,\{\del_2,\del_5,\del_6\}$$ with 
\beq
\label{chiomepp}
\mathfrak X^{\omega'}(M')\,=\,\{x_4\del_1+\ker\omega'\}.
\eeq
The comparison between \eqref{chiomep} and \eqref{chiomepp} give an example of the inclusion \eqref{ciocio}.  If we identify $\mathcal F(M')=\{f\in\mathcal F(M):\del_3f=0\}$, then a function $f\in\mathcal F(M')$ is a constant of the motion for the given system if and only if $$\del_1f=\del_2f=\del_5f=\del_6f=0\quad\Leftrightarrow\quad f=f(x_4).$$ The corresponding  infinitesimal Cartan symmetries are given by the set of vector fields $$\mathfrak X(M')\ni X=(\del_4f)\del_1+\ker\omega'.$$ 
 \end{example}

\noindent We conclude this description by noticing that, if $f\in\mathcal F(M')$ is invariant along the integral curves of $\mathfrak X^{\omega'}(M')$, then it is invariant also along the integral curves of $\mathfrak X^{\omega}(M')$. Moreover, one can prove (see \cite{dLdD96}) that, if $X\in\mathfrak X(M)$ such that $i_X\omega=\dd f$ for $f\in\mathcal F(M)$ and $i_X\alpha=0$, then the vector field $X$  turns to be tangent to $M'$ (i.e. $X\in\mathfrak X(M')$  and  its restriction $X_{\mid_{M'}}$ is an infinitesimal  Cartan symmetry for $(M',\omega', \alpha'))$.

\subsection*{Acknowledgments}

We acknowledge the financial support from theSpanish Ministry of Economy and Competitiveness through the Severo Ochoa Programme for Centres of Excellence in RD (SEV-2015/0554), from the Santander/UC3M Excellence Chair Programme 2019/2020.

\appendix
\section{The exterior differential calculus on a manifold}
\label{app1}
\subsection{Cartan calculus on a manifold}
We assume the reader to be familiar with the notion of (finite) $N$ dimensional (smooth) manifold $M$, (local) charts and transition maps, so that one can write the local coordinates of a point $m\in M$ as $x=(x^1,\dots,x^N)$. For a more complete  analysis of the topics of this appendix we refer to \cite{AM78, AMP82,michor-book}.

Given a point $m\in M$, the set of curves through $m$ is the set of functions $\gamma\,:\,\R\supseteq I\,\to\,M$ such that there exists an element (say $s=0$) in $I$ such that $\gamma(0)=m$. Two such curves $\gamma, \gamma'$ are identified if 
$$
\frac{\dd (f\circ\gamma)}{\dd s}\mid_{s=0}\,=\,\frac{\dd (f\circ\gamma')}{\dd s}\mid_{s=0}
$$
for any differentiable function $f:M\to\R$. 
The corresponding quotient space $T_mM$ defines the tangent space to the manifold $M$ at the point $m$. Elements of $T_mM$ are denoted by $v_{(m)}$, or equivalently by the equivalence class of curves $[\gamma]_{m}$.
The set of pairs $(m\in M, T_mM)$ is given a suitable atlas coming from the one defining $M$, so to have the tangent bundle manifold $TM$ together with the projection $\pi_{TM}\,:\,TM\to M$. Analogously, the set of pairs $(m\in M,(T_mM)^*)$ is given the structure of the cotangent bundle $T^*M$ together with the projection $\pi_{T^*M}\,:\,T^*M\,\to\,M$.

Given a smooth (paracompact, as we assumed throughout this paper) manifold $M$, the set $\mathcal F(M)$ of smooth real valued functions on $M$ is a commutative algebra with respect to the standard (local) pointwise product. The action of any element $X\in\mathfrak X(M)$, the set of  derivations for $\mathcal F(M)$ (i.e. the set of linear operators on $\mathcal F(M)$ which satisfies the Leibniz rule) can be locally represented as
\beq
\label{appe1}
X(f)\,=\,X^{j}\frac {\del f}{\del x^j}
\eeq
for any $f\in\mathcal F(M)$, with $X^j\in\mathcal F(M)$. Elements in $\mathfrak X(M)$ are called vector fields on $M$, and they are proven to give all the smooth sections $\sigma\,:\, M\,\to\,TM$ of the tangent bundle $TM$ (with $\pi_{TM}\circ\sigma\,=\,{\rm id}_M$), which can be locally written as 
\beq
\label{appe2}
\sigma_X\,:\,x\,\to\,(x, X=(X^1,\dots,X^N)).
\eeq
Notice that the composition of two derivations $X,Y\in\mathfrak X(M)$ is not a derivation, i.e. the composition of two vector fields is not a vector field. The skew-symmetrised expression $[X,Y]=X\circ Y-Y\circ X$ is indeed a vector field, with a local representation given by 
\beq
\label{appecomm}
[X,Y](f)\,=\,(X^a\del_aY^b-Y^a\del_aX^b)\del_bf.
\eeq
The vector field $[X,Y]\in\mathfrak X(M)$ is the commutator of $X,Y$. Equipped with such a bilinear operation, $(\mathfrak X(M), [~,~])$ is an infinite dimensional Lie algebra.

The set $\mathfrak X(M)$ is a (left) $\mathcal F(M)$-module. Its dual module $\Lambda^1(M)$ is the set of (differential) 1-forms on $M$, which is proven to be representable as 
\beq
\label{app7}
\Lambda^1(M)\,=\,\{f_a\,\dd g_a\,=\,(\dd g_a)f_a\,:\,f_a,g_a\in\mathcal F(M)\}
\eeq
where the differential $\dd f\,:\,\mathfrak X(M)\,\to\,\mathcal F(M)$ of a function  is defined \eqref{appe1} by 
\beq
\label{apped1}
\dd f(X)=X(f).
\eeq 
From \eqref{app7} it is easy to see that one can locally represent the set of 1-forms as $\Lambda^1(M)\,=\,\{f_a\dd x^a\}$ with $f_a\in\mathcal F(M)$ and $\dd x^a$ giving the dual basis to the set of derivations $\del_a$, i.e. $\dd x^a(\del_b)=\delta^a_b$.  Dually  to the case of the tangent bundle,  the set of 1-forms is proven to coincide with the set of sections of the cotangent bundle $T^*M$, so that one can write (compare with \eqref{appe2})
\beq
\label{appe3}
\alpha\,:\,x\,\to\,(x, \alpha=(\alpha_1,\dots,\alpha_N))
\eeq
for  $\alpha=\alpha_j\dd x^j$. Given such a 1-form, the duality between $\Lambda^1(M)$ and $\mathfrak X(M)$ is written as 
\beq
\label{appe4}
\alpha(X)\,=\,X(\alpha)\,=\,\alpha_jX^j.
\eeq
The tensor product 
\beq
\label{appe10}
(\Lambda^1(M))^{\otimes k}\,=\,\Lambda^1(M)\otimes_{\mathcal F(M)}\Lambda^1(M)\otimes_{\mathcal F(M)}\cdots\otimes_{\mathcal F(M)}\Lambda^1(M)
\eeq 
provides a $\mathcal F(M)$-bimodule, which is dual to the tensor product
\beq
\label{appe11}
(\mathfrak X(M))^{\otimes k}\,=\,\mathfrak X(M)\otimes_{\mathcal F(M)}\mathfrak X(M)\otimes_{\mathcal F(M)}\cdots\otimes_{\mathcal F(M)}\mathfrak X(M)
\eeq
which provides, since $\mathcal F(M)$ is clearly a commutative algebra, a left $\mathcal F(M)$-module after the identification $$fX_1\otimes X_2\otimes\cdots\otimes X_k=X_1\otimes fX_2\otimes\cdots\otimes X_k=X_1\otimes X_2\otimes\cdots fX_k$$
for any $f\in\mathcal F(M)$. The duality between such modules can be represented as 
\beq
\label{appe9}
(\alpha_{(1)}\otimes\alpha_{(k)})(X_{(1)}\otimes X_{(k)})\,=\,\Pi_{j=1}^k\{\alpha_{(j)}(X_{(j)})\},
\eeq
thus generalising the previous \eqref{appe4}.
The totally anti-symmetric  subset of the tensor products $(\Lambda^1(M))^{\otimes k}$ from \eqref{appe10} provides the $\mathcal F(M)$-bimodules $\Lambda^k(M)$ of exterior $k$-forms on $M$. 
The associativity of the tensor product descends to the associativity of the totally anti-symmetric wedge product\footnote{Notice that the same construction can be developped starting from the vector fields $\mathfrak (M)$. One obtains the set of multi vector fields. An example of a bivector field, as discussed in section \ref{sec:poisson}, is the Poisson tensor.} 
$$
\wedge\,:\,\Lambda^{k}(M)\,\times\,\Lambda^{k'}(M)\,\to\,\Lambda^{k+k'}(M),
$$
so one has the graded\footnote{the grading given by $k=0, 1, \dots, N$.} exterior algebra $(\Lambda(M)\,=\,\oplus_{k=0}^N\Lambda^k(M), \wedge)$, with $\Lambda^0(M)=\mathcal F(M)$ and elements having the local representation
$$
\Lambda^k(M)\,\ni\,\alpha\,=\,\frac{1}{k!}\,\alpha_{j_1\cdots j_k}\dd x^{j_1}\wedge\dd x^{j_2}\wedge\cdots\wedge \dd x^{j_k}.
$$
Given a vector field $X\in\mathfrak X(M)$, one can define the \emph{interior} product, or contraction, of an \emph{exterior} form $$i_X\,:\,\Lambda^{k}(M)\,\to\,\Lambda^{k-1}(M)$$ by the conditions
$i_Xf=0$ for any $f\in\mathcal F(M)$ and $i_X\alpha=\alpha(X)$ (see \eqref{appe4}) for a 1-form $\alpha$ in terms of the duality between 1-forms and vector fields. The operator $i_X$ is extended to $\Lambda^k(M)$ by recursion, requiring it to satisfy a graded Leibniz rule with respect to the wedge product, i.e.
\beq
\label{appe5}
i_X(\alpha\wedge\beta)\,=\,(i_X\alpha)\wedge\beta\,+\,(-1)^a\alpha\wedge i_X\beta 
 \eeq
for any $\alpha\in\Lambda^a(M)$. Equivalently, the interior product for $\alpha\in\Lambda^a(M)$ can be defined by
\beq
\label{appe12}
(i_X\alpha)(X_1,\ldots,X_{a-1})\,=\,\alpha(X,X_1, \ldots,X_{a-1}).
\eeq
The linear operator $\dd\,:\,\mathcal F(M)\to\Lambda^{1}(M)$ already defined as $f\,\mapsto\,\dd f$ is extended to the whole exterior algebra as $\dd\,:\,\Lambda^k(M)\,\to\,\Lambda^{k+1}(M)$ by requiring it to satisfy a graded Leibniz rule with respect to the wedge product, i.e. 
$$
\dd(\alpha\wedge\beta)\,=\,(\dd\alpha)\wedge\beta\,+\,(-1)^{a}\alpha\wedge \dd\beta
$$
for $\alpha\in\Lambda^{a}(M)$ and the condition $\dd^2\alpha=\dd(\dd\alpha)=0$ for any $\alpha\in\Lambda(M)$. 
Equivalently, the exterior differential can be defined on $\alpha\in\Lambda^a(M)$ by the expression
\begin{align}
\label{appe13}
\dd\alpha(X_1,\ldots,X_{a+1})&=\,\sum_{j=1}^{a+1}(-1)^jX_j(\alpha(X_1,\ldots, \hat X_j,\ldots,X_{a+1}))\, \\ &\qquad +\,\sum_{j<k}(-1)^{j+k}\alpha([X_j,X_k], X_1,\ldots,\hat X_j, \ldots, \hat X_k,\ldots, X_{a+1})
\end{align}
where the terms $\hat X_j$ are missing.
The set $$(\Lambda(M), \wedge, \dd)$$ is the differential graded de Rham algebra on $M$. 
The contraction operator $i_X$ and the de Rham differential $\dd$ allow to define the Lie derivative $L_X\,:\,\Lambda^k(M)\,\to\,\Lambda^k(M)$ of an exterior form by the formula
\beq
\label{appe6}
L_X\alpha\,=\,i_X\dd\alpha\,+\,\dd(i_X\alpha).
\eeq
Equipped with such operators, the set
$$
(\Lambda(M), \wedge, \dd, i_X, L_X)
$$
is referred to as the exterior algebra with the Cartan exterior differential calculus. Notice that the following identities have been often implicitly used in the main text:
\begin{align}
&L_Xf\,=\,i_X(\dd f); \nn \\
&L_X\dd\alpha\,=\,\dd(L_X\alpha); \nn \\
&i_{[X,Y]}\alpha\,=\,L_Xi_Y\alpha-i_YL_X\alpha 
\label{appe7}
\end{align}
for any $X,Y\in\mathfrak X(M)$, any $f\in\mathcal F(M)$  and any exterior form $\alpha\in\Lambda(M)$.

\subsection{Tangent maps and Pull-backs} Let $$\phi\,:\,M\,\to\,M'$$ be a smooth map between the $N$-dimensional manifold $M$ and the $N'$-dimensional smooth manifold $M'$. The pull-back  $\phi^*\,:\,\mathcal F(M')\,\to\,\mathcal F(M)$ associated to $\phi$ is defined by 
$$
\phi^*f'\,=\,f'\circ\phi
$$
for any $f'\in\mathcal F(M')$. Upon setting the conditions \begin{align}
&\phi^*(\alpha'\wedge\beta')=(\phi^*\alpha')\wedge(\phi^*\beta'), \nn \\ &\dd\circ\phi^*=\phi^*\circ\dd \label{appe14}
\end{align} for any $\alpha',\beta'\in\Lambda(M')$, one extends the pull-back as $$\phi^*\,:\,\Lambda^k(M')\,\to\,\Lambda^k(M).$$ 
Fix now a point $m\in M$, with $m'=\phi(m)\in M'$. Each smooth curve $\gamma$ with $\gamma(0)=m$ is mapped under $\phi$ into the smooth curve $\gamma'=\phi\circ\gamma$ with $\gamma'(0)=m'$. It is immediate to prove that $\phi$ induces a meaningful map $T\phi\,:\,(m, T_mM)\,\to\,(\phi(m), T_{\phi(m)}M)$ such that the action on the tangent vector space $$T_m\phi\,:\,T_mM\,\to\,T_{\phi(m)}M'$$ is linear, defined  by $[\gamma]_{m}\,\mapsto\,[\gamma']_{\phi(m)}$. Such a map is usually referred to as the \emph{tangent} map at $m$ associated to $\phi$. Its action is represented, with respect to local charts,  by the Jacobian matrix of $\phi$ at each point $m\in M$. Although the tangent map $T\phi$ is well defined at each point $m\in M$, it does not necessarily map a vector field $X\in\mathfrak X(M)$ into a vector field $X'\in\mathfrak X(M')$. Given $X\in\mathfrak X(M)$, if an element $X'\in\mathfrak X(M')$ exists such that 
\beq
\label{appe15}
X(\phi^*f')\,=\,\phi^*(X'f')
\eeq
for any $f'\in\mathcal F(M')$, then we say that $X'$ is $\phi$-\emph{related} to $X$, or that $\phi_*X=X'$. The map $\phi_*$ is also called  \emph{push-forward} associated to $\phi$. When $\alpha'\in\Lambda^k(M')$, then the action of the  element $\phi^*\alpha'\in\Lambda^k(M)$ can be written as
$$
(\phi^*\alpha')(X_1,\ldots, X_k)\,=\,\alpha'(\phi_*X_1,\ldots,\phi_*X_k)
$$
only if the vector fields $X_j$ on $M$ have a $\phi$-related vector field $X_j'=\phi_*X_j$ on $M'$.

\section{Distributions and foliations. The Frobenius theorem}
\label{app2}
\noindent In order to introduce the Frobenius theorem, we recall the notions of \emph{distributions} and \emph{foliations} on a manifold, which we have extensively used in the paper, and refer the reader to chapter 18 in \cite{mssv85} for a more complete analysis of the subject.

Let $M$ be a smooth $N$-dimensional manifold. A distribution $\mathcal D$ on it is a mapping assigning to each point $m\in M$ a vector subspace $D_{(m)}\subset T_mM$. 
The dimension of $\mathcal{D}_{(m)}$ gives the \emph{rank} of $\mathcal{D}$ in $m$. Within this paper we have focussed on constant rank $k$  (smooth) distributions, namely those distributions $\D$ whose rank does not depend on $m$ (for the more general case we refer also  to \cite{mks_natural,michor-book}) and such that  for each open neighbourhood $U\subseteq M$ there exists a set $\{X_j\}_{j=1,\ldots,k}$ of vector fields on $U$ such that $\D_{(m)}\,=\,(m, {\rm span}\{X_j\}_{j=1,\ldots,k})$ at each $m\in U$. Equivalently, a smooth distribution can be given as  the intersection 
\beq
\label{appe202}
\ker\alpha_{1}\cap\ker\alpha_2\cap\ldots\cap\ker\alpha_{N-k}
\eeq
for a suitable set $\{\alpha_i\}_{i=1,\ldots,N-k}$ of 1-forms on $U\subseteq M$. 

Recall that by a smooth map $\phi\,:\,Q\,\to\,M$ is defined to be an \emph{immersion} at the point $q\in Q$ if the tangent map $T_q\phi:T_qQ\,\to\,T_{\phi(q)}M$ is injective (which requires $\dim Q\leq \dim M$). If $\phi:Q\to M$ is an immersion at a point $q\in Q$, then it is immediate to see that there exists an open subset $V\subseteq Q$ such that $\phi$ is an immersion at each point in $V$. An immersion is said \emph{global} if the map $T_q\phi$ is injective at each point $q\in Q$. An immersion which is also an homeomorphism (i.e. it has a continuous inverse) on its image is called an \emph{embedding}. For a local immersion $\phi:V\subseteq Q\,\to\,\phi(V)\subseteq M$ it is possible to prove that local coordinate chart $\{x^j\}_{j=1,\ldots,k}$ for $V$ and $\{y^s\}_{s=1,\ldots,N}$ for $\phi(V)$ exist, such that the action of the restriction $\phi_{\mid V}$ can be written as 
\beq
\label{appe200}
\phi_{\mid V}\,:\,(x^1,\ldots,x^k) \mapsto (y^1=x^1,\dots,y^k=x^k, y^{k+1}=0,\ldots, y^N=0) 
\eeq
with $\dim Q=k$ and $\dim M=N$. If $Q$ is a submanifold of $M$, then a local immersion $i\,:\,Q\,\hookrightarrow\,M$ (which depends on the submanifold $Q$, and in such a respect it is therefore usually called \emph{canonical}) can always be defined. 

Given a smooth distribution $\D$ in $M$, a submanifold  $Q$ with the immersion $i\,:\,Q\,\hookrightarrow\,M$ is called an integral submanifold for $\D$ if, for any $q\in Q$,  the range of the tangent map 
$T_qi\,:\,T_qQ\,\to\,T_{i(q)}M$ is a linear subset included in $\D_{(i(q))}$. When the tangent map $T_qi$ at each point $q\in Q$ transforms $T_qQ$ surjectively (and then bijectively, the map $i$ being an immersion) onto $\D_{i(q)}$, the submanifold $Q$ is called an integral manifold of maximal dimension for  the distribution $\D$.   It is then natural to define a distribution $\D$ \emph{completely integrable} if for each point $m\in M$  there exists a submanifold $i:Q\hookrightarrow M$ with $m\in i(Q)$ which is an integral submanifold of maximal dimension for $\D$. 

Strictly related with the notion of completely integrable distribution one has the notion of \emph{foliation}. In order to describe what a foliation is, we start by recalling that, given the smooth $N$-dimensional manifold $M$ and the smooth $k$ dimensional manifold $Q$,  a smooth map $\phi\,:\,M\,\to\,Q$ is a (local) \emph{submersion} at $m\in M$ if the tangent map $T_m\phi\,:\,T_mM\,\to\,T_{\phi(m)}Q$ is surjective (notice that this is possible only if $N\geq k$).  If such a condition holds for any element $m\in M$, then the foliation is said \emph{global}. For a local submersion $\phi:U\subseteq M\,\to\,\phi(U)\subseteq Q$ it is possible to prove that local coordinate chart $\{x^j\}_{j=1,\ldots,N}$ for $U$ and $\{y^s\}_{s=1,\ldots,k}$ for $\phi(U)$ exist, such that the action of the restriction $\phi_{\mid U}$ can be written as 
\beq
\label{appe201}
\phi_{\mid V}\,:\,(x^1,\ldots,x^N) \mapsto (y^1=x^1,\dots,y^k=x^k). 
\eeq 
A set $\{S_\alpha\}_{\alpha\in J}$ (with $J$ an index set) of disjoint connected subsets in $M$, one passing through each point $m\in M$,  such that the map $i_{\alpha}\,:\,S_{\alpha}\hookrightarrow M$ is an embedding, gives a \emph{foliation} of $M$. Each element $S_{\alpha}$ is a leaf of such a foliation. For each $m\in S_{\alpha}$ there exist an integer $k$,  an open subset $U\subset M$ and a local chart $\{x^j\}_{j=1,\ldots,N}$ such that there exists a map  $\pi\,:\,U\cap S_{\alpha}\,\to\,\R^{N-k}$ given by $m=(x^1,\ldots, x^N)\,\mapsto\,c=(c^{k+1},\ldots,c^N)$ with constant $c\in\R^{N-k}$, which turns to be a submersion. Each intersection $U\cap S_{\alpha}$ is identified with the inverse image $\pi^{-1}(c)$ for a suitable $c\in\R^{N-k}$. 
 If the integer $k$ does not depend on $\alpha$, then the foliation is said  \emph{regular}, and each leave is a $k$ dimensional manifold embedded in $M$. 
 
 One can then say that a regular foliation defines a local submersion. 
The implicit function theorem allows to prove that, if $\phi:M\to Q$ is a (local) submersion, the inverse image $\phi^{-1}(q)\subset M$ is, with $q$ any element within the range of $\phi$, a submanifold in $M$. This reads that a submersion defines a regular foliation.  

The notion of foliation is strictly related to that of distribution. If $\{S_\alpha\}_{\alpha\in J}$ is a regular foliation in $M$, then 
$$\D_{(m)}\,=\,T_m\mathcal{S}_\alpha \,\subset\, T_mM$$
clearly gives a completely integrable distribution in $M$. On the other hand, to every completely integrable distribution is associated a foliation, namely the set of all its integral manifold. The Frobenius theorem characterises completely integrable distributions. 

In this setting, the following theorem characterizing a completely integrable distribution holds.
\begin{prop}
\label{prop.frob}
Given a smooth $N$-dimensional manifold $M$, the regular $k$-dimensional distribution generated by the set of vector fields $\{X_1,\ldots, X_k\}$ is completely integrable if and only if it is \emph{involutive}, i.e. the commutator $[X_a,X_b]$ 
is in the span of the vector fields $\{X_j\}_{j=1,\ldots,k}$ for any $a,b\in1,\ldots,k$.
\end{prop}
\noindent If the distribution $\D$ is given as in \eqref{appe202} in terms of the 1-forms $\{\alpha_j\}_{j=1,\ldots,N-k}$, then the Frobenius theorem can be written as follows.
\begin{prop}
\label{prop.frofor}
 Given a smooth $N$-dimensional manifold $M$, the regular $k$ dimensional distribution $\D$ spanned by the vector fields $X$ such that $i_X\alpha_j=0$ for a suitable set $\{\alpha_j\}_{j=1,\ldots,N-k}$ of 1-forms is integrable if and only if one of the following equivalent conditions holds:
\begin{itemize}
\item $\dd\alpha_j \wedge \theta \,=\, 0 \;\;\; \forall \,\,j=1,...,N-k$, with $\theta=\alpha_1 \wedge\dots \wedge \alpha_{N-k}$;
\item there exists a one form $\alpha$  such that $\dd\theta=  \alpha \wedge \theta$;
\item there exists a set $\{\,\beta^i_j\}_{i,j = 1,\ldots,N-k}$ of 1-forms,  such that $\dd\alpha_j \,=\, \beta^i_j \wedge \alpha_i$;
\item there exist elements $\{\,f_i\}_{i = 1,\ldots,N-k}$ and $\{g_i^j\}_{i, j = 1,\ldots,N-k}$ in $\mathcal F(M)$ such that one can locally write $\alpha_i \,=\, g_i^j \dd f_j$.
\end{itemize}
\end{prop}


\end{document}